\documentstyle[rep12]{report}
\begin{document}
\large
\begin{center}{\large\bf EXACT CONSTRUCTION OF THE
ELECTROMAGNETIC CURRENT OPERATOR FOR RELATIVISTIC COMPOSITE
SYSTEMS}\end{center}
\vskip 1em \begin{center} {\large Felix M. Lev} \end{center}
\vskip 1em \begin{center} {\it Laboratory of Nuclear
Problems, Joint Institute for Nuclear Research, Dubna, Moscow region
141980 Russia (E-mail:  lev@nusun.jinr.dubna.su)} \end{center}
\vskip 1em
\tableofcontents
\newpage
\vskip 1em {\it Abstract:}
\vskip 0.5em
The electromagnetic current operator of a composite system must
be a relativistic vector operator satisfying current conservation,
cluster separability and the condition that interactions between the
constituents do not renormalize the total electric charge.
Assuming that these interactions are described in the framework of
relativistic quantum mechanics of systems with a fixed number of
particles we explicitly construct the current operators satisfying the
above properties in cases of two and three particles and prove that
solutions exist for any number of particles. The consideration is
essentially based on the method of packing operators in the point form of
relativistic dynamics developed by Sokolov. Using the method developed by
Sokolov and Shatny we also construct the current operators in the instant
and front forms and prove that the corresponding results are physically
equivalent. The paper is self-contained what makes it possible for the
reader to learn both, relativistic quantum mechanics of systems with a
fixed number of particles and the problem of constructing the
electromagnetic current operator for such systems.

\chapter{Introduction}
\label{C1}
\section{The statement of the problem}
\label{S1.1}

 The present and future experiments on powerful electron
accelerators will yield an important information about nuclear forces
at small distances and the quark structure of matter. However to
extract this information from the experimental data it is necessary
to know the structure of the electromagnetic current operator (ECO)
for the system under consideration. The problem of constructing the
ECO is essentially model-dependent since the ECO depends on the
interactions between the constituents and on the way of introducing
these interactions into the ECO. Nevertheless any model for the ECO
must necessarily be such that the ECO is a relativistic vector
operator satisfying current conservation, cluster separability
and the condition that interactions between the constituents comprising
our system do not renormalize the total electric charge.
Let us formulate these conditions mathematically.

 The wave function of any relativistic system must transform
according to a unitary representation of the Poincare group in some
Hilbert space $H$. Let ${\hat U}(a)=exp(\imath {\hat
P}_{\mu}a^{\mu})$ be the representation operator corresponding to the
displacement of the origin in spacetime translation of Minkowski
space by the 4-vector $a$. Here ${\hat P}=({\hat P}^0,{\hat {\bf
P}})$ is the operator of the 4-momentum, ${\hat P}^0={\hat E}$ is the
Hamiltonian, and ${\hat {\bf P}}$ is the operator of ordinary
momentum. Let also ${\hat U}(l)$ be the representation operator
corresponding to $l\in SL(2,C)$ and $L(l)$ be the Lorentz
transformation corresponding to $l$. Then the ECO
${\hat  J}^{\mu}(x)$ must be the selfadjoint relativistic vector
operator such that
\begin{eqnarray}
&&{\hat U}(a)^{-1}{\hat  J}^{\mu}(x){\hat U}(a)=
{\hat J}^{\mu}(x-a),\nonumber\\
&& {\hat U}(l)^{-1}{\hat J}^{\mu}(x){\hat U}(l)=L(l)^{\mu}_{\nu}
{\hat J}^{\nu}(L(l)^{-1}x)
\label{1}
\end{eqnarray}
where a sum over repeated indices $\mu,\nu=0,1,2,3$ is assumed. Since
at least some of the operators ${\hat U}(a)$ and ${\hat U}(l)$ depend
on interactions in the system under consideration, the immediate
consequence of Eq. (\ref{1}) is that ${\hat  J}^{\mu}(x)$ also depends
on these interactions and thus ${\hat  J}^{\mu}(x)$ cannot be written
only as a sum of the constituent ECO's. This fact was first pointed out
by Siegert~\cite{Sieg}.

 Another important condition is the continuity equation or current
conservation
\begin{equation}
  \frac{\partial {\hat J}^{\mu}(x)}{\partial x^{\mu}}=0
\label{2}
\end{equation}

\begin{sloppypar}
 The cluster separability condition for ${\hat J}^{\mu}(x)$ can be
formulated along with this condition for the representation
operators. Namely, it is assumed that the representation space is the
same in all cases (not depending on whether interactions are present
or not) and if $\alpha_1,...\alpha_n$ are arbitrary subsystems
comprising the system under consideration and all interactions
between these subsystems are turned off then the representation
${\hat U}$ must become the tensor product of the representations
${\hat U}_{\alpha_i}$ describing the subsystems and
${\hat J}^{\mu}(x)$ must become a sum of the ECO's
${\hat J}_{\alpha_i}^{\mu}(x)$ for the subsystems. In other words the
cluster separability condition for the current operator is formulated
in full analogy with this condition for the generators of the
representation ${\hat U}$ (for more details see refs.
\cite{sok1,sok2,CP,mutze,lev1}).
\end{sloppypar}

 Let
\begin{equation}
\hat Q=\int\nolimits {\hat J}^{\mu}(x)d\sigma_{\mu}(x)
\label{3}
\end{equation}
be the system electric charge operator where $d\sigma_{\mu}(x)=
\lambda_{\mu}\delta(\lambda x-\tau)d^4x$ is the volume element of the
space-like hypersurface defined by the time-like vector $\lambda
\quad (\lambda^2=1)$ and the evolution parameter $\tau$. Then the
important physical condition is that the interactions do not
renormalize the electric charge, i.e. ${\hat Q}$ does not depend on
the choice of $\lambda$ and $\tau$ and has only one eigenvalue
equal to the sum of electric charges of constituents. It is well
known that Eq. (\ref{2}) ensures that ${\hat Q}$ does not depend on
$\tau$ and $\lambda$ but this condition does not ensure that
${\hat Q}$ has the same value as for noninteracting particles.

 In addition, if we consider the problem of constructing the ECO for
a system of strongly interacting particles, then we must require that
\begin{eqnarray}
&&{\hat U}_P({\hat J}^0(x^0,{\bf x}),{\hat {\bf J}}(x^0,{\bf x}))
{\hat U}_P^{-1}=({\hat J}^0(x^0,-{\bf x}),-{\hat {\bf J}}(x^0,-{\bf
x})),\nonumber\\
&&{\hat U}_R{\hat J}^{\mu}(x){\hat U}_R^{-1}={\hat J}^{\mu}(-x)
\label{3a}
\end{eqnarray}
where ${\hat U}_P$ is the unitary operator corresponding to the space
reflection $P$ and ${\hat U}_R$ is the antiunitary operator
corresponding to the space-time reflection $R=PT$. The operator
${\hat U}_R$ must be antiunitary according to Wigner~\cite{Wig1}; an
alternative definition of ${\hat U}_R$ proposed by Schwinger
\cite{Sch} involves transposed operators, but we shall not discuss
this question.

\section{Brief review of the literature on constructing
the ECO and related topics}
\label{S1.2}

 In the framework of local quantum field theory there exist the well
known methods of introducing electromagnetic interactions into the
system of strongly interacting quantized fields (for example, the
minimal substitution method). Each electromagnetic process is described
by many Feynman diagrams including those obtained from some diagrams of
the strong interaction theory by coupling the virtual or real photons to
all possible places of the latter diagrams . To explicitly construct the
ECO in such an approach it is necessary to explicitly construct first the
underlying theory of strong interactions. Therefore a realistic
construction of the ECO can be carried out only at some additional
assumptions.

 A vast amount of literature is devoted to the problem of
constructing the ECO for few nucleon systems assuming that the
underlying theory of strong interactions is the meson-nucleon theory.
It is clear that the most detailed study can be carried out for
two-nucleon systems. As it has been shown by several authors (see for
example, refs.~\cite{Bentz,GrRi,NaKo,Ohta,KorSh,Anis}, the corresponding
ECO satisfying the properties specified in Sec.~\ref{S1.1} can be
constructed assuming that the interaction between the nucleons is
described by the Bethe-Salpeter (BS) or other covariant equation. This
is achieved by choosing the longitudinal part of the ECO in the form
satisfying the two-particle Ward-Takahashi identity. It has been also
shown that the knowledge of the kernel of the two-body equation is not
sufficient for a unique determination of the ECO since the transversal
part of the ECO remains unconstrained by current conservation.

 The ECO considered in the above references cannot be written only
as a sum of one-nucleon ECO's and necessarily contains the
contribution of interaction currents usually associated with meson
exchange currents (MEC). These currents may give contributions to
different observables even in the nonrelativistic approximation. The
first detailed study of MEC was carried out apparently in
ref.~\cite{GarHu}. The present status of nonrelativistic MEC is
described in refs.~\cite{Ris,GrHen} and in references cited therein.
The role of MEC in three and four-nucleon systems is discussed in
refs.~\cite{HenSa,UnkHof} and in references cited therein. The problem
of reformulating the theory of MEC in the framework of constituent quark
model is discussed in ref. \cite{Gian} and in references cited therein.

 In the nonrelativistic case the interaction between  nucleons is
described by a nucleon-nucleon potential. By analogy with QED and
other realistic theories it is reasonable to assume that the
potential description of nucleon-nucleon interactions exists not only
in the nonrelativistic approximation but at least to order $(v/c)^2$.
A detailed investigation of MEC in first order in $(v/c)^2$ was
carried out in ref.~\cite{Friar}. The present status of this theory has
been described in refs.~\cite{TsuRi} and in references cited
therein. A fully covariant approach to MEC has been considered in
refs.~\cite{HuTj1,HuTj2}. A detailed investigation of all possible
contributions to the deuteron photodisintegration to order $(v/c)^3$
has been carried out in refs.~\cite{Adam,ArSa,GolAr} and in
references cited therein. The latest development in the description
of the deuteron electrodisintegration is described in
refs.~\cite{Tamura,MoRi,MoPaRi} and in references cited therein. The
photo and electrodisintegration of three-nucleon systems have been
considered so far only in the plane wave impulse approximation
{}~\cite{KleBr,SchSa}.

 The approach in which the problem of constructing the ECO is tackled
starting from the underlying theory of strong interactions is of
course most fundamental. However this approach encounters
considerable practical difficulties. For example, using the two-body BS
equation we have either to assume some dependence of the kernel of this
equation on the relative time (or relative energy) or to use some
three-dimensional reduction of this equation. Meanwhile it is not
clear how to control the validity of approximations used to solve the
BS equation and even the crucial property of relativistic invariance
may be lost at some approximations. Let us note that in local quantum
field theory any description of an interacting system can be
consistent only in terms of infinite number of degrees of freedom
since the number of particles is not the conserving physical
quantity. In particular the Hilbert space $H$ necessarily contains
subspaces with all possible numbers of particles, and the
representation operators of the Poincare group necessarily have
transitions between these subspaces. Therefore any truncation of the
Hilbert space leaving only a finite number of particles necessarily
breaks relativistic invariance.

 For systems of three and more particles the description in terms
of the BS equation becomes much more complex. In particular the cluster
separability condition is not trivial in this case. One might think
that in nuclear physics different observables at intermediate energies
can be reliably calculated using the $v/c$ expansion, though, as noted by
some authors (see, for example, refs.~\cite{Giusti,CoGlLee}), the
convergence of the $v/c$ expansion may be poor. However, as shown in
refs.~\cite{GodIs,CaIs}, the unified description of the meson and
baryon spectra is possible only in fully relativistic constituent
quark models. Let us also note that for systems of constituent quarks
the underlying theory of strong interactions is "soft" QCD, i.e. the
theory which is not elaborated so far. For this reason even in
calculations of the electromagnetic form factor of the pion (considered
as a system of two constituent quarks) the interaction currents are
usually not taken into account (see, for example,
refs.~\cite{IsSm,JaKi,Ito,KrTr}) and the electromagnetic properties of
baryons as systems of three constituent quarks are also described
assuming that the baryon ECO is the sum of the quark ECO's \cite{Gian}.

 It is often reasonable to assume that interactions between constituents
can be reliably described in the framework of relativistic quantum
mechanics (RQM) of systems with a fixed number of particles.  This theory
is expected to be a reasonable approach in different applications. The
well known examples are the
atom consisting of a fixed number of electrons, the nucleus
consisting of a fixed number of nucleons, and the meson and baryon
consisting of a fixed number of constituent quarks. The first results in
RQM was obtained in the work by Dirac \cite{Dir} and the important
contribution was made by Coester, Sokolov and other scientists. The
reviews of RQM can be found, for example, in refs. \cite{KP,lev2}.
The practical advantage of RQM is that in models based on this theory
such fundamental properties as relativistic invariance and cluster
separability are satisfied automatically.

 The statement of the problem considered in the present work can now
be formulated as follows. We have to investigated whether the ECO
satisfying the properties described above can be explicitly
constructed in the framework of RQM. In the remainder of this section
we discuss some results of RQM and the literature on constructing the
ECO in this theory.

 The problem of constructing the operator ${\hat J}^{\mu}(x)$ in the
framework of RQM was investigated by several authors. The general
properties of the ECO and the constraints imposed on the matrix element
of ${\hat J}^{\mu}(x)$ by the above conditions has been studied in detail
in Ref. \cite{pol1}.  As noted by different authors \cite{O1,CO,C,pol1},
the above conditions are not sufficient for a unique determination of
${\hat J}^{\mu}(x)$, (in agreement with the results obtained in the
framework of local quantum field theory) but these conditions are
necessary, since a consideration of electromagnetic processes using
the current operator not satisfying these conditions is, generally
speaking, inconsistent.  The above conditions are also necessary in
order to be sure that the low-energy theorem for the Compton
scattering and the Drell-Hearn-Gerasimov sum rules are
satisfied\cite{O2,Ger,DH}.

 In the literature there exists a rather complete investigation of the
problem under consideration in first order in $1/c^2$
\cite{BP,ClCo,ClOs,FK,Fr}. For example,
it has been known for a long time that the nontrivial effect in this
order is the appearance of the nonadditive Foldy-Wauthuysen
corrections. It has been also shown by Osborn and
Foldy\cite{OF} and Coester and Ostebee\cite{CO} that (in agreement
with the results by Gross\cite{G} obtained in the framework of
quantum field theory) interaction terms indeed may be present already
in the nonrelativistic current-density operator as was first noted in
ref.~\cite{Sieg}.

 In the nonrelativistic approximation the charge-density operator
${\hat J}^{0}(x)$ is written in zero order in $1/c$ and the
current-density operator ${\hat {\bf J}}(x)$ in first order. In order
$1/c^2$ the first relativistic corrections to ${\hat J}^0(x)$ are
taken into account. As noted above, this approximation is
insufficient to describe the existing data in nuclear and particle
physics, and in the literature there have been proposed different
approaches for calculating relativistic effects in electromagnetic
processes to orders more high than $1/c^2$. For example, in
refs.\cite{FS,GrKo,CCKP,GrKoFS} the deuteron electromagnetic form
factors at large momentum transfer were calculated using the front
form of dynamics and the assumption that some matrix elements can be
calculated with a good accuracy using the additivity of the +
components of the current operator. A similar approach was used for
deriving the light-front sum rules \cite{Keister} and for calculations of
the pion and nucleon electromagnetic form factors and the photoproduction
of the delta isobar in the framework of constituent quark model
\cite{CCP1,CC,W}.

 The approach of refs.~\cite{FS,GrKo,CCKP,GrKoFS} was criticized in
refs.~\cite{KS} since, as shown in these references, the expressions
for the form factors used in refs.~\cite{FS,GrKo,CCKP,GrKoFS} contained
the admixture of nonphysical terms depending on the direction in
which the observer passes to the infinite-momentum-frame. It was proposed
to use instead the expressions which do not contain this dependence but
again no interaction terms were taken into account. As the result, the
authors of refs.~\cite{KS} have come to the prescription
proposed in ref.~\cite{Gl}. To substantiate neglecting the interaction
terms in their expressions the authors of refs.~\cite{KS} consider some
class of Feynman diagrams and show that these terms can be present only
in the contribution of multiparticle states while the contribution of the
two-nucleon component in the deuteron does not contain the interaction
terms. However, as noted above, it is difficult to justify the
two-nucleon approximation in the diagrammatic approach.

 In the just described calculations no expansion in powers of $1/c$ was
used. On the other hand, in refs.\cite{WS1,CL,WS2,LC,CAP1,CAP2,Li} the
elastic electromagnetic nucleon form factors, the electro and
photoproduction of nucleon resonances and the nucleon Compton
scattering were calculated in constituent quark model taking into account
some interaction terms, but the current operator was expanded only to
order $1/c^3$. In ref.~\cite{lev3} the problem has been solved to
order $1/c^4$ but the resulting expressions have turned out to be
very cumbersome.

 In the present paper we explicitly find a class of exact solutions
for systems of two and three particles and prove that solutions exist
for any number of particles. Therefore the results can be applied to
the description of various electromagnetic processes involving the
lightest nuclei, mesons (as systems of two constituent quarks) and
baryons (as systems of three constituent quarks).

 As pointed out by Dirac \cite{Dir}, any physical system can be
described in different forms of relativistic dynamics. Let
${\hat M}^{\mu\nu}$ (${\hat M}^{\mu\nu}=-{\hat M}^{\nu\mu}$) be the
generators of the Lorentz group. We use $P$ and $M^{\mu\nu}$ to
denote the 4-momentum operator and the generators of the Lorentz
group in the case when all interactions are turned off. By definition,
the description in the point form implies that the operators
${\hat U}(l)$ are the same as for noninteracting particles, i.e.
${\hat U}(l)=U(l)$ and ${\hat M}^{\mu\nu}=M^{\mu\nu}$, and thus
interaction terms can be present only in the 4-momentum operators
${\hat P}$ (i.e. in the general case ${\hat P}^{\mu}\neq P^{\mu}$ for
all $\mu$). The description in the instant form implies that the
operators of ordinary momentum and angular momentum do not depend on
interactions, i.e. ${\hat {\bf P}}={\bf P}$, ${\hat {\bf M}}={\bf M}$
$({\hat {\bf M}}=({\hat M}^{23},{\hat M}^{31},{\hat M}^{12}))$ and
therefore interactions may be present only in $\hat E$ and the
generators of the Lorentz boosts ${\hat {\bf N}}=({\hat M}^{01},
{\hat M}^{02},{\hat M}^{03})$. In the front form with the marked $z$
axis we introduce the + and - components of the 4-vectors as $p^+=
(p^0+p^z)/\sqrt{2}$, $p^-=(p^0-p^z)/\sqrt{2}$. Then we require that
the operators ${\hat P}^+,{\hat P}^j,{\hat M}^{12},{\hat M}^{+-},
{\hat M}^{+j}$ $(j=1,2)$ are the same as the corresponding free
operators and therefore interaction terms may be present only in the
operators ${\hat M}^{-j}$ and ${\hat P}^-$. We see that the front form
contains 3 generators depending on interactions while the point and
instant form contain 4 such generators. However, in the front form
the operators ${\hat U}_P$ and ${\hat U}_R$ necessarily depend on
interactions while in the point and instant forms we can choose
representations with ${\hat U}_P=U_P$ and ${\hat U}_R=U_R$.

 The most popular forms are the instant and front ones since it is
clear how these forms are related to quantum field theory (see,
for example, refs.~\cite{KP,lev2}). In the case of the point form the
relation is less obvious \cite{sok3}. However from the group theory
point of view the point form looks as most natural. In the literature
different authors argue in favor of one or other form. However, as
proved by Sokolov and Shatny \cite{SoSh} all the three basic forms
are unitarily equivalent (unfortunately this important result is not
widely known). Therefore the choice of the form is only the matter of
convenience but not the matter of principle.

 Recently Karmanov ~\cite{Karm} and Fuda
{}~\cite{Fuda1} considered in detailed such a version of the front form
where the direction of the motion of the infinite momentum frame is
defined by the arbitrary unit vector ${\mbox{\boldmath $\xi$}}$. In
ref.~\cite{Fuda1} it is proved that the "$\xi$-picture" is unitarily
equivalent to the ordinary front form at least for systems of two
particles and there are all grounds to believe that this is also the
case for any number of particles.

 We shall see below that the natural solution for the ECO exists in the
point form. The method of packing operators was first
developed by Sokolov also in the point form \cite{sok2}. Then this method
was used in other forms and the full solution of the N-body problem in
RQM was first given by Coester and Polyzou~\cite{CP} and Mutze
\cite{mutze} in the instant form. Therefore, using the construction
of ref. \cite{SoSh} it can be shown that the full solution of the N-body
problem in RQM and the solution for the ECO exist in all forms.

 In Ref. \cite{sok2} the explicit expressions for the three-particle
packing operators were demonstrated on the example when one of the
particles has spin 1/2 and two others are spinless. The explicit
expressions for particles with arbitrary spins in the point form were
derived in ref.~\cite{lev1} (see also ref.~\cite{lev2}), but the
derivation was rather complicated since the explicit expressions for
the Lorentz group generators in the light cone variables were used.
Let us note that in refs.~\cite{CP,mutze,lev4} the expressions
for the packing operators were also derived using the expressions for
the Lorentz group generators.

\section{The outline of this paper}
\label{S1.3}

 In Secs.~\ref{S2.2} and ~\ref{S2.3} following the ideas of
ref.~\cite{sok2} we show that the expressions for the three-particle
packing operators in the
point form for particles with arbitrary spins can be obtained in a much
more simple way than in previous publications if one works directly
with the operators $U(l)$ and not with the generators of the Lorentz
group. The main results of Chap.~\ref{C2} are derived explicitly,
and therefore this chapter can be used even for the first
acquaintance with RQM. The general properties of the ECO are analyzed
in Chap.~\ref{C3}, and in Chap.~\ref{C4}, using the results
of Chaps.~\ref{C2} and~\ref{C3} we explicitly describe a
class of solutions when the ECO satisfies the properties specified in
Sec.~\ref{S1.1}. In Chap.~\ref{C5}, using the results of
ref.~\cite{SoSh} we construct the ECO in the instant form assuming
that the ECO in the point form has been already constructed. In
Chap.~\ref{C6} a similar procedure is used for constructing the ECO
in the front form.

 As it has been already mentioned, the explicit expressions for the
representation generators are not needed to solve the problem.
Nevertheless it is useful to write down these expressions in some
cases. In the general case the commutation relations for the
representation generators can be realized in the form
$$[{\hat P}^{\mu},{\hat P}^{\nu}]=0, \quad [{\hat M}^{\mu\nu},
{\hat P}^{\rho}]= -\imath({\eta}^{\mu\rho}{\hat P}^{\nu}-
{\eta}^{\nu\rho}{\hat P}^{\mu}),$$
\begin{equation}
[{\hat M}^{\mu\nu},{\hat M}^{\rho\sigma}]=-\imath ({\eta}^{\mu\rho}
{\hat M}^{\nu\sigma}+{\eta}^{\nu \sigma}{\hat M}^{\mu\rho}-{\eta}^
{\mu\sigma}{\hat M}^{\nu\rho}-{\eta}^{\nu\rho}{\hat M}^{\mu\sigma})
\label{4}
\end{equation}
where $\mu,\nu,\rho,\sigma=0,1,2,3$, the metric tensor in Minkowski
space has the nonzero components $\eta^{00}=-\eta^{11}=-\eta^{22}=
-\eta^{33}=1$, and we use the system of units with $\hbar=c=1$.

\chapter{Systems with a fixed number of particles in the point form of
relativistic dynamics}
\label{C2}
\section{Systems of two particles}
\label{S2.1}

  To describe a relativistic system of interacting particles it is
necessary to choose first the explicit form of the unitary irreducible
representation (UIR) of the Poincare group describing an elementary
particle of mass $m>0$ and spin s. There are many equivalent ways to
construct an explicit realization of such a representation~\cite{Wig,Nov}.
For our purposes it is convenient to choose the realization in the
following form.

 Let $p$ be the particle 4-momentum, $g=p/m$ be the particle 4-velocity,
${\bf s}$ be the spin operator, ${\cal D}({\bf s})$ be the space of the
UIR of the group SU(2) with the spin ${\bf s}$, $||...||$ be the norm in
${\cal D}({\bf s})$ and
\begin{equation}
d\rho (g)=\frac{d^3{\bf g}}{2(2\pi)^3g^0}=
\frac{d^3{\bf p}}{2(2\pi)^3m^2\omega({\bf p})}
\label{5}
\end{equation}
where $\omega({\bf p})=(m^2+{\bf p}^2)^{1/2}$ and $g^0=(1+{\bf
g}^2)^{1/2}$ (since $g^2=1$, only three components of $g$ are
independent). Then the Hilbert space $H$ can be chosen as the space
of functions $\varphi(g)$ with the range in ${\cal D}(\bf s)$ and such
that
\begin{equation}
(\varphi,\varphi)=\int\nolimits ||\varphi(g)||^2d\rho(g)\quad < \infty
\label{6}
\end{equation}

 Let $\alpha(g)\in$ SL(2,C) be the matrix
\begin{equation}
\alpha(g)=\frac{g^0+1+{\mbox {\boldmath $\sigma$} }{\bf g}}
{[2(g^0+1)]^{1/2}}
\label{7}
\end{equation}
where ${\mbox {\boldmath $\sigma$} }$
are the Pauli matrices, $D[{\bf s};u]$ is the representation operator of
the group SU(2) corresponding to the element $u\in$SU(2) for the
representation with the generators ${\bf s}$, and $g'=L(l)^{-1}g$.
Then if an element of the Poincare group is defined by $a$ and $l$
then the corresponding representation operator acts as follows
\begin{equation}
U(a,l)\varphi(g)=exp(\imath mg'a)D[{\bf s};\alpha(g)^{-1}l\alpha(g')]
\varphi(g')
\label{8}
\end{equation}
(it is easy to verify that $\alpha(g)^{-1}l\alpha(g')\in$SU(2)).

 The generators of the UIR given by Eq. (\ref{8}) have the form
\begin{equation}
P=mg,\qquad {\bf M}={\bf l}({\bf g})+{\bf s}, \qquad
{\bf N}=-\imath g^0\frac{\partial}{\partial {\bf g}}+
\frac{{\bf s}\times{\bf g}}{1+g^0}
\label{9}
\end{equation}
where ${\bf g}$ is the operator of multiplication by ${\bf g}$ and
${\bf l}({\bf g})=-\imath {\bf g}\times (\partial / \partial {\bf g})$
is the orbital angular-momentum operator.

 Let us now consider a system of two free particles with the masses
$m_i>0$ and the spin operators ${\bf s}_i$ $(i=1,2)$. As follows from
cluster separability (see Sec.~\ref{S1.1}) the representation describing
the free system (1,2) must be the tensor product of the single-particle
representations for particles 1 and 2. Therefore the representation
space $H$ is the space of functions $\varphi(g_1,g_2)$ with the range in
the tensor product ${\cal D}({\bf s}_1)\bigotimes {\cal D}({\bf s}_2)$
and such that
\begin{equation}
\int\nolimits ||\varphi(g_1,g_2)||^2d\rho(g_1)d\rho(g_2)\quad < \infty
\label{10}
\end{equation}
where $||...||$ is the norm in ${\cal D}({\bf s}_1)\bigotimes
{\cal D}({\bf s}_2)$ and, as follows from Eq. (\ref{8}), the operators of
this representation act as
\begin{eqnarray}
&&U(a,l)\varphi(g_1,g_2)=exp[\imath (m_1g_1'+m_2g_2')a]\cdot\nonumber\\
&&\prod_{i=1}^{2}D[{\bf s}_i;\alpha(g_i)^{-1}
l\alpha(g_i')]\varphi(g_1',g_2')
\label{11}
\end{eqnarray}

 As follows from Eq. (\ref{11}) the total 4-momentum is now equal to
$P=m_1g_1+m_2g_2$. By analogy with the single-particle case we define
the system mass and 4-velocity as $M=|P|$ and $G=PM^{-1}$
respectively where the modulus of a 4-vector is taken in the the Lorentz
metric, i.e. as $|P|=(P^2)^{1/2}$. It is easy to
see that $M>0$. Let us also define the 4-vectors $q_i$ $(i=1,2)$ as
\begin{equation}
q_i=L[\alpha(G)]^{-1}m_ig_i
\label{12}
\end{equation}
It is easy to verify that ${\bf q}_1=-{\bf q}_2$, i.e.
${\bf q}_i$ are the momenta of the particles in their c.m.frame.

 The above expressions define ${\bf G}$ and ${\bf q}\equiv {\bf q}_1$
as the functions of ${\bf g}_i$ $(i=1,2)$. Conversely, the ${\bf g}_i$
are uniquely defined by ${\bf G}$ and ${\bf q}$:
\begin{equation}
m_i{\bf g}_i={\bf q}_i+{\bf G}\omega_i({\bf q})+
\frac{({\bf G}{\bf q}_i){\bf G}}{1+G^0}
\label{13}
\end{equation}
where $\omega_i({\bf q})=(m_i^2+{\bf q}^2)^{1/2}$, $G^0=(1+{\bf
G}^2)^{1/2}$. It is easy to see that $M$ as the function of ${\bf q}$
is equal to $M=\omega_1({\bf q})+\omega_2({\bf q})$. A direct
calculation using Eq. (\ref{13}) shows that
\begin{eqnarray}
&&d\rho(g_1)d\rho(g_2)=d\rho(G)d\rho(int),\nonumber\\
&&d\rho(int)=\frac{M({\bf q})^3d^3{\bf q}}{2(2\pi)^3\omega_1({\bf q})
\omega_2({\bf q})m_1^2m_2^2}
\label{14}
\end{eqnarray}

 Let us define the "internal" Hilbert space $H_{int}$ as the space of
functions $\chi=\chi({\bf q})$ with the range in ${\cal D}({\bf s}_1)
\bigotimes {\cal D}({\bf s}_2)$ and such that
\begin{equation}
||\chi||^2=\int\nolimits ||\chi({\bf q})||^2 d\rho(int)\quad < \infty
\label{15}
\end{equation}
We also define the Hilbert space ${\tilde H}$ as the space of
functions ${\tilde \varphi}(G)$ with the range in $H_{int}$ and such that
\begin{equation}
\int\nolimits ||{\tilde \varphi}(G)||^2_{int} d\rho (G)\quad <\infty
\label{16}
\end{equation}
As easily follows from Eqs. (\ref{15}) and (\ref{16}), the operator
\begin{equation}
U_{12}=U_{12}(G,{\bf q})=\prod_{i=1}^{2} D[{\bf
s}_i;\alpha(g_i)^{-1}\alpha(G)\alpha(q_i/m_i)]
\label{17}
\end{equation}
is the unitary operator from ${\tilde H}$ to $H$.

 Our nearest goal is to determine the explicit form of the unitary
representation ${\tilde U}$ in ${\tilde H}$ such that
$U(a,l)=U_{12}{\tilde U}(a,l)U_{12}^{-1}$. The form of the operators
${\tilde U}(a)$ corresponding to the spacetime translations is obvious
from Eq. (\ref{11}) since $U_{12}$ commutes with $U(a)$:
\begin{equation}
{\tilde U}(a){\tilde \varphi}(G)=exp[\imath M(Ga)]{\tilde \varphi}(G)
\label{18}
\end{equation}

 To determine the form of the operators ${\tilde U}(l)$ let us note
that if $g'_i=L(l)^{-1}g_i$, then, as follows from the definition of G
as the function of $g_1$ and $g_2$: $G(g'_1,g'_2)\equiv G'=L(l)^{-1}G$.
Taking into account that $\alpha(G)^{-1}l\alpha(G')\in$SU(2) we
get from Eq. (\ref{12}) that the vector functions ${\bf q}'\equiv
{\bf q}(g'_1,g'_2)$ and ${\bf q}={\bf q}(g_1,g_2)$ are related to
each other by a three-dimensional rotation:
\begin{equation}
{\bf q}'=L[\alpha(G')^{-1}l^{-1}\alpha(G)]{\bf q}
\label{19}
\end{equation}

 Let us define the unitary representation of the group SU(2) in
$H_{int}$ as follows. If $u\in$SU(2) then
\begin{equation}
R(u)\chi({\bf q})=\chi[L(u)^{-1}{\bf q}]
\label{20}
\end{equation}
It is well known that ${\bf l}({\bf q})$ are just the generators of
the representation defined by Eq. (\ref{20}) and therefore $R(u)=
D[{\bf l}({\bf q});u]$. If $\varphi=\varphi(g_1,g_2)\in H$,
${\tilde \varphi}={\tilde \varphi}(G)\in {\tilde H}$ and
$\varphi=U_{12}{\tilde \varphi}$, then, as follows from
Eqs. (\ref{11}), (\ref{17}), (\ref{19}) and (\ref{20})
\begin{eqnarray}
&&U(l)\varphi(g_1,g_2)=U_{12}\prod_{i=1}^{2}D[{\bf s}_i;
\alpha(q_i/m_i)^{-1}\alpha(G)^{-1}\cdot\nonumber\\
&&\cdot l\alpha(G')\alpha(q'_i/m_i)] D[{\bf l}
({\bf q});\alpha(G)^{-1}l\alpha(G')]{\tilde \varphi}(G')
\label{21}
\end{eqnarray}
Now we take into account the well known property (which can be
verified directly) that if $u\in$SU(2) and $g$ is some 4-vector such
that $g^2=1$ then
\begin{equation}
u\alpha(g)=\alpha(L(u)g)u
\label{22}
\end{equation}
Using this property and taking into account Eq. (\ref{19}) we derive
from Eq. (\ref{21}) the final expression for the action of ${\tilde U}(l)$
in ${\tilde H}$:
\begin{equation}
{\tilde U}(l){\tilde \varphi}(G)=D[{\bf S};\alpha(G)^{-1}l\alpha(G')]
{\tilde \varphi}(G')
\label{23}
\end{equation}
where ${\bf S}={\bf l}({\bf q})+{\bf s}_1+{\bf s}_2$.

 Comparing Eqs. (\ref{18}) and (\ref{23}) on the one hand and
Eq. (\ref{8}) on the other we see that indeed $M$ can be interpreted as
the mass of the two-particle system, G as its 4-velocity and
$H_{int}$ as the internal space of the two-particle system since the
operators ${\tilde U}(a,l)$ in ${\tilde H}$ have the same form as the
single-particle generators in Eq. (\ref{8}) if $m$ is replaced by $M$,
$g$ by $G$ and ${\bf s}$ by ${\bf S}$. Therefore the generators
${\tilde \Gamma}^i$ $(i=1,2...10)$ of the representation ${\tilde U}$
in ${\tilde H}$ can be immediately written by analogy with Eq. (\ref{9})
\begin{equation}
{\tilde P}=MG,\qquad {\tilde {\bf M}}={\bf l}({\bf G})+{\bf S}, \qquad
{\tilde {\bf N}}=-\imath G^0\frac{\partial}{\partial {\bf G}}+
\frac{{\bf S}\times{\bf G}}{1+G^0}
\label{24}
\end{equation}
Here $M$ is the operator of multiplication by the function $M({\bf
q})$. This operator and the operator ${\bf S}$ act only through the
variables of the space $H_{int}$.

 Since the representation $U$ in $H$ is the tensor product of the
single-particle representations, the generators of U are sums of the
corresponding single-particle generators. We conclude that the sums
of the single-particle generators can be written in the form of
Eq. (\ref{24}) only for the case of spinless particles since only in
this case $U_{12}=1$. In the general case, if $\Gamma^i$ are the
generators of the representation $U$ then
\begin{equation}
\Gamma^i=U_{12}{\tilde \Gamma}^iU_{12}^{-1}
\label{25}
\end{equation}

 If particles 1 and 2 interact with each other then cluster
separability (see Sec.~\ref{S1.1}) implies that the
representation ${\hat U}$ describing the system (1,2) should be
constructed in such a way that it becomes $U$ when the interaction is
turned off. The condition that ${\hat U}$ is the representation of
the Poincare group is equivalent to the condition that the generators
${\hat \Gamma}^i$ $(i=1,2...10)$ of this representation satisfy the
commutation relations in the form of Eq. (\ref{4}). In the point form
the simplest way of introducing the interaction into the two-particle
system is to write ${\hat \Gamma}^i=U_{12}{\hat {\tilde \Gamma^i}}
U_{12}^{-1}$ where ${\hat {\tilde \Gamma^i}}$ are the generators
defined by Eq. (\ref{24}) but with $M$ replaced by a selfadjoint
operator ${\hat M}^{int}$ in $H$ such that ${\hat M}^{int}$ commutes
with ${\bf S}$ and becomes $M$ when the interaction is turned off.
Therefore, as follows from Eq. (\ref{24}), the representation
generators in $H$ have the form
\begin{eqnarray}
&&{\hat P}={\hat M}G,\qquad {\bf M}=U_{12}[{\bf l}({\bf G})+{\bf
S}]U_{12}^{-1}, \nonumber\\
&&{\bf N}=U_{12}[-\imath G^0\frac{\partial}{\partial {\bf G}}+
\frac{{\bf S}\times{\bf G}}{1+G^0}]U_{12}^{-1}
\label{26}
\end{eqnarray}
where ${\hat M}=U_{12}{\hat M}^{int}U_{12}^{-1}$.

 A similar way of introducing the interaction into the two-particle
system was first proposed by Bakamdjian and Thomas \cite{BT} in the
instant form. The Bakamdjian-Thomas (BT) procedure has been discussed
in detail in refs.~\cite{lev2,lev5} where it has been shown that in
the general case interaction terms should be also introduced into the
operator $U_{12}$, i.e. we should replace $U_{12}$ by a unitary
operator ${\hat U}_{12}=A_{12}U_{12}$ while the choice $A_{12}=1$ can
be substantiated only for particles of equal mass. Since the most
interesting applications of the results concerning the ECO are
expected to be in cases when constituents have equal masses (in
nuclear physics we usually can neglect the difference of masses
between the proton and the neutron and in constituent quark models
the same can be assumed for the masses of the $u$ and the $d$
quarks), we shall consider in this paper only the case $A_{12}=1$.

 We can write ${\hat M}^{int}=M+v$, and if the interaction operator
$v$ is an integral operator then its action can be written as (see
Eq. (\ref{14}))
\begin{equation}
v\chi({\bf q})=\int\nolimits v({\bf q},{\bf q}')\chi({\bf q}')
\frac{M({\bf q}')^3d^3{\bf q}'}{2(2\pi)^3\omega_1({\bf
q}')\omega_2({\bf q}')}
\label{27}
\end{equation}
where the kernel $v({\bf q},{\bf q}')$ is an operator acting only
through the spin variables.

\begin{sloppypar}
\section{Sokolov's method of packing operators for systems of three
particles}
\label{S2.2}
\end{sloppypar}

 Let us consider a system of three particles with the masses $m_i>0$
and the spin operators ${\bf s}_i$ $(i=1,2,3)$. The three-particle
Hilbert space $H$ is the tensor product of the single-particle spaces
$H_i$. Now we supply the operators relating to the subsystem $ij$
with the corresponding index, for example ${\hat P}_{ij}$, ${\hat
M}_{ij}$, ${\bf q}_{ij}$ etc. We use $P$ to denote the 4-momentum
operator when all the three particles do not interact with each
other, $P_{(ij)k}$ to denote this operator when only particles $i$
and $j$ interact with each other ($i\neq j\neq k$) and ${\hat P}$ to
denote this operator when all the three particles interact with each
other. The corresponding mass operators will be written as $M$,
$M_{(ij)k}$ and $\hat M$ and the corresponding 4-velocity operators
as $G$, $G_{(ij)k}$ and $\hat G$. As follows from cluster
separability and Eqs. (\ref{9}) and (\ref{26}), $P_{(ij)k}=
{\hat M}_{ij}G_{ij}+m_kg_k$ and therefore
\begin{equation}
M_{(ij)k}=|{\hat M}_{ij}G_{ij}+m_kg_k|,\qquad
G_{(ij)k}=\frac{{\hat M}_{ij}G_{ij}+m_kg_k}
{|{\hat M}_{ij}G_{ij}+m_kg_k|}
\label{28}
\end{equation}
while
\begin{eqnarray}
&&M=|M _{ij}G_{ij}+m_kg_k|=|m_ig_i+m_jg_j+m_kg_k|,\nonumber\\
&&G=\frac{M _{ij}G_{ij}+m_kg_k}{|M _{ij}G_{ij}+m_kg_k|}=
\frac{m_ig_i+m_jg_j+m_kg_k}{|m_ig_i+m_jg_j+m_kg_k|}
\label{29}
\end{eqnarray}

 In the point form the operators ${\hat U}(a)=exp[\imath({\hat P}a)]$
and $U(l)$ define a unitary representation of the Poincare group or
in other words, the operator $\hat P$ and the generators of the
representation $l\rightarrow U(l)$ must form the system of
generators satisfying the commutation relations (\ref{4}). In
addition, as follows from cluster separability, the operator
$\hat P$ must become $P_{(ij)k}$ when all interactions involving
particle $k$ are turned off.

 By analogy with the way of introducing the interaction in
Sec.~\ref{S2.1} and with ordinary quantum mechanics one might think
that a possible way of writing the three-body mass operator is to
represent it in the form
\begin{equation}
{\hat M}=M+(M_{(12)3}-M)+(M_{(13)2}-M)+(M_{(23)1}-M)
\label{30}
\end{equation}
since this operator indeed becomes $M_{(ij)k}$ when all interactions
involving any particle $k$ are turned off. However Eq. (\ref{30})
cannot be correct since it leads to the breaking of the commutation
relations (\ref{4}). The matter is that $M$ commutes with $G$,
$M_{(ij)k}$ commutes with $G_{(ij)k}$ and all the operators $G$,
$G_{(12)3}$, $G_{(13)2}$, $G_{(23)1}$ differ each other.

 The idea of the Sokolov method of packing operators~\cite{sok2} is
that the composition of interactions in the three-body mass operator
should involve only mass operators commuting with one and the same
4-velocity operator, for example with $G$. To realize this idea
Sokolov notes first that, as follows from physical considerations,
all the operators $G$ and $G_{(ij)k}$ should have the same spectrum
and therefore these operators should be unitarily equivalent to each
other even if the operators ${\hat M}_{ij}$ and $M_{ij}$ have
different spectra. Let $A_{ij,k}$ be such a unitary operator that
\begin{equation}
G_{(ij)k}=A_{ij,k}GA_{ij,k}^{-1}
\label{31}
\end{equation}
Then the operators
\begin{equation}
M_{ij,k}=A_{ij,k}^{-1}M_{(ij)k}A_{ij,k}
\label{32}
\end{equation}
commute with $G$. Suppose also that the unitary operator $A$ is
constructed from the operators $A_{12,3}$, $A_{13,2}$ and $A_{23,1}$
in such a way that it becomes $A_{ij,k}$ when all interactions
involving particle $k$ are turned off. Then we can construct the
mass operator $\hat M$ as
\begin{eqnarray}
&&{\hat M}=AM_{123}A^{-1},\quad M_{123}=M+(M_{12,3}-M)+\nonumber\\
&&+(M_{13,2}-M)+(M_{23,1}-M)+V_{123}
\label{33}
\end{eqnarray}
where $V_{123}$ is a three-particle interaction operator which
commutes with $G$, with $U(l)$ and becomes zero if all interactions
involving any of three particles are turned off.

 The difference between Eqs. (\ref{30}) and (\ref{33}) is that the
"auxiliary" mass operators $M_{ij,k}$ entering into the expression
for the "auxiliary" mass operator $M_{123}$ commute with one and the
same 4-velocity operator $G$. The operator $\hat M$ in Eq. (\ref{33})
satisfies cluster separability since if all interactions
involving any particle $k$ are turned off then $M_{123}$ becomes
$M_{ij,k}$, $A$ becomes $A_{ij,k}$ and therefore, as follows from
Eq. (\ref{32}), $\hat M$ becomes $M_{(ij)k}$. At the same time, if the
operators $A_{ij,k}$ and $A$ commute with $U(l)$ then the commutation
relations and cluster separability will be satisfied if
\begin{equation}
{\hat P}={\hat M}{\hat G},\qquad {\hat G}=AGA^{-1}
\label{34}
\end{equation}

 This statement can be proved if we note the following. The set
($P_{(ij)k}$,${\bf M}$,${\bf N}$), where ${\bf M}$ and ${\bf N}$ are the
generators of the three-particle representation $U(l)$, satisfies the
commutation relations (~\ref{4}) by construction. Therefore the set
($P_{ij,k}=M_{ij,k}G,{\bf M},{\bf N}$) also satisfies these relations
as follows from Eqs. (\ref{31}), (\ref{32}) and the fact that $A_{ij,k}$
commutes with ${\bf M}$ and ${\bf N}$. In turn this means that
$M_{ij,k}$ commutes with $G$, ${\bf M}$ and ${\bf N}$. Therefore, as
follows from Eq. (\ref{33}), $M_{123}$ also commutes with $G$, ${\bf M}$
and ${\bf N}$. In turn this means that the set
($P_{123}=M_{123}G$,${\bf M}$,${\bf N}$) satisfies the conditions
(\ref{4}). Therefore the set (${\hat P}={\hat M}G$,${\bf M}$,${\bf N}$)
also satisfies these conditions since $A$ commutes with
${\bf M}$ and ${\bf N}$. This set satisfies cluster separability by
construction since when all interactions involving any particle $k$ are
turned off then $\hat P$ becomes $P_{(ij)k}$.

 We see that the sense of the operators $A_{ij,k}$ is that they
"pack" the operators $M_{(ij)k}$ to the operators $M_{ij,k}$ (see
Eq. (\ref{32})). The latter enter into the expression determining the
composition of interactions in Eq. (\ref{33}). That is why Sokolov
referred to his method as that of packing operators.

 Following Sokolov~\cite{sok2} we shall seek the operators $A_{ij,k}$
in the form
\begin{equation}
A_{ij,k}=U_{ij}B_{ij,k}({\hat M}_{ij}^{int})^{-1}B_{ij,k}(M_{ij})
U_{ij}^{-1}
\label{35}
\end{equation}
where the unitary operators $B_{ij,k}({\hat M}_{ij}^{int})$ and
$B_{ij,k}(M_{ij})$ commute with the operators of the tensor product
${\tilde U}_{ij}\bigotimes U_k$ reduced on SL(2,C). As usual, the
dependence on a selfadjoint operator is understood in the sense of
the spectral decomposition. Namely, if ${\hat e}_{ij}^{int}(m)$ is
the spectral function of the operator ${\hat M}_{ij}^{int}$ and
$e_{ij}$ is the spectral function of the operator $M_{ij}$ then
\begin{eqnarray}
&&B_{ij,k}({\hat M}_{ij}^{int})=
\int\nolimits B_{ij,k}(m)d{\hat e}_{ij}^{int}(m),\nonumber\\
&&B_{ij,k}(M_{ij})=
\int\nolimits B_{ij,k}(m)de_{ij}(m)
\label{36}
\end{eqnarray}
Here the integrals are understood as the strong limits of the
corresponding Riemann sums. The operators (\ref{36}) are correctly
defined and are unitary if the operators $B_{ij,k}(m)$ are unitary and
commute with ${\hat e}_{ij}^{int}(m')$ and $e_{ij}(m')$ for all $m$
and $m'$ belonging to the spectra of the operators
${\hat M}_{ij}^{int}$ and $M_{ij}$.

 As follows from Eqs. (\ref{26}), (\ref{28}) and (\ref{29})
\begin{eqnarray}
&&G_{(ij)k}=U_{ij}\{\int\nolimits G(G_{ij},g_k,m)
d{\hat e}_{ij}^{int}(m)\}U_{ij}^{-1},\nonumber\\
&&G=U_{ij}\{\int\nolimits G(G_{ij},g_k,m)de_{ij}(m)\}U_{ij}^{-1}
\label{37}
\end{eqnarray}
where
\begin{equation}
G(G_{ij},g_k,m)=\frac{mG_{ij}+m_kg_k}{|mG_{ij}+m_kg_k|}
\label{38}
\end{equation}
Then, as follows from Eqs. (\ref{35}-\ref{37}), Eq. (\ref{31}) will be
satisfied if the operator
\begin{equation}
C(G_{ij},g_k)=B_{ij,k}(m)G(G_{ij},g_k,m)B_{ij,k}(m)^{-1}
\label{39}
\end{equation}
does not depend on $m$. In refs.~\cite{sok2} there were explicitly
considered the cases $C(G_{ij},g_k)=g_k$ and $C(G_{ij},g_k)=G_{ij}$
and in ref.~\cite{lev1} there was explicitly considered an infinite
number solutions corresponding to different choices of
$C(G_{ij},g_k)$. It has been shown that only the solution with
$C(G_{ij},g_k)=G_{ij}$ has physical sense. Therefore the problem of finding
the operators $A_{ij,k}$ will be solved if we succeed in finding the
unitary operators $B_{ij,k}(m)$ commuting with
${\hat e}_{ij}^{int}(m')$, $e_{ij}(m')$, with the operators
${\tilde U}_{ij}\bigotimes U_k$ reduced on SL(2,C) and such that
\begin{equation}
B_{ij,k}(m)^{-1}C(G_{ij},g_k)B_{ij,k}(m)=G(G_{ij},g_k,m)
\label{40}
\end{equation}
\vskip 1em
\section{Explicit expressions for the packing operators and the
operators $M_{ij,k}$}
\label{S2.3}

 As follows from Eqs. (\ref{11}) and (\ref{23}), the tensor product of
the representations ${\tilde U}_{ij}$ and $U_k$ reduced on SL(2,C) is
realized in the space of functions $\varphi(G_{ij},g_k)$ with the
range in $H_{ij}^{int}\bigotimes {\cal D}({\bf s}_k)$ and such that
\begin{equation}
\int\nolimits ||\varphi(G_{ij},g_k)||^2d\rho(G_{ij})d\rho(g_k)\quad <
\infty
\label{41}
\end{equation}
where the norm is taken in the space $H_{ij}^{int}\bigotimes
{\cal D}({\bf s}_k)$. The operators of this representation act as
\begin{eqnarray}
&&[{\tilde U}_{ij}(l)\bigotimes U_k(l)] \varphi(G_{ij},g_k)=
D[{\bf S}_{ij};\alpha(G_{ij})^{-1}l\alpha(G_{ij}')]\cdot \nonumber\\
&&\cdot D[{\bf s}_k;\alpha(G_k)^{-1}l\alpha(G_k')]\varphi(G_{ij}',g_k')
\label{42}
\end{eqnarray}
where $G_{ij}'=L(l)^{-1}G_{ij}$, $g_k'=L(l)^{-1}g_k$ and ${\bf
S}_{ij}$ is the spin operator of the system $ij$ defined in
Sec.~\ref{S2.1}.

 We introduce the 4-vector
\begin{equation}
K_{ij}(m)=K_{ij}(G_{ij},g_k,m)=L[\alpha (G(m))]^{-1}mG_{ij}
\label{43}
\end{equation}
where for simplicity we write $G(m)$ instead of $G(G_{ij},g_k,m)$.
The vector defined by Eq. (\ref{43}) has the sense of the 4-momentum of
the system $ij$ with the mass $m$ in the c.m.frame of the system
consisting of $ij$ as the subsystem and particle $k$ as the other
subsystem (compare with Eq. (\ref{11})). We use $K_{ij}'(m)$ to denote
the quantity $K_{ij}(G_{ij}',g_k',m)$. By analogy with Eq. (\ref{19})
it follows from Eq. (\ref{43}) that the quantities $K_{ij}(m)$ and
$K_{ij}'(m)$ are related to each other by a usual three-dimensional
rotation:
\begin{equation}
K_{ij}(m)=L[\alpha (G(m))^{-1}l\alpha (G'(m))]K_{ij}'(m)
\label{44}
\end{equation}
where $G'(m)=L(l)^{-1}G(m)$.

 Instead of $G_{ij}$ and $g_k$ we can choose as independent variables
$G(m)$ and $g_k$. Then a direct calculation using Eq. (\ref{38}) yields
{}~\cite{sok2}
\begin{eqnarray}
&&d\rho(G(m))=J(G_{ij},g_k,m)d\rho(G_{ij}),\nonumber\\
&&J(G_{ij},g_k,m)=\frac{m^3[m+m_k(G_{ij},g_k,m)]}{|mG_{ij}+m_kg_k|^4}
\label{45}
\end{eqnarray}
Now using Eqs. (\ref{43})-(\ref{45}) it is easy to verify that the
operator $B_{ij,k}(m)^{-1}$ defined as
\begin{eqnarray}
&&B_{ij,k}(m)^{-1}\varphi(G_{ij},g_k)=J(G_{ij},g_k,m)^{1/2}
D[{\bf S}_{ij};\alpha(G_{ij})^{-1}\cdot\nonumber\\
&&\cdot \alpha(G(m))\alpha(K_{ij}(m)/m)]\varphi(G(m),g_k)
\label{47}
\end{eqnarray}
is indeed unitary and commutes with the operators
${\tilde U}_{ij}(l)\bigotimes U_k(l)$ defined by Eq. (\ref{42}).

Let us introduce the 4-vectors \cite{sok2}
\begin{eqnarray}
&&H(m)=H(G_{ij},g_k,m)=\frac{1}{m}\{G_{ij}[m_k(G_{ij},g_k)+\nonumber\\
&&+(m^2+m_k^2(G_{ij},g_k)^2-m_k^2)^{1/2}]-m_kg_k\},\nonumber\\
&&R(m)=R(G_{ij},g_k,m)=L[\alpha(G_{ij})]^{-1}mH(m)
\label{48}
\end{eqnarray}
The operations $G_{ij}\rightarrow H(m)$ and $G_{ij}\rightarrow G(m)$
are inverse to each other since it is easy to verify that
$$H(G(m),g_k,m)=G(H(m),g_k,m)=G_{ij}.$$ Therefore using Eq.
(\ref{47}) it is easy to verify that
\begin{eqnarray}
&&B_{ij,k}(m)\varphi(G_{ij},g_k)=J(H(m),g_k,m)^{-1/2}
D[{\bf S}_{ij};\alpha (R(m)/m)^{-1}\cdot\nonumber\\
&&\alpha(G_{ij})^{-1}\alpha(H(m)/m)]\varphi(H(m),g_k)
\label{49}
\end{eqnarray}

 Finally, it is easy to verify that the operators defined by
Eqs. (\ref{47}) and (\ref{49}) satisfy Eq. (\ref{40}). The action of these
operators through the variables of the space $H_{ij}^{int}$ is fully
defined by the operator ${\bf S}_{ij}$ and the quantities $G_{ij}$ and
$g_k$.  Since ${\bf S}_{ij}$, $G_{ij}$ and $g_k$ commute with ${\hat
M}_{ij}^{int}$ and $M_{ij}$ (see Sec.~\ref{S2.1}) then $B_{ij,k}(m)$
indeed commutes with ${\hat e}_{ij}^{int}(m')$ and $e_{ij}(m')$ for
all $m$ and $m'$. Therefore we conclude that the operators
$B_{ij,k}(m)^{-1}$ and $B_{ij,k}(m)$ defined by Eqs. (\ref{47})
and (\ref{49}) satisfy all the necessary conditions.

 The three-particle representation of the Poincare group is
realized in the tensor product of the three-particle spaces, i.e. the
Hilbert space $H$ is now the space of functions
$\varphi(g_1,g_2,g_3)$ with the range in ${\cal D}({\bf
s}_1)\bigotimes {\cal D}({\bf s}_2)\bigotimes {\cal D}({\bf s}_3)$ and
such that
\begin{equation}
\int\nolimits ||\varphi(g_1,g_2,g_3)||^2 \prod_{i=1}^{3} d\rho(g_i)
\quad <\quad \infty
\label{50}
\end{equation}
Instead of the variables $g_1$,$g_2$,$g_3$ we introduce the variables
$G$,$k_1$,$k_2$,$k_3$, where $G$ id defined by Eq. (\ref{29}) and the
$k_i$ $(i=1,2,3)$ are formally defined as $q_i$ in Eq. (\ref{12}) but
for the case of three-particles:
\begin{equation}
q_i=L[\alpha(G)]^{-1}m_ig_i,
\label{51}
\end{equation}
while Eq. (\ref{12}) should be rewritten as
\begin{equation}
k_i^{(ij)}=L[\alpha(G_{ij})]^{-1}m_ig_i
\label{52}
\end{equation}
and ${\bf q}_{ij}$ should be understood as the spatial part of
$q_i^{(ij)}$. Now $q_i^{(ij)}$ has the sense of the 4-momentum of
particle $i$ in the c.m.frame of the system $ij$ and the quantities
$k_i$ are the 4-momenta in the c.m.frame of the three-particle
system. These 4-momenta are not independent since ${\bf k}_1+{\bf
k}_2 +{\bf k}_3=0$ as it should be. A direct calculation using
Eqs. (\ref{29}) and (\ref{51}) yields
\begin{eqnarray}
&&\prod_{i=1}^{3}d\rho(g_i)=d\rho(G)d\rho(int),\qquad
\rho(int)=2(2\pi)^3M^3\cdot\nonumber\\
&&\cdot \delta^{(3)}({\bf k}_1+{\bf k}_2+{\bf k}_3)
\prod_{i=1}^{3}d\rho(k_i/m_i)
\label{53}
\end{eqnarray}
where the mass of the three-particle system is expressed in terms
of the $k_i$ as $M=\omega_1({\bf k}_1)+\omega_2({\bf k}_2)+
\omega_3({\bf k}_3)$.

 In the three-particle case we introduce $H_{int}$ as the space of
functions $\chi=\chi({\bf k}_1,{\bf k}_2,{\bf k}_3)$
with the range in ${\cal D}({\bf s}_1)\bigotimes {\cal D}({\bf s}_2)
\bigotimes {\cal D}({\bf s}_3)$ and such that
\begin{equation}
||\chi||^2=\int\nolimits ||\chi({\bf k}_1,{\bf k}_2,{\bf k}_3)||^2
d\rho(int)\quad < \infty
\label{54}
\end{equation}
We also introduce the Hilbert space ${\tilde H}$ as the space of
functions ${\tilde \varphi}(G)$ with the range in $H_{int}$ and such that
\begin{equation}
\int\nolimits ||{\tilde \varphi}(G)||^2_{int} d\rho (G)\quad <\infty
\label{55}
\end{equation}
(formally this expression looks as Eq. (\ref{16}), but it is clear that
in the cases of two and three particles the spaces $H_{int}$ are
different). Then, as easily follows from Eqs. (\ref{50}) and
(\ref{53}-\ref{55}), the operator
\begin{equation}
U_{123}=\prod_{i=1}^{3} D[{\bf s}_i;\alpha(g_i)^{-1}\alpha(G)
\alpha(k_i/m_i)]
\label{56}
\end{equation}
is the unitary operator from ${\tilde H}$ to $H$.

 Since the vectors $k_i$ are not independent, it is often convenient
to choose in $H_{int}$ any two independent vectors, for example ${\bf
K}_{ij}$ and ${\bf k}_{ij}$ where ${\bf K}_{ij}$ is the spatial part
of the 4-vector $K_{ij}=k_i+k_j$ and ${\bf k}_{ij}$ is the spatial part
of the 4-vector
\begin{equation}
k_i^{(ij)}=L[\alpha(K_{ij}/M_{ij})]^{-1}k_i
\label{57}
\end{equation}
As follows from Eq. (\ref{51}), $K_{ij}$ coincides with the 4-vector
introduced in Eq. (\ref{43}) for the case $m=M_{ij}$. A direct
calculation shows that instead of Eq. (\ref{53}) the volume element
$d\rho(int)$ in the variables ${\bf K}_{ij}$ and ${\bf k}_{ij}$ has
the form
\begin{equation}
d\rho(int)=\frac{M^3M_{ij}d^3{\bf K}_{ij}d^3{\bf k}_{ij}}
{4(2\pi)^6E_{ij}\omega_i({\bf
k}_{ij})\omega_j({\bf k}_{ij})\omega_k({\bf K}_{ij})m_1^2m_2^2m_3^2}
\label{58}
\end{equation}
where $M_{ij}=\omega_i({\bf k}_{ij})+\omega_j({\bf k}_{ij})$,
$E_{ij}=(M_{ij}^2+{\bf K}_{ij}^2)^{1/2}$, $M=E_{ij}+
\omega_k({\bf K}_{ij})$. Let us note that the dependence of $M_{ij}$
on ${\bf k}_{ij}$ is the same as on ${\bf q}_{ij}$ (see
Sec.~\ref{S2.1}) since, as follows from Eqs. (\ref{51}), (\ref{52})
and (\ref{57}), these vectors are related to each other by the
three-dimensional rotation:
\begin{equation}
{\bf q}_{ij}=L[\alpha(G_{ij})^{-1}\alpha(G)\alpha(K_{ij}/M_{ij})]
{\bf k}_{ij}
\label{59}
\end{equation}

 Let $u\rightarrow R(u)$ be the representation of the group SU(2) in
$H_{int}$ which acts as
\begin{equation}
R(u)\chi({\bf k}_1,{\bf k}_2,{\bf k}_3)=
\chi(L(u)^{-1}{\bf k}_1,L(u)^{-1}{\bf k}_2,L(u)^{-1}{\bf k}_3)
\label{60}
\end{equation}
It is easy to show that the generators of this representation can be
written as ${\bf l}({\bf K}_{ij})+{\bf l}({\bf k}_{ij})$ for any
$ij$. Then by analogy with Eq. (\ref{23}) it is easy to show that if
$U(a,l)=U_{123}{\tilde U}(a,l)U_{123 }^{-1}$ then the action of
${\tilde U}(l)$ in ${\tilde H}$ is given by
\begin{equation}
{\tilde U}(l){\tilde \varphi}(G)=D[{\bf S};\alpha(G)^{-1}l\alpha(G')]
{\tilde \varphi}(G')
\label{61}
\end{equation}
where $G'=L(l)^{-1}G$ and ${\bf S}={\bf l}({\bf K}_{ij})+{\bf l}({\bf
k}_{ij})+{\bf s}_1+{\bf s}_2+{\bf s}_3$. Comparing Eqs. (\ref{8})
and (\ref{61}) we see that the operator ${\bf S}$ has the sense of the
spin operator for the three-particle system, and therefore the
operators ${\tilde {\bf M}}$ and ${\tilde {\bf N}}$ for the
representation defined by Eq. (\ref{61}) indeed have the canonical form
given by Eq. (\ref{24}) but now $G$ is the three-particle 4-velocity
and ${\bf S}$ is the three-particle spin operator.

 Let us begin to calculate the operator $M_{ij,k}$. As follows from
Eqs. (\ref{26}), (\ref{28}), (\ref{35}), (\ref{47}) and (\ref{49})
\begin{eqnarray}
&&M_{ij,k}=U_{ij}B_{ij,k}(M_{ij})^{-1}B_{ij,k}({\hat M}_{ij}^{int})
|{\hat M}_{ij}^{int}G_{ij}+m_kg_k|\cdot\nonumber\\
&&\cdot B_{ij,k}({\hat M}_{ij}^{int})^{-1}B_{ij,k}(M_{ij})U_{ij}^{-1}
\label{62}
\end{eqnarray}
and, as follows from Eqs. (\ref{48}) and (\ref{49})
\begin{eqnarray}
&&B_{ij,k}({\hat M}_{ij}^{int})|{\hat M}_{ij}^{int}G_{ij}+m_kg_k|
B_{ij,k}({\hat M}_{ij}^{int})^{-1}=m_k(G_{ij},g_k)+\nonumber\\
&&+[({\hat M}_{ij}^{int})^2+m_k^2(G_{ij},g_k)^2-m_k^2]^{1/2}
\label{63}
\end{eqnarray}
If ${\hat M}_{ij}^{int}=M_{ij}+v_{ij}$ then we introduce the operator
$$v_{ij}^B=U_{ij}B_{ij,k}(M_{ij})^{-1}v_{ij}B_{ij,k}(M_{ij})U_{ij}^{-1}$$
and, as follows from Eqs. (\ref{47}) and (\ref{49})
\begin{equation}
M_{ij,k}=\omega_k({\bf K}_{ij})+[(M_{ij}+v_{ij}^B)^2+{\bf
K}_{ij}^2]^{1/2}
\label{64}
\end{equation}
where we have taken into account that, as follows from Eq. (\ref{51})
and the definition of ${\bf K}_{ij}$
\begin{equation}
\frac{m_k[M_{ij}(G_{ij},g_k)+m_k]}{|M_{ij}G_{ij}+m_kg_k|}=
\omega_k({\bf K}_{ij})
\label{65}
\end{equation}

 The action of $M_{ij,k}$ in $\tilde H$ is determined by the operator
${\tilde M}_{ij,k}$ such that $M_{ij,k}=U_{123}{\tilde M}_{ij,k}
U_{123}^{-1}$. Let us introduce the operator ${\tilde v}_{ij}$ such
that $v_{ij}^B=U_{123}{\tilde v}_{ij}U_{123}^{-1}$. Then as follows
from Eq. (\ref{64})
\begin{equation}
{\tilde M}_{ij,k}=\omega_k({\bf K}_{ij})+[(M_{ij}+{\tilde v}_{ij})^2+
{\bf K}_{ij}^2]^{1/2}
\label{66}
\end{equation}

 The operators of multiplication by $\omega_k$ and $M_{ij}$ do not
depend on the variable $G$ and act only through the variables of the
space $H_{int}$. A direct calculation using the definitions of the
operators $v_{ij}^B$, ${\tilde v}_{ij}$ and Eqs. (\ref{17}), (\ref{27}),
(\ref{47}), (\ref{49}) (\ref{56}) shows that ${\tilde v}_{ij}$ has
the same properties and its action in $H_{int}$ is given by
\begin{eqnarray}
&&{\tilde v}_{ij}\chi({\bf K}_{ij},{\bf k}_{ij})=\int\nolimits
\{\prod_{l=i,j} D[{\bf s}_l;\alpha(k_l/m_l)
\alpha(K_{ij}/M_{ij})\alpha(k_l^{(ij)}/m_l)]\}\cdot\nonumber\\
&&v_{ij}({\bf k}_{ij},{\bf k}_{ij}')
\{\prod_{l=i,j} D[{\bf s}_l;\alpha(k_l^{'({ij})}/m_l)^{-1}
\alpha(K'_{ij}/M'_{ij})^{-1}\alpha(k'_l/m_l)]\}\cdot\nonumber\\
&&(\frac{M'}{M})^{3/2}(\frac{K_{ij}^0}{K_{ij}^{'0}})^{1/2}
\chi({\bf K}_{ij},{\bf k}'_{ij})\frac{M_{ij}M_{ij}^{'2}d^3{\bf k}'_{ij}}
{2(2\pi)^3m_i^2m_j^2\omega_i({\bf k}'_{ij})\omega_j({\bf k}'_{ij})}
\label{67}
\end{eqnarray}
where $K_{ij}$ is the 4-vector $((M_{ij}^2+{\bf K}_{ij}^2)^{1/2},
{\bf K}_{ij})$ and in all the primed functions of ${\bf K}_{ij}$ and
${\bf k}_{ij}$ the argument ${\bf k}_{ij}$ is replaced by
${\bf k}'_{ij}$. We conclude that ${\tilde M}_{ij,k}$ actually acts
only in $H_{int}$ and therefore we can write $M_{ij,k}^{int}$ instead
of ${\tilde M}_{ij,k}$. This result can be also proved when
$A_{ij}\neq 1$ (see Sec.~\ref{S2.2}).

 Using Eq. (\ref{58}) one can verify that ${\tilde v}_{ij}$ is the
Hermitian operator in $H_{int}$, and, as easily seen from
Eq. (\ref{58}), $[{\tilde v}_{ij},{\bf S}]=0$. Comparing Eqs. (\ref{27})
and (\ref{67}) we see that while $v_{ij}$ does not act through
$G_{ij}$ and the variables of particle $k$ and acts only through
${\bf q}_{ij}$ and the spin variables of particles $i$ and $j$ (i.e.
acts only in $H_{ij}^{int}$), the operator ${\tilde v}_{ij}$ does not
act through $G$ and the spin variables of particle $k$, commutes with
${\bf K}_{ij}$ and acts only through ${\bf k}_{ij}$ and the spin
variables of particles $i$ and $j$. In the general case, if $O_{ij}$
is an operator which acts only in $H_{ij}^{int}$, does not act
through the other variables and
\begin{equation}
{\tilde O}_{ij}=U_{123}^{-1}U_{ij}B_{ij,k}(M_{ij})^{-1}O_{ij}
B_{ij,k}(M_{ij})U_{ij}^{-1}U_{123}
\label{68}
\end{equation}
Then by analogy with Eq. (\ref{67}) it can be shown that
\begin{eqnarray}
&&{\tilde O}_{ij}=\frac{(K_{ij}^0)^{1/2}M_{ij}}{M^{3/2}}
\{\prod_{l=i,j} D[{\bf s}_l;\alpha(k_l/m_l)^{-1}
\alpha(K_{ij}/M_{ij})\cdot\nonumber\\
&&\cdot\alpha(k_l^{(ij)}/m_l)]\}F(O_{ij})
 \{\prod_{l=i,j} D[{\bf s}_l;\alpha(k_l/m_l)^{-1}\cdot\nonumber\\
&&\cdot\alpha(K_{ij}/M_{ij})
\alpha(k_l^{(ij)}/m_l)]\}^{-1}\frac{M^{3/2}}{(K_{ij}^0)^{1/2}M_{ij}}
\label{69}
\end{eqnarray}
where the operator $F(O_{ij})$ acts through ${\bf k}_{ij}$ in the
same manner as $O_{ij}$ acts through ${\bf q}_{ij}$.

 The operator $V_{123}$ in Eq. (\ref{33}) will satisfy all the needed
properties if
$$V_{123}=U_{123}v_{123}U_{123}^{-1}$$
and $v_{123}$ is an
operator which acts only in $H_{int}$ and commutes with ${\bf S}$.
This follows from the fact that, as pointed out above, the generators
of the representation defined by Eq. (\ref{61}) has the form of
Eq. (\ref{24}). Therefore the operator $M_{123}$ can be written as
$M_{123}=U_{123}M_{123}^{int}U_{123}^{-1}$ where
\begin{equation}
M_{123}^{int}=M+(M_{12,3}^{int}-M)+(M_{13,2}^{int}-M)+(M_{23,1}^{int}-M)
+v_{123}
\label{71}
\end{equation}

 If we are interested in calculating only the spectrum of the
three-body system and the internal three-body wave function, we can
consider only the eigenvalue problem for the operator $M_{123}^{int}$
in $H_{int}$ since $M_{123}^{int}$ is unitarily equivalent to the
three-particle operator $\hat M$ by construction. Using Eqs. (\ref{66})
and (\ref{67}) it can be easily shown that the operator defined by
Eq. (\ref{71}) coincides with the three-body mass operator in the
instant form derived in ref.~\cite{lev5} if $A_{ij}=1$ and the
normalization in both cases is chosen to be the same. In turn, as
shown in ref.~\cite{lev2}, such a mass operator is unitarily
equivalent to the mass operators derived in
refs.~\cite{Coest2,sok2,BKT,GKM,Coest3,CP,mutze,lev1,lev4} from
different considerations in different forms of dynamics, and the
choice of such a solution for the three-body mass operator has a
physical substantiation.

 To completely describe the representation of the Poincare group for
the three-body system we have to choose an explicit expression for
the operator $A$ in terms of $A_{ij,k}$. This can be done in
different ways~\cite{sok2,CP,mutze}. However, as we shall see in
Sec.~\ref{S3.4}, the physical observables for the three-particle
system do not depend on the choice of the expression for $A$ in terms
of $A_{12,3}$, $A_{13,2}$ and $A_{23,1}$.

 The results of Secs.~\ref{S2.2} and~\ref{S2.3} can be summarized as
follows. The operators ${\hat U}(a)$ and $U(l)$ describing the
representation of the Poincare group for the three-particle system
can be written as
$${\hat U}(a)=AU_{123}{\hat{\tilde U}}(a)U_{123}^{-1}
A^{-1},\quad U(l)=U_{123}{\tilde U}(l)U_{123}^{-1}$$
where $A$ is a
unitary operator commuting with $U(l)$. The "auxiliary"
representation defined by the operators ${\hat{\tilde U}}(a)$ and
${\tilde U}(l)$ has the same form as the single-particle
representation but the role of the "external" variable is played by
$G$, the role of the internal space---by the space $H_{int}$, and the
role of the mass and spin operators---by the operators
$M_{123}^{int}$ and ${\bf S}$ which act only in $H_{int}$ and commute
with each other. Therefore the three-particle generators in the
three-particle Hilbert space $H$ can be written as (compare with
Eq. (\ref{26}))
\begin{eqnarray}
&&{\hat P}=AU_{123}M_{123}^{int}GU_{123}^{-1}A^{-1},
\qquad {\bf M}=U_{123}[{\bf l}({\bf G})+{\bf
S}]U_{123}^{-1},\nonumber\\
&&{\bf N}=U_{123}[-\imath G^0\frac{\partial} {\partial {\bf G}}+
+\frac{{\bf S}\times{\bf G}}{1+G^0}]U_{123}^{-1}
\label{72}
\end{eqnarray}
\vskip 1em

\section{On the problem of constructing the packing operators for
systems with any number of particles}
\label{S2.4}

 The representation of the Poincare group for a system of $N$
particles with the masses $m_i>0$ and the spin operators ${\bf s}_i$
$(i=1,...N)$ is realized in the space of functions
$\varphi(g_1,...g_N)$ with the range in ${\cal D}({\bf s}_1)
\bigotimes\cdots \bigotimes {\cal D}({\bf s}_N)$ and such that
\begin{equation}
\int\nolimits ||\varphi(g_1,...g_N)||^2 \prod_{i=1}^{N} d\rho(g_i)
\quad <\quad \infty
\label{2.41}
\end{equation}
Instead of the variables $g_1$,...$g_N$ we introduce the variables
$G$,$k_1$,...$k_N$ where $G=(m_1g_1+...+m_Ng_N)/|m_1g_1+...+m_Ng_N|$
and the $k_i$ are formally defined as in Eq. (\ref{51}). Then by analogy
with Eq. (\ref{53}) one can show that
\begin{eqnarray}
&&\prod_{i=1}^{N}d\rho(g_i)=d\rho(G)d\rho(int),\quad
 d\rho(int)=2(2\pi)^3M^3\cdot\nonumber\\
&&\cdot \delta^{(3)}({\bf k}_1+...+{\bf k}_N)\prod_{i=1}^{N}d\rho(k_i/m_i)
\label{2.42}
\end{eqnarray}
where $M=\omega_1({\bf k}_1)+...+\omega_N({\bf k}_N)$. The "internal"
space $H_{int}$ can be defined by analogy with Eq. (\ref{54}), i.e. as
the space of functions $\chi({\bf k}_1,...{\bf k}_N)$ with the range
in ${\cal D}({\bf s}_1)\bigotimes...\bigotimes {\cal D}({\bf s}_N)$
and such that
\begin{equation}
||\chi||^2=\int\nolimits ||\chi({\bf k}_1,...{\bf k}_N)||^2
d\rho(int)\quad < \infty
\label{2.43}
\end{equation}
and the space ${\tilde H}$ can be defined as the space of functions
${\tilde \varphi}(G)$ with the range in $H_{int}$ and such that
Eq. (\ref{55}) is satisfied. Then by analogy with Eq. (\ref{56}) one can
show that
\begin{equation}
{\cal U}=\prod_{i=1}^{N} D[{\bf s}_i;\alpha(g_i)^{-1}\alpha(G)
\alpha(k_i/m_i)]
\label{2.44}
\end{equation}
is the unitary operator from ${\tilde H}$ to $H$.

 Our goal is to construct a unitary operator $A$ in $H$ such that $A$
commutes with $U(l)$, the generators $\hat {\Gamma^i}$ of the
representation under consideration $(i=1,...10)$ have the form
$\hat {\Gamma^i}=A{\cal U}{\hat {\tilde \Gamma^i}}{\cal U}^{-1}A^{-1}$,
where the operators ${\hat {\tilde \Gamma^i}}$ in $\tilde H$ have the
following "canonical" form (compare with Eqs. (\ref{26}) and (\ref{72}))
\begin{equation}
{\hat P}={\hat M}_{int}G, \quad {\bf M}=[{\bf l}({\bf G})+{\bf S}],
\quad {\bf N}=-\imath G^0\frac{\partial}{\partial {\bf G}}+
\frac{{\bf S}\times{\bf G}}{1+G^0},
\label{73}
\end{equation}
${\hat M}_{int}$ and ${\bf S}$ act only in $H_{int}$, ${\bf S}$
satisfies the commutation relation for the spin operators and
$[{\hat M}_{int},{\bf S}]=0$.

 As explicitly shown above, such a construction does
exist for systems of two and three particles. In the case of four and
more particles a possible way of constructing the system of generators
in the form of Eq. (\ref{73}) is the following.

 First it is necessary to find unitary operators $A_{\alpha\beta}$
such that (compare with Eqs. (\ref{28}) and (\ref{31})) ${\hat
G}_{\alpha\beta}=A_{\alpha\beta}GA_{\alpha\beta}^{-1}$ where
$\alpha$ and $\beta$ are any subsystems comprising our system and
\begin{equation}
{\hat G}_{\alpha\beta}=\frac{{\hat M}_{\alpha}{\hat G}_{\alpha}+
{\hat M}_{\beta}{\hat G}_{\beta}}{|{\hat M}_{\alpha}{\hat G}_{\alpha}+
{\hat M}_{\beta}{\hat G}_{\beta}|}
\label{74}
\end{equation}
where ${\hat M}_{\alpha}$ and ${\hat G}_{\alpha}$ are the mass and
the 4-velocity operators of the system $\alpha$ and analogously for
${\hat M}_{\beta}$ and ${\hat G}_{\beta}$. If we know the operators
$A_{\alpha\beta}$ for any partition of our system into two
noninteracting subsystems then we also know the operators
$A_{\alpha_1...\alpha_n}$ for the case when our system is partitioned
into $n$ noninteracting subsystems $\alpha_1,...\alpha_n$. These
operators satisfy the property
\begin{equation}
A_{\alpha_1...\alpha_n}GA_{\alpha_1...\alpha_n}^{-1}=
{\hat G}_{\alpha_1...\alpha_n}
\label{75}
\end{equation}
where ${\hat G}_{\alpha_1...\alpha_n}$ is the 4-velocity in the case
when the subsystems $\alpha_1,...\alpha_n$ do not interact with each
other (in this case ${\hat G}_{\alpha_1...\alpha_n}$ depends only on
interactions inside the subsystems $\alpha_1,...\alpha_n$).

 Let $M_{(\alpha_1),...(\alpha_n)}$ be the system mass operator in
this case. According to the idea of the Sokolov method of packing
operators, we can introduce the "auxiliary" mass operators
\begin{equation}
M_{\alpha_1...\alpha_n}={\cal U}^{-1}A_{\alpha_1...\alpha_n}^{-1}
M_{(\alpha_1)...(\alpha_n)}A_{\alpha_1...\alpha_n}{\cal U}
\label{76}
\end{equation}
which act in $H_{int}$ and commute with $G$ by construction. Then we
construct the operator ${\hat M}_{int}$ from the operators
$M_{\alpha_1...\alpha_n}$ and the unitary operator $A$ from the
operators $A_{\alpha_1...\alpha_n}$ in such a way that ${\hat
M}_{int}$ becomes $M_{\alpha_1...\alpha_n}$ and $A$ becomes
$A_{\alpha_1...\alpha_n}$ when all interactions between the
subsystems $\alpha_1,...\alpha_n$ are turned off. If this program is
carried out then the generators given by Eq. (\ref{73}) will satisfy
all the required properties.

 This program has been proposed and partially carried out by Sokolov
{}~\cite{sok4,sok5}. The major technical difficulty in realizing the program
is the following. The operators $A_{\alpha\beta}$ should be
constructed in such a way that the operators
$A_{\alpha_1,...\alpha_n}$ do not depend on the order in which the
interactions between the subsystems are turned off (see the
discussion in refs. ~\cite{sok5,CP,mutze,lev6,lev2}. If this is the
case then the operator ${\hat M}_{int}$ can be constructed as (see
refs.~\cite{sok5,CP,mutze,lev6,lev2} for details)
\begin{equation}
{\hat M}_{int}=\sum_{k=2}^{N} (-1)^k(k-1)!M_{(k)}^{int}
+ v_N,
\qquad M_{(k)}^{int}=\sum_{\alpha_1...\alpha_k}
M_{\alpha_1...\alpha_k}^{int}
\label{77}
\end{equation}
where $v_N$ is some
"N-particle" interaction and the sum in the second expression is
taken over all partitions of the indices 1,2,...N into $k$ groups
$\alpha_1...\alpha_k$. In particular, this expression becomes
Eq. (\ref{71}) if $N=3$.

 One might try to seek explicit solutions for the operators
$A_{\alpha\beta}$ not leaving the frames of the point form. The
operators satisfying Eq. (\ref{75}) have been found by Sokolov
in ref.~\cite{sok4}, but these operators do not satisfy the symmetry
condition mentioned above if $N>3$. Nevertheless the construction by
Sokolov and Shatny~\cite{SoSh} which establishes the unitary
equivalence of three basic forms for any number of particles makes it
possible to find the packing operators in the point form if they are
found in some other form. Since the N-body problem is solved by Coester
and Polyzou~\cite{CP} and Mutze~\cite{mutze} in the instant form (see
also refs.~\cite{lev4,KP,lev2}), then the solution for the operators
$A_{\alpha\beta}$ exist also in the instant form. For this reason,
considering the problem of constructing the ECO we shall assume that
for any $N$ the generators of the Poincare group can be written in
the form of Eq. (\ref{73}). Some aspects of the unitary equivalence of
three basic forms are considered in Secs.~\ref{S5.2} and ~\ref{S6.2}.

 In this chapter we did not consider the conditions imposed by the P
and T invariance, but it is easy to show \cite{CP,KP,mutze,lev1,lev2}
that they are not too restrictive. Therefore considering the problem
of constructing the ECO we shall assume that RQM can be constructed
in such a way that it is invariant under the extended Poincare group
containing the operations P and T. As noted in Sec.~\ref{S1.2}, the
operators ${\hat U}_P$ and ${\hat U}_R$ can be chosen as for the case
of free particles and therefore the action of these operators in $H$
can be written in the form \cite{Nov}
\begin{eqnarray}
U_P\varphi({\bf g}_1,...{\bf g}_N)&=&(\prod_{i=1}^{N}\eta_{iP})
\varphi(-{\bf g}_1,...-{\bf g}_N), \nonumber\\
U_R\varphi({\bf g}_1,...{\bf g}_N)&=&\{\prod_{i=1}^{N}\eta_{iR}D[{\bf
s}_i;C]\}\bar{\varphi}({\bf g}_1,...{\bf g}_N)
\label{77a}
\end{eqnarray}
Here $\eta_{iP}$ and $\eta_{iR}$ are the P and R parities of the i-th
particle, the bar means the complex conjugation, $C=\imath \sigma_2$
and we write ${\bf g}_i$ instead of $g_i$ in the arguments of the
function $\varphi$. It is easy to show that the operators $U_P$ and
$U_R$ reduced onto the space $H_{int}$ have the form
\begin{eqnarray}
{\tilde U}_P\chi({\bf k}_1,...{\bf k}_N)&=&(\prod_{i=1}^{N}\eta_{iP})
\chi(-{\bf k}_1,...-{\bf k}_N), \nonumber\\
{\tilde U}_R\chi({\bf k}_1,...{\bf k}_N)&=&
\{\prod_{i=1}^{N}\eta_{iR}D[{\bf s}_i;C]\}\bar{\chi}({\bf
k}_1,...{\bf k}_N)
\label{77b}
\end{eqnarray}

\chapter{General properties of the electromagnetic current operator}
\label{C3}
\section{Relativistic invariance and current conservation}
\label{S3.1}

 Let $x=0$ be the origin in Minkowski space. Then, as follows from
Eq. (\ref{1})
\begin{equation}
{\hat J}^{\mu}(x)=exp(\imath {\hat P}x){\hat J}^{\mu}(0)
exp(-\imath {\hat P}x)
\label{78}
\end{equation}
and, as follows from Eqs. (\ref{1}-\ref{3}) and (\ref{78}), the
operator ${\hat J}^{\mu}(0)$ must satisfy the properties
\begin{equation}
U(l)^{-1}{\hat J}^{\mu}(0)U(l)=L(l)^{\mu}_{\nu}{\hat J}^{\nu}(0),
\qquad [{\hat P}_{\mu},{\hat J}^{\mu}(0)]=0
\label{79}
\end{equation}
while the charge operator is expressed in terms of
${\hat J}^{\nu}(0)$ as
\begin{equation}
\hat Q=\int\nolimits exp(\imath {\hat P}x){\hat J}^{\mu}(0)
exp(-\imath {\hat P}x)d\sigma_{\mu}(x)
\label{80}
\end{equation}

 On the contrary, suppose we have found the operator
${\hat J}^{\mu}(0)$ satisfying Eq. (\ref{79}), cluster separability
and the condition that the operator defined by Eq. (\ref{80}) does not
depend on $\lambda$ and $\tau$ and has only one eigenvalue equal to
the sum of the electric charges of constituents. In this case we
treat Eq. (\ref{78}) as {\it the definition} of ${\hat J}^{\mu}(x)$.
Then, using the well known properties of representations of the
Poincare group, it is easy to verify that the operator
${\hat J}^{\mu}(x)$ defined in such a way indeed satisfies all the
necessary conditions. We see that the problem of seeking the operator
${\hat J}^{\mu}(x)$ can be reduced to the problem of seeking the
operator ${\hat J}^{\mu}(0)$.

 The action of ${\hat J}^{\mu}(0)$ in $\tilde H$ is defined by the
operator ${\hat J}^{\mu}$ such that
\begin{equation}
{\hat J}^{\mu}(0)=A{\cal U}{\hat J}^{\mu}{\cal U}^{-1}A^{-1}
\label{81}
\end{equation}
It will be convenient to define the action of ${\hat J}^{\mu}$ in
$\tilde H$ as follows
\begin{equation}
{\hat J}^{\mu}{\tilde \varphi}(G)=2\int\nolimits {\hat M}_{int}^{3/2}
{\hat J}^{\mu}(G,G'){\hat M}_{int}^{3/2}{\tilde \varphi}(G')d\rho(G')
\label{82}
\end{equation}
where the kernel ${\hat J}^{\mu}(G,G')$ is an operator in $H_{int}$
for any fixed values of $G$ and $G'$. Since ${\hat J}^{\mu}$ is the
selfadjoint operator in $\tilde H$, the kernel must satisfy  the
property ${\hat J}^{\mu}(G,G')^*={\hat J}^{\mu}(G',G)$ where * means
the Hermitian conjugation in $H_{int}$ (in the general case the
property of an operator to be selfadjoint is stronger than to be
Hermitian but we shall not discuss this question).

 It is obvious that the first expression in Eq. (\ref{79}) will be
satisfied if
\begin{equation}
{\tilde U}(l)^{-1}{\hat J}^{\mu}{\tilde U}(l)=L(l)^{\mu}_{\nu}
{\hat J}^{\nu}
\label{83}
\end{equation}
The action of the operators ${\tilde U}(l)$ is defined by
Eq. (\ref{61}) and therefore, using Eq. (\ref{82}), one can show that
Eq. (\ref{83}) is satisfied if
\begin{eqnarray}
&&{\hat J}^{\mu}(G,G')=L(l)^{\mu}_{\nu}D[{\bf S};\alpha(G)^{-1}l
\alpha(L(l)^{-1}G)]{\hat J}^{\nu}(L(l)^{-1}G,\nonumber\\
&&L(l)^{-1}G') D[{\bf S};\alpha(G')^{-1}l \alpha(L(l)^{-1}G')]^{-1}
\label{84}
\end{eqnarray}
Note that in contrast with the notations of Chap.~\ref{C2} the
quantities $G$ and $G'$ in this expression are arbitrary and $l$ is
an arbitrary element of the group SL(2,C).

 We use $\alpha(G,G')$ to denote $\alpha((G+G')/|G+G'|)\in$SL(2,C)
and $L(G,G')$ to denote the Lorentz transformation $L[\alpha(G,G')]$.
We also introduce the 4-vectors
\begin{equation}
f=L(G,G')^{-1}G, \qquad f'=L(G,G')^{-1}G'
\label{85}
\end{equation}
These 4-vectors are constructed as the c.m.frame 4-velocities of two
particles with unit masses and the 4-velocities $G$ and $G'$
(compare with Eq. (\ref{12})). Let us note that this is only a formal
construction since $G$ and $G'$ in Eq. (\ref{82}) have the sense of the
4-velocities of {\it one and the same} system in the final and
initial states. Nevertheless, as follows from Eq. (\ref{85}), the
4-vectors $f$ and $f'$ are such that
\begin{equation}
f^2=f'^2=1,\qquad {\bf f}+{\bf f}'=0,\qquad f^0=f'^0=(1+{\bf f}^2)^{1/2}
\label{86}
\end{equation}
Therefore the 4-vectors $f$ and
$f'$ are fully determined by one three-dimensional vector ${\bf h}=
{\bf f}/f^0$. Note that ${\bf h}<1$.

  If the operator ${\hat J}^{\mu}$ satisfies Eq. (\ref{83}) then,
as follows from Eq. (\ref{84}),
\begin{eqnarray}
{\hat J}^{\mu}(G,G')&=&L(G,G')^{\mu}_{\nu}D[{\bf S};\alpha(G)^{-1}
\alpha(G,G')\alpha(f)]{\hat j}^{\nu}({\bf h})\cdot\nonumber\\
&\cdot&D[{\bf S};\alpha(G')^{-1}\alpha(G,G')\alpha(f')]^{-1}
\label{87}
\end{eqnarray}
where we use ${\hat j}^{\nu}({\bf h})$ to denote ${\hat
J}^{\mu}(f,f')$.  As follows from Eq. (\ref{85}), the condition for the
Hermiticity of the operator ${\hat J}^{\mu}(G,G')$ (see the remark
after Eq. (\ref{82})) will be satisfied if and only if
${\hat j}^{\nu}({\bf h})^*={\hat j}^{\nu}(-{\bf h})$. We see that if
Eq. (\ref{83}) is satisfied then the kernel of the operator ${\hat
J}^{\mu}$ is fully determined by an operator the action of which in
$H_{int}$ depends only on ${\bf h}$. As follows from Eq. (\ref{84})
and the fact that $u\alpha(G)=\alpha(L(u)G)u$ if $u\in$SU(2), the
operator ${\hat j}^{\nu}({\bf h})$ has the property
\begin{equation}
{\hat j}^{\nu}({\bf h})=L(u)^{\nu}_{\rho}D[{\bf S};u]
{\hat j}^{\rho}(L(u)^{-1}{\bf h})D[{\bf S};u]^{-1}
\label{88}
\end{equation}
In particular, since $L(u)$ is the three-dimensional rotation
corresponding to $u\in$SU(2), Eq. (\ref{87}) yields
\begin{eqnarray}
&&{\hat j}^0({\bf h})=D[{\bf S};u]{\hat j}^0(L(u)^{-1}{\bf h})
D[{\bf S};u]^{-1},\nonumber\\
&&{\hat {\bf j}}({\bf h})=L(u)D[{\bf S};u]
{\hat {\bf j}}(L(u)^{-1}{\bf h})D[{\bf S};u]^{-1}
\label{89}
\end{eqnarray}

 On the contrary, let ${\hat j}^{\nu}({\bf h})$ be an operator
satisfying Eq. (\ref{89}) and Eq. (\ref{87}) be treated as {\it
the definition} of ${\hat J}^{\mu}(G,G')$. Then we have to verify
that Eq. (\ref{84}) is satisfied. This can be done directly taking into
account that the 4-vectors $f(l)$ and $f'(l)$ which are constructed
from $L(l)^{-1}G$ and $L(l)^{-1}G'$ in the same manner as $f$ and
$f'$ are constructed from $G$ and $G'$, are connected with $f$ and
$f'$ by the three-dimensional rotation:
\begin{eqnarray}
&&f=L[\alpha(G,G')^{-1}l\alpha(L(l)^{-1}G,L(l)^{-1}G')]f(l),\nonumber\\
&&f'=L[\alpha(G,G')^{-1}l\alpha(L(l)^{-1}G,L(l)^{-1}G')]f'(l)
\label{90}
\end{eqnarray}

 We conclude that the first expression in Eq. (\ref{79}) will be
satisfied if ${\hat J}^{\mu}(G,G')$ is expressed in terms of
${\hat j}^{\nu}({\bf h})$ according to Eq. (\ref{87}) and
${\hat j}^{\nu}({\bf h})$ satisfies Eq. (\ref{89}).

 Let us now consider the second expression in Eq. (\ref{79}) which
is the consequence of the continuity equation. As easily follows from
Eq. (\ref{82}), this expression is satisfied if and only if
\begin{equation}
{\hat M}_{int}G_{\mu}{\hat J}^{\mu}(G,G')-{\hat J}^{\mu}(G,G')
{\hat M}_{int}G_{\mu}'=0
\label{90a}
\end{equation}
In turn, as follows from Eqs. (\ref{82}) and (\ref{85}-\ref{87}),
Eq. (\ref{90a}) is satisfied if and only if
\begin{equation}
[{\hat M}^{int},{\hat j}^0({\bf h})]={\bf h}\{{\hat M}^{int},
{\hat {\bf j}}({\bf h})\}
\label{91}
\end{equation}
where we use curly brackets to denote the
anticommutator. Therefore Eq. (\ref{91}) is the sufficient condition
ensuring the continuity equation.

 Let us consider some simple consequences of Eq. (\ref{91}). If ${\bf
h}=0$ then
\begin{equation}
[{\hat M}^{int},{\hat j}^0(0)]=0
\label{V1}
\end{equation}
It will be shown in Sec.~\ref{S3.2} that ${\hat j}^0(0)=e$ where $e$ is
the electric charge of the system under consideration. Therefore
${\hat j}^0(0)$ must indeed commute with ${\hat M}^{int}$.

 Taking the derivative of Eq. (\ref{91}) over ${\bf h}$ at ${\bf h}=0$
we get
\begin{equation}
[{\hat M}^{int},\frac{\partial {\hat j}^0(0)}{\partial {\bf h}}]=
\{{\hat M}^{int},{\hat{\bf j}}(0)\}
\label{V2}
\end{equation}
Let $\Pi_i$ be the projector onto the subspace in $H_{int}$
corresponding to the discrete eigenvalue $m_i$ of the operator
${\hat M}^{int}$. Then as follows from Eq. (\ref{V2})
\begin{equation}
\Pi_i{\hat{\bf j}}(0)\Pi_j=\frac{m_i-m_j}{m_i+m_j}
\Pi_i\frac{\partial {\hat{\bf j}}(0)}{\partial {\bf h}}\Pi_j
\label{V3}
\end{equation}
In particular
\begin{equation}
\Pi_i{\hat{\bf j}}(0)\Pi_i=0
\label{V4}
\end{equation}

 Assuming that ${\hat j}^0({\bf h})$ and ${\hat {\bf j}}({\bf h})$
are analytic functions of ${\bf h}$ in the vicinity of the point
${\bf h}=0$ we can expand both parts of Eq. (\ref{91}) in powers of
${\bf h}$ and equate the terms containing equal powers of $|{\bf h}|$.
Then we get
\begin{equation}
\bigl[ {\hat M}^{int},\frac{\partial^{n+1}{\hat j}^0(0)}{\partial
h_{i_1} \cdots \partial h_{i_{n+1}}} \bigr] =\bigl\{ {\hat M}^{int},
\sum_{l=1}^{n} \frac{\partial^n {\hat j}_{i_l}(0)}{\partial
h_{i_1}\cdots \partial h_{i_{l-1}} \partial h_{i_{l+1}}\cdots \partial
h_{i_{n+1}}} \bigr\}
\label{V5}
\end{equation}
where $i_1$,...$i_{n+1}$=1,2,3, $n=0,1\ldots$. To investigate the
consequences of Eq. (\ref{91}) in more details it is convenient to use
the spectral decomposition of the ECO. This is done in the next
section.

\section{Spectral decomposition of the ECO}
\label{S3.1a}

 Let ${\hat e}_{int}(m)$ be the spectral function of the operator
${\hat M}^{int}$. In the general case ${\hat J}^{\mu}(G,G')$ does not
commute with ${\hat M}_{int}$ and therefore we cannot write the
spectral decomposition of ${\hat J}^{\mu}(G,G')$ by analogy with
Eq. (\ref{36}). However we can define ${\hat J}^{\mu}(G,G')$ using a
set of operators ${\hat J}^{\mu}(G,m;G',m')$ such that
\begin{equation}
{\hat J}^{\mu}(G,G')=\int\nolimits\int\nolimits d{\hat e}_{int}(m)
{\hat J}^{\mu}(G,m;G',m')d{\hat e}_{int}(m')
\label{91a}
\end{equation}
where the integrals are understood as the strong limits of the
corresponding Riemann sums constructed as follows. Let $S$ be the
spectrum of the operator ${\hat M}^{int}$ and $\{S_{\alpha}\}$ be a
partition of $S$ such that
\begin{equation}
\bigcap_{\alpha} S_{\alpha}=\oslash,\qquad \bigcup_{\alpha}
S_{\alpha}=S
\label{91aa}
\end{equation}
Let $\Pi_{\alpha}$ be the projector onto the subspace of $H_{int}$
corresponding to $S_{\alpha}$. Then ${\hat J}^{\mu}(G,G')$ can be
approximated by the sums
\begin{equation}
{\hat J}^{\mu}(G,G')=\sum_{\alpha,\beta}\Pi_{\alpha}
{\hat J}^{\mu}(G,m_{\alpha};G',m_{\beta}) \Pi_{\beta}
\label{91ab}
\end{equation}
where $m_{\alpha} \in S_{\alpha}$.

 It is easy to see that Eq. (\ref{90a}) will be satisfied if
\begin{equation}
(mG_{\mu}-m'G'_{\mu}){\hat J}^{\mu}(G,m;G',m')=0
\label{91b}
\end{equation}

 Analogously we can write the spectral decomposition of
${\hat j}^{\nu}({\bf h})$ in the form
\begin{equation}
{\hat j}^{\nu}({\bf h})=\int\nolimits\int\nolimits d{\hat e}^{int}(m)
{\hat j}^{\nu}({\bf h};m,m')d{\hat e}^{int}(m')
\label{92}
\end{equation}
and the relation between ${\hat J}^{\mu}(G,m;G',m')$ and
${\hat j}^{\nu}({\bf h};m,m')$ is obvious from Eq. (\ref{87}). As
follows from Eqs. (\ref{87}), (\ref{90}) and (\ref{91b}), the continuity
equation will be satisfied if
\begin{equation}
(m-m'){\hat j}^0({\bf h};m,m')=(m+m'){\bf h}{\hat {\bf j}}({\bf h};m,m')
\label{93}
\end{equation}
and this result is also obvious from Eq. (\ref{91}).

 The operators ${\hat j}^{\nu}({\bf h};m,m')$ are simply related to
${\hat j}^{\nu}({\bf h})$ if $m$ and $m'$ belong to the discrete
spectrum of the operator ${\hat M}^{int}$:
\begin{equation}
{\hat j}^{\nu}({\bf h};m_i,m_j)=\Pi_i{\hat j}^{\nu}({\bf h}) \Pi_j
\label{93a}
\end{equation}
The analogous relation takes place for the operator
$${\hat J}^{\mu}(G,m;G',m').$$
However if  $m$ and $m'$ belong to the
continuous spectrum of the operator ${\hat M}^{int}$, the operators
${\hat J}^{\mu}(G,m;G',m')$ and ${\hat j}^{\nu}({\bf h};m_i,m_j)$
should be handled with care since these operators are the operatorial
distributions. For example, as follows from Eq. (\ref{93})
\begin{equation}
(m-m'){\hat j}^0(0;m,m')=0
\label{93b}
\end{equation}
(compare with Eq. (\ref{V1})) but this does not necessarily imply that
${\hat j}^0(0;m,m')=0$ since, as follows from Eq. (\ref{92})
\begin{equation}
\int\nolimits\int\nolimits d{\hat e}^{int}(m)
{\hat j}^0(0;m,m')d{\hat e}^{int}(m')=e
\label{93c}
\end{equation}
We see that ${\hat j}^0(0;m,m')$ is the operatorial analog of
$e\delta(m-m')$.

 Assuming that ${\hat j}^{\nu}({\bf h};m,m')$ is the analytic
function of ${\bf h}$ we obtain from Eq. (\ref{93}) (compare with
Eq. (\ref{V5}))
\begin{equation}
\frac{m-m'}{m+m'}\frac{\partial^{n+1}{\hat j}^0(0;m,m')}{\partial
h_{i_1} \cdots \partial h_{i_{n+1}}} =
\sum_{l=1}^{n} \frac{\partial^n {\hat j}_{i_l}(0;m,m')}{\partial
h_{i_1}\cdots \partial h_{i_{l-1}} \partial h_{i_{l+1}}\cdots \partial
h_{i_{n+1}}}
\label{93d}
\end{equation}
In particular (compare with Eq. (\ref{V2}))
\begin{equation}
\frac{m-m'}{m+m'}\frac{\partial {\hat j}^0(0;m,m')}{\partial {\bf
h}}={\hat{\bf j}}(0;m,m')
\label{93e}
\end{equation}
If $m$ and $m'$ belong to the discrete spectrum of the operator
${\hat M}^{int}$, this expression is equivalent to Eq. (\ref{V3}) if we
take into account Eq. (\ref{93a}). However Eq. (\ref{93e}) does not
necessarily imply that ${\hat{\bf j}}(0;m,m)=0$ if $m$ belongs to the
continuous spectrum of the operator ${\hat M}^{int}$. The analog of
such a situation is the equality $(m-m')\frac{d\delta(m-m')}{dm'}=
\delta(m-m')$. We shall see in Sec.~\ref{S4.2} that for systems of two
free particles
\begin{eqnarray}
&&{\hat{\bf j}}(0;m,m')=0\,\, \mbox{if} \,\, m\neq m';\nonumber\\
&&\int\nolimits\int\nolimits d{\hat e}^{int}(m)
{\hat{\bf j}}(0;m,m')d{\hat e}^{int}(m')={\hat{\bf j}}(0)
\label{93f}
\end{eqnarray}

 The theory of operatorial integrals is given in several monographs
(see, for example, refs.\cite{Dix,Naim}). To avoid singularities in
practical calculations of the sums defined by Eq. (\ref{91ab}) it
is desirable to choose the points $m_{\alpha}$ and $m_{\beta}$ in
such a way that $m_{\alpha}\neq m_{\beta}$ when $m_{\alpha}$ and
$m_{\beta}$ belong to the continuous spectrum of the operator
${\hat M}^{int}$.

\section{Charge operator and cluster separability}
\label{S3.2}

 Our next goal is to express the action of the charge operator in
$\tilde H$ in terms of ${\hat j}^{\nu}({\bf h})$. As follows from
Eqs. (\ref{80}), (\ref{82}) and (\ref{87}) this action is given by
\begin{eqnarray}
&&{\tilde Q}{\tilde \varphi}(G)=2\int\nolimits\int\nolimits
{\hat M}_{int}^{3/2}exp[\imath {\hat M}_{int}(Gx)]L(G,G')^{\mu}_{\nu}
D[...]{\hat j}^{\nu}({\bf h})D[...]^{-1}\cdot\nonumber\\
&&{\hat M}_{int}^{3/2}exp[-\imath {\hat M}_{int}(G'x)]
{\tilde \varphi}(G')\lambda_{\mu}\delta((\lambda x)-\tau)d^4xd\rho(G')
\label{94}
\end{eqnarray}
where for brevity we do not write the arguments of the D-operators.
Instead of $x$ we introduce the 4-vector $y=L(G,G')^{-1}x$ and use
$z=z(G,G')$ to denote the 4-vector $L(G,G')^{-1}\lambda$. Then using
Eqs. (\ref{85}), (\ref{86}) and (\ref{92}) we can integrate over $y$ and
obtain
\begin{eqnarray}
&&{\tilde Q}{\tilde \varphi}(G)=2(2\pi)^3\int\nolimits\int\nolimits
\int\nolimits (mm')^{3/2}
exp[\imath (m-m')f^0\frac{\tau}{z^0}]\cdot\nonumber\\
&&\delta^{(3)}[(m-m')f^0\frac{{\bf z}}{z^0}-
(m+m'){\bf f}]D[...]d{\hat e}^{int}(m)
[{\hat j}^0({\bf h};m,m')-\nonumber\\
&&\frac{{\bf z}}{z^0}{\hat {\bf j}}({\bf h};m,m')]
d{\hat e}^{int}(m')D[...]^{-1}{\tilde \varphi}(G')d\rho(G')
\label{95}
\end{eqnarray}
Note that ${\hat e}_{int}(m)$ and $D[...]$ commute with each other
since ${\hat M}_{int}$ commutes with ${\bf S}$.

 It is easy to see that only those $m'$ contribute to Eq. (\ref{95})
which are in the infinitely small vicinity of the point $m'=m$.
Indeed, if $m'\neq m$ then the argument of the delta-function can be
equal to zero only if ${\bf z}/z^0=(m+m'){\bf h}/(m-m')$, and the
integrand in Eq. (\ref{95}) becomes zero as follows from Eq. (\ref{93}).
We see that the exponent with $\tau$ in Eq. (\ref{95}) can be dropped
and indeed one of the consequence of the continuity equation in the
form of Eq. (\ref{93}) is that $\hat Q$ does not depend on $\tau$. The
second consequence is that only an infinitely small vicinity of the
point ${\bf h}$ contributes to Eq. (\ref{95}). Since $G'$ can be
expressed as the function of $G$ and ${\bf h}$ and it is obvious from
Eq. (\ref{85}) that $G=G'$ if ${\bf h}=0$, we can replace $G'$ by $G$
in the arguments of the D-operators, the function $\varphi$ and the
4-vector $z$ entering into the integrand of Eq. (\ref{95}). Then, as
follows from Eq. (\ref{87}), the D-operators in Eq. (\ref{95}) can be
dropped and $z$ becomes the 4-vector $z=z(G)=L[\alpha(G)]^{-1}\lambda$.

 By analogy with Eq. (\ref{13}) we can write neglecting the
order-$o({\bf h})$ terms
\begin{equation}
{\bf G}'=-{\bf h}+\frac{{\bf G}+{\bf G}'}{2}-\frac{({\bf G}{\bf h})
{\bf G}}{1+G^0}
\label{96}
\end{equation}
where we have taken into account that $|G+G'|=2(1+f^0)\approx 2$. As
easily follows from this expression, we can replace $d\rho(G')$ by
$4d^3{\bf h}/(2\pi)^3$ in Eq. (\ref{95}). Therefore Eq. (\ref{95}) can be
written in the form
\begin{eqnarray}
&&{\tilde Q}{\tilde \varphi}(G)=\int\nolimits\int\nolimits
d{\hat e}^{int}(m)
[{\hat j}^0(\frac{(m-m'){\bf z}(G)}{(m+m')z^0(G)};m,m')-\nonumber\\
&-&\frac{{\bf z}(G)}{z^0(G)}
{\hat {\bf j}}(\frac{(m-m'){\bf z}(G)}{(m+m')z^0(G)};m,m')]
d{\hat e}^{int}(m')d^3{\bf h} {\tilde \varphi}(G)
\label{97}
\end{eqnarray}
We again see that the integrand is equal to zero if $m\neq m'$ as
easily follows from Eq. (\ref{93}).

 The contribution of the discrete spectrum to the integral (\ref{97})
can be easily determined since, as follows from
Eqs. (\ref{V4}) and (\ref{93a}), each point $m_i$ of the discrete
spectrum contributes $\Pi_i {\hat j}^0(0) \Pi_i$ to this
integral. In the general case, expanding ${\hat j}^0$ and ${\hat{\bf
j}}$ in Eq. (\ref{97}) in powers of the first argument and using
Eq. (\ref{93d}) we see that only ${\hat j}^0(0;m,m')$ survives in this
expansion while all other terms cancel. Therefore we conclude that
\begin{equation}
{\tilde Q}=\int\nolimits\int\nolimits d{\hat e}^{int}(m)
{\hat j}^0(0;m,m')d{\hat e}^{int}(m')={\hat j}^0(0)
\label{98}
\end{equation}

 If the particles do not interact with each other then, as follows
from cluster separability (see Sec.~\ref{S1.1}), the operator
${\hat J}^{\mu}(x)$ is a sum of the constituent ECO's and
therefore, as follows from Eq. (\ref{3}), the system electric charge
$e$ is the sum of the electric charges of constituents as it
should be. Therefore, as follows from Eq. (\ref{98}), the operator
${\hat j}^0({\bf h})$ must depend on interactions only in such a way
that ${\hat j}^0(0)=j^0(0)=e$.

 Let us now consider the conditions imposed on
${\hat j}^{\nu}({\bf h})$ by cluster separability in the general
case. If our system consists of $n$ noninteracting subsystems
$\alpha_1,...\alpha_n$ and the ECO's ${\hat J}_{\alpha_i}^{\mu}(x)$
for them are known then, as follows from
cluster separability, the ECO for the whole system is equal to
\begin{equation}
{\hat J}_{\alpha_1...\alpha_n}^{\mu}(x)=\sum_{i=1}^{n}
{\hat J}_{\alpha_i}^{\mu}(x)
\label{103}
\end{equation}
Therefore the operator ${\hat J}_{\alpha_1...\alpha_n}^{\mu}(0)$ is
also known and the action of this operator in $\tilde H$ is defined
by the operator ${\hat J}_{\alpha_1...\alpha_n}^{\mu}$ such that
(see Eq. (\ref{81}))
\begin{equation}
{\hat J}_{\alpha_1...\alpha_n}^{\mu}(0)=A_{\alpha_1...\alpha_n}
{\cal U}{\hat J}_{\alpha_1...\alpha_n}^{\mu}{\cal U}^{-1}
A_{\alpha_1...\alpha_n}^{-1}
\label{104}
\end{equation}
In turn, if ${\hat J}_{\alpha_1...\alpha_n}^{\mu}$ is known, we can
determine the operator $${\hat J}_{\alpha_1...\alpha_n}^{\mu}(G,G')$$
such that (see Eq. (\ref{82}))
\begin{eqnarray}
&&{\hat J}_{\alpha_1...\alpha_n}^{\mu}{\tilde
\varphi}(G)=2\int\nolimits (M_{\alpha_1...\alpha_n}^{int})^{3/2}
{\hat J}_{\alpha_1...\alpha_n}^{\mu}(G,G')\cdot\nonumber\\
&&\cdot (M_{\alpha_1...\alpha_n}^{int})^{3/2}
{\tilde \varphi}(G')d\rho(G')
\label{105}
\end{eqnarray}
where $M_{\alpha_1...\alpha_n}$ is defined by Eq. (\ref{76}). Finally,
the operator $${\hat J}_{\alpha_1...\alpha_n}^{\mu}(G,G')$$ determines
the operator ${\hat j}_{\alpha_1...\alpha_n}^{\nu}({\bf h})$ such
that (see Eq. (\ref{87}))
\begin{eqnarray}
&&{\hat J}_{\alpha_1...\alpha_n}^{\mu}(G,G')=L(G,G')^{\mu}_{\nu}
D[{\bf S};\alpha(G)^{-1}\alpha(G,G')\alpha(f)]\cdot\nonumber\\
&& {\hat j}_{\alpha_1...\alpha_n}^{\nu}({\bf h})
D[{\bf S};\alpha(G')^{-1}\alpha(G,G')\alpha(f')]^{-1}
\label{106}
\end{eqnarray}

 We see that if the ECO's for all subsystems are known then it is
possible to determine the operators
${\hat j}_{\alpha_1...\alpha_n}^{\nu}({\bf h})$ corresponding to the
case when the subsystems $\alpha_1,...\alpha_n$ do not interact with
each other. Let us now compare Eqs. (\ref{81}), (\ref{82}) and (\ref{87})
on the one hand and Eqs. (\ref{104}-\ref{106}) on the other. If all
interactions between the subsystems $\alpha_1,...\alpha_n$ are turned
off then $A$ becomes $A_{\alpha_1...\alpha_n}$ and ${\hat M}_{int}$
becomes $M_{\alpha_1...\alpha_n}$ by construction (see
Sec.~\ref{S2.4}); therefore cluster separability will be satisfied if
${\hat j}^{\nu}({\bf h})$ becomes
${\hat j}_{\alpha_1...\alpha_n}^{\nu}({\bf h})$.

\section{General method of constructing the ECO}
\label{S3.3}

 The results of Secs.~\ref{S3.1} and ~\ref{S3.2} can be formulated as
follows. The operator ${\hat J}^{\mu}(x)$ satisfies all the
conditions described in Sec.~\ref{S1.1} if the operator ${\hat
j}^{\nu}({\bf h})$ satisfies Eqs. (\ref{89}) and (\ref{91}),
${\hat j}^0(0)$ is the operator of multiplication by $e$,
${\hat j}^{\nu}({\bf h})$
becomes ${\hat j}_{\alpha_1...\alpha_n}^{\nu}({\bf h})$ if for any set
$(\alpha_1,...\alpha_n)$ the interactions between the subsystems
$\alpha_1,...\alpha_n$ are turned off and ${\hat j}^{\nu}({\bf h})^*
={\hat j}^{\nu}(-{\bf h})$ (in some cases it is convenient to use
Eq. (\ref{91}) in the form of Eq. (\ref{93})). Therefore the problem of
constructing the operator ${\hat J}^{\mu}(x)$ will be solved if we
succeed in constructing the operator ${\hat j}^{\nu}({\bf h})$
satisfying the above properties.

 In turn the latter problem can be reduced as follows.
First, using Eqs. (\ref{V1}) and (\ref{V2}), we can write Eq. (\ref{91})
in the form
\begin{equation}
[{\hat M}^{int},{\hat j}^0({\bf h})-{\hat j}^0(0)-{\bf h}
\frac{\partial {\hat j}^0(0)}{\partial {\bf h}}]={\bf h}\{{\hat M}^{int},
{\hat {\bf j}}({\bf h})-{\hat {\bf j}}(0)\}
\label{106a}
\end{equation}
Since ${\hat {\bf j}}({\bf h})-{\hat {\bf j}}(0)$ becomes zero when
${\bf h}\rightarrow 0$, we can uniquely decompose this operator into
longitudinal and transverse parts:
\begin{equation}
{\hat {\bf j}}({\bf h})={\hat {\bf j}}(0)+\frac{{\bf h}}{|{\bf h}|}
{\hat j}_{||}({\bf h})+{\hat {\bf j}}_{\bot}({\bf h})
\label{108}
\end{equation}
where ${\bf h}{\hat {\bf j}}_{\bot}({\bf h})=0$ and
\begin{eqnarray}
&&{\hat j}_{||}({\bf h})=\frac{1}{|{\bf h}|}
({\bf h},{\hat {\bf j}}({\bf h})-{\hat {\bf j}}(0)),\nonumber\\
&&{\hat {\bf j}}_{\bot}({\bf h})={\hat {\bf j}}({\bf h})-
{\hat {\bf j}}(0)-\frac{{\bf h}}{|{\bf h}|^2}({\bf h},
{\hat {\bf j}}({\bf h})-{\hat {\bf j}}(0))
\label{107}
\end{eqnarray}

 Now, as follows from Eqs. (\ref{106a}) and (\ref{108}),
Eqs. (\ref{91}) and (\ref{93}) can be written in the form
\begin{eqnarray}
&&[{\hat M}^{int},{\hat j}^0({\bf h})-{\hat j}^0(0)-{\bf h}
\frac{\partial {\hat j}^0(0)}{\partial {\bf h}}]=
{\bf h}\{{\hat M}^{int},{\hat j}_{||}({\bf h})\},\nonumber\\
&&{\hat j}_{||}({\bf h};m,m')=
\frac{(m-m')}{(m+m')|{\bf h}|}[{\hat j}^0({\bf h};m,m')-\nonumber\\
&&-{\hat j}^0(0;m,m')-\frac{\partial {\hat j}^0(0;m,m')}
{\partial{\bf h}}]
\label{109}
\end{eqnarray}
Equations (\ref{93e}) and (\ref{109}) show that if ${\hat j}^0({\bf h})$
is known then ${\hat{\bf j}}(0)$ and ${\hat j}_{||}({\bf h})$ can be
uniquely determined from the continuity equation while, as
follows from Eqs. (\ref{108}) and (\ref{109}), the continuity equation
does not impose any constraint on the transverse part
${\hat {\bf j}}_{\bot}({\bf h})$ of the operator
${\hat {\bf j}}({\bf h})$.

 We conclude that the problem of constructing the operators
${\hat J}^{\mu}(x)$ will be solved if we succeed in constructing the
operators ${\hat j}^0({\bf h})$ and ${\hat {\bf j}}_{\bot}({\bf h})$
such that (see Eqs. (\ref{89}) and (\ref{98}))
\begin{eqnarray}
{\hat j}^0({\bf h})&=&D[{\bf S};u]{\hat j}^0(L(u)^{-1}{\bf h})
D[{\bf S};u]^{-1},\nonumber\\
{\hat {\bf j}}_{\bot}({\bf h})&=&L(u)D[{\bf S};u]
{\hat {\bf j}}_{\bot}(L(u)^{-1}{\bf h})D[{\bf S};u]^{-1},
\label{111}
\end{eqnarray}
${\hat j}^0(0)=e$, the operators
${\hat j}^0({\bf h})$ and ${\hat {\bf j}}_{\bot}({\bf h})$ satisfy
cluster separability in the sense described above, and
${\hat j}^0({\bf h})^*={\hat j}^0(-{\bf h})$,
${\hat {\bf j}}_{\bot}({\bf h})^*={\hat {\bf j}}_{\bot}(-{\bf h})$.

 In addition, it is easy to show that if the theory should be P and T
invariant then, as follows from Eq. (\ref{3a})
\begin{eqnarray}
{\tilde U}_P({\hat j}^0({\bf h}),{\hat {\bf j}}_{\bot}({\bf h}))
{\tilde U}_P^{-1}&=&({\hat j}^0(-{\bf h}),-{\hat {\bf j}}_{\bot}(-{\bf
h})),\nonumber\\
 {\tilde U}_R {\hat j}^{\nu}({\bf h}){\tilde U}_R^{-1}&=&
{\hat j}^{\nu}({\bf h})
\label{112a}
\end{eqnarray}
where the operators ${\tilde U}_P$ and ${\tilde U}_R$ are defined by
Eq. (\ref{77b}).

 In Chap.~\ref{C4} we consider how the problem of constructing such
operators can be solved in the cases of one, two, three and many
particles.

\section{Matrix elements of the ECO}
\label{S3.4}

  Let us define the one-particle states with a definite 4-momentum
$p'=mg'$ and the spin projections $\sigma'$. In the scattering theory
such states are usually normalized as
\begin{equation}
\langle p",\sigma"|p',\sigma'\rangle =2(2\pi)^3\omega({\bf p}')
\delta^{(3)}({\bf p}"-{\bf p}')\delta_{\sigma"\sigma'}
\label{113}
\end{equation}
where $\delta_{\sigma"\sigma'}$ is the Cronecker symbol. Therefore if
the scalar product is chosen according to Eq. (\ref{6}) then the state
$|p',\sigma'\rangle$ depends on $g$ and $\sigma$ as follows
\begin{equation}
|p',\sigma'\rangle =\frac{2}{m} (2\pi)^3g^{'0}\delta^{(3)}
({\bf g}-{\bf g}')\delta_{\sigma\sigma'}
\label{114}
\end{equation}

 To define the scattering states in the N-particle case we have to
solve the eigenvalue problem for the operator ${\hat M}_{int}$ in the
space $H_{int}$. Let $\chi'\in H_{int}$ be the internal wave function
of a bound state with the mass $M'$ and $G'$ be the 4-velocity of
this bound state. Then, as follows from Eq. (\ref{73}), the wave
function of such a state in the space $H$ can be written as
\begin{equation}
|P',\chi'\rangle =A{\cal U}\frac{2}{M'}(2\pi)^3G^{'0}\delta^{(3)}
({\bf G}-{\bf G}')\chi'
\label{3.43}
\end{equation}
where $P'=M'G'$. If $P"=M"G"$ and $\chi"$ are the total 4-momentum
and the internal wave function in the final state then it is obvious
from Eq. (\ref{3.43}) that (compare with Eq. (\ref{113}))
\begin{equation}
\langle P",\chi"|P',\chi'\rangle =2(2\pi)^3(M^{'2}+{\bf P}^{'2})^{1/2}
\delta^{(3)}({\bf P}"-{\bf P}')\langle \chi"|\chi' \rangle
\label{3.44}
\end{equation}
Here the last scalar product is taken in the space $H_{int}$. It
is obvious that $\langle \chi"|\chi' \rangle =0$ if $M"\neq M'$.

 In the general case the particles in the initial or final state can
be not only bound but the system under consideration may consist of
$n$ bound subsystems $(1\leq n \leq N)$. Then the system wave
function can be also written in the form of Eq. (\ref{3.43}) but
$\chi'$ should be a generalized eigenfunction of the operator ${\hat
M}_{int}$. The construction of such functions ($|in\rangle$ or
$|out\rangle$ states) requires a special consideration in each
concrete case, and we shall not discuss this problem.

 Since the functions defined by Eq. (\ref{3.43}) form a (generalized)
basis in $H$, all matrix elements of the operator ${\hat J}^{\mu}(x)$
can be expressed in terms of the quantities
$\langle P",\chi"|{\hat J}^{\mu}(x)|P',\chi'\rangle$. As follows from
Eqs. (\ref{78}), (\ref{81}), (\ref{82}), (\ref{87}) and (\ref{3.43})
\begin{eqnarray}
&&\langle P",\chi"|{\hat J}^{\mu}(x)|P',\chi'\rangle =
2(M"M')^{1/2}exp(\imath \Delta x)L(G",G')^{\mu}_{\nu}\cdot\nonumber\\
&&\cdot \langle \chi"|D[{\bf S};\alpha(G")^{-1}
\alpha(G",G')\alpha(f)]{\hat j}^{\nu}({\bf h})\cdot\nonumber\\
&&\cdot D[{\bf S};\alpha(G')^{-1}\alpha(G",G')\alpha(f')]^{-1}
|\chi' \rangle
\label{3.45}
\end{eqnarray}
where $\Delta=P"-P'$, $f$ and $f'$ are defined by Eq. (\ref{85}) with
$G$ replaced by $G"$, and the matrix element on the right-hand-side
must be calculated only in the space $H_{int}$.

 One of the consequences of Eq. (\ref{3.45}) is that the matrix
elements of the ECO in the N-particle case do not depend on the way
of constructing the N-particle operator $A$. In particular, the
matrix elements in the three-particle case indeed do not depend on
the way of constructing the operator $A$ in terms of $A_{12,3}$,
$A_{13,2}$ and $A_{23,1}$ (see Sec.~\ref{S2.3}). However this does
not imply that the operators $A$ play only a formal role since the
operator ${\hat M}_{int}$ for the N-particle system depends on the
operators $A$ for its subsystems.

 The main conclusion of this section is that the matrix elements of
the ECO are fully defined by the operator ${\hat j}^{\nu}({\bf h})$
which must satisfy the properties mentioned in Sec.~\ref{S3.3}. Let
us also note that the matrix element given by Eq. (\ref{3.45}) has
especially simple form if ${\bf G}"+{\bf G}'=0$ and in this case
\begin{equation}
\langle P",\chi"|{\hat J}^{\mu}(x)|P',\chi'\rangle =
2(M"M')^{1/2}exp(\imath \Delta x)\langle \chi"|
{\hat j}^{\mu}({\bf h})|\chi' \rangle
\label{3.46}
\end{equation}
It is interesting to note that the Breit frame is defined by the
condition ${\bf P}"+{\bf P}'=0$ which is not equivalent to
${\bf G}"+{\bf G}'=0$ if $M"\neq M'$.

\chapter{Explicit construction of the electromagnetic current
operator for different systems}
\label{C4}
\section{One-particle ECO}
\label{S4.1}

 In the one-particle case the method of constructing the operator
$J_{\mu}(x)$ given in Sec.~\ref{S3.3} leads to the well-known
results. Nevertheless we briefly describe some of these results since
they are used in the following.

 Now the role of $H_{int}$ is played by ${\cal D}({\bf s})$ (see
Sec.~\ref{S2.1}). Therefore $j^{\nu}({\bf h})$ is the operator which
acts only through the spin variables. Since the spectrum of the mass
operator consists only of one point corresponding to the particle
mass $m$, it follows from Eqs. (\ref{V4}) and (\ref{109}) that
${\bf j}(0)=j_{||}({\bf h})=0$ and, as follows from Eq. (\ref{108})
${\bf j}({\bf h})={\bf j}_{\bot}({\bf h})$.

 The number of independent functions determining the structure of the
operator $j^{\nu}({\bf h})$ depends on whether in addition to
Poincare invariance the P and T invariance are also assumed (see
ref.~\cite{pol1} for details). We suppose that this is the case.

 For the spinless particle the most general choice of $j^{\nu}({\bf h})$
under the above conditions is the following
\begin{equation}
j^0({\bf h})=\frac{e}{(1-{\bf h}^2)^{1/2}}
F_E(-\frac{4m^2{\bf h}^2}{(1-{\bf h}^2)^{1/2}}),\qquad
{\bf j}({\bf h})=0
\label{115}
\end{equation}
where $F_E$ is some real function such that $F_E(0)=1$. Then as
follows from Eqs. (\ref{6}), (\ref{81}), (\ref{82}) and (\ref{87}) the
action of the operator $J^{\mu}(0)=J^{\mu}$ is given by
\begin{equation}
J^{\mu}\varphi (g)=em^2\int\nolimits F_E((p-p')^2)(p+p')^{\mu}
\varphi (g') d\rho(g')
\label{116}
\end{equation}
where $p=mg$ and we take into account that, as follows from
Eqs. (\ref{12}), (\ref{13}) and (\ref{86})
\begin{equation}
|g+g'|=\frac{2}{(1-{\bf h}^2)^{1/2}},
\qquad (p-p')^2=-\frac{4m^2{\bf h}^2}{(1-{\bf h}^2)^{1/2}}
\label{117}
\end{equation}
In turn, as follows from Eqs. (\ref{6}), (\ref{79}), (\ref{113}) and
(\ref{116})
\begin{equation}
\langle p",\sigma"|J^{\mu}(x)|p',\sigma'\rangle =
eF_E(\Delta^2)exp(\imath \Delta x)(p"+p')^{\mu}
\label{118}
\end{equation}
where $\Delta=p"-p'$. This is the well known form of the matrix
element of the ECO in the spinless case and it is well-known that
$F_E$ has the sense of the electric form factor.

 If the spin of the particle under consideration is equal to 1/2,
then the most general choice of $j^{\nu}({\bf h})$ is the following
\begin{eqnarray}
&&j^0({\bf h})=eF_E(-\frac{4m^2{\bf h}^2}{(1-{\bf h}^2)^{1/2}}),\nonumber\\
&&{\bf j}({\bf h})=-\frac{2\imath e}{(1-{\bf h}^2)^{1/2}}
F_M(-\frac{4m^2{\bf h}^2}{(1-{\bf h}^2)^{1/2}})({\bf h}\times {\bf s})
\label{119}
\end{eqnarray}
where $F_E$ and $F_M$ are some real functions and $F_E(0)=1$.

\begin{sloppypar}
Using
Eqs. (\ref{6}), (\ref{81}), (\ref{82}) and (\ref{87}) one can verify
that instead of Eq. (\ref{116})
\begin{equation}
J^{\mu}\varphi(g)=2m^3\int\nolimits J^{\mu}(g,g')\varphi(g')d\rho(g')
\label{120}
\end{equation}
where $J^{\mu}(g,g')$ acts in ${\cal D}({\bf s})$ as
\begin{eqnarray}
&&J^{\mu}(g,g')\chi (\sigma)=\sum_{\sigma'} {\bar u}(p,\sigma)
[(F_E((p-p')^2)-F_M((p-p')^2))\cdot\nonumber\\
&&\frac{(p+p')^{\mu}}{(p+p')^2}+
\frac{1}{2m}F_M((p-p')^2))\gamma^{\mu}]u(p',\sigma') \chi(\sigma'),
\label{121}
\end{eqnarray}
$u(p,\sigma)$ is the Dirac byspinor describing the particle with the
4-momentum $p$ and the spin projection $\sigma$, $\gamma^{\mu}$ is
the Dirac $\gamma$ matrix and ${\bar u}=u^*\gamma^0$. In turn, as
follows from Eqs. (\ref{6}), (\ref{79}), (\ref{113}), (\ref{120}) and
(\ref{121})
\begin{eqnarray}
&&\langle p",\sigma"|J^{\mu}(x)|p',\sigma'\rangle =
e\cdot exp(\imath \Delta x){\bar u}(p",\sigma")
[2m\big(F_E(\Delta^2)-\nonumber\\
&&-F_M(\Delta^2))\frac{(p"+p')^{\mu}}
{(p"+p')^2}\big)+F_M(\Delta^2)\gamma^{\mu}]u(p',\sigma')
\label{122}
\end{eqnarray}
and, since this is the well known expression, we conclude that $F_E$
and $F_M$ are just the Sachs electric and magnetic form factors.
Therefore $F_M(0)$ is the particle magnetic moment in units $e/2m$.
\end{sloppypar}

 For the particle with unit spin a possible choice of $j^{\nu}({\bf
h})$ is
\begin{eqnarray}
j^0({\bf h})&=&\frac{e}{(1-{\bf h}^2)^{1/2}}
\{F_E(-\frac{4m^2{\bf h}^2}{(1-{\bf h}^2)^{1/2}})+\nonumber\\
&+&\frac{2}{(1-{\bf h}^2)^{1/2}}F_Q(-\frac{4m^2{\bf h}^2}
{(1-{\bf h}^2)^{1/2}})
[\frac{2}{3}{\bf h}^2-({\bf s}{\bf h})^2]\},\nonumber\\
{\bf j}({\bf h})&=&-\frac{\imath e}{1-{\bf h}^2}
F_M(-\frac{4m^2{\bf h}^2}{(1-{\bf h}^2)^{1/2}})
({\bf h}\times {\bf s})
\label{123}
\end{eqnarray}
where $F_E$, $F_M$ and $F_Q$ are some real functions and $F_E(0)=1$.
Using Eqs. (\ref{6}), (\ref{79}), (\ref{81}), (\ref{82}), (\ref{87}),
(\ref{113}) and (\ref{123}) it can be shown that
\begin{eqnarray}
&& \langle p",\sigma"|J^{\mu}(x)|p',\sigma'\rangle = e\cdot
exp(\imath \Delta x)E_{\alpha}^{"*}E'_{\beta}\cdot\nonumber\\
&&\cdot \{-(p"+p')^{\mu}[F_0(\Delta^2)\eta^{\alpha\beta}+
\frac{1}{2m^2}F_2(\Delta^2)\Delta^{\alpha}\Delta^{\beta}]+\nonumber\\
&&+F_1(\Delta^2)(\eta^{\alpha\mu}\Delta^{\beta}-
\eta^{\beta\mu}\Delta^{\alpha})\}
\label{124}
\end{eqnarray}
where $\alpha,\beta,\mu=0,1,2,3$, $E'$ is the polarization vector in
the initial state, $E"$ is the polarization vector in the final state
and the quantities $(F_0,F_1,F_2)$ are expressed in terms of
$(F_E,F_M,F_Q)$ as follows:
\begin{eqnarray}
&&F_E=\frac{1}{3}[(3+2\eta)F_0-2\eta F_1 -2\eta(1+\eta)F_2],\quad
F_M=F_1, \nonumber\\
&&F_Q=F_0-F_1-(1+\eta)F_2
\label{125}
\end{eqnarray}
where $\eta={\bf h}^2/(1-{\bf h}^2)$. Equations (\ref{124}) and
(\ref{125}) are
well known, and it is well known that $F_M(0)$ is the particle magnetic
moment in units $e/2m$ and $F_Q(0)$ is the particle quadrupole moment in
units $e/m^2$.

\section{ECO for systems of two particles}
\label{S4.2}

 Let us first consider the case when the particles do not interact
with each other. Then as follows from Eq. (\ref{103})
\begin{equation}
J^{\mu}(0)=J_1^{\mu}(0)+J_2^{\mu}(0)
\label{126}
\end{equation}
where $J_i^{\mu}(0)$ $(i=1,2)$ is the one-particle ECO for particle
$i$. The action of $J_1^{\mu}(0)$ in the two-particle Hilbert space
$H$ (see Sec.~\ref{S2.1}) is given by (see Eqs. (\ref{82}))
\begin{equation}
J_1^{\mu}(0)\varphi(g_1,g_2)=2m_1^3\int\nolimits J_1^{\mu}(g_1,g_1')
\varphi(g_1',g_2) d\rho(g_1')
\label{127}
\end{equation}
and the analogous expression can be written for the action of
$J_2^{\mu}(0)$. As follows from Eqs. (\ref{5}) and (\ref{14}), the
integral in this expression can be written in the form
\begin{eqnarray}
&&\int\nolimits J_1^{\mu}(g_1,g_1')
\varphi(g_1',g_2) d\rho(g_1')=\int\nolimits\int\nolimits
J_1^{\mu}(g_1,g_1')\varphi(g_1',g_2')\cdot\nonumber\\
&&2(2\pi)^3g_2^{'0}\delta^{(3)}({\bf g}_2-{\bf g}_2')d\rho(g_1')
d\rho(g_2')=\int\nolimits\int\nolimits
J_1^{\mu}(g_1,g_1')\varphi(G',{\bf q}')\cdot\nonumber\\
&&g_2^{'0}\delta^{(3)}({\bf g}_2-{\bf g}_2')\frac{M({\bf
q}')^3d^3{\bf q}'}{\omega_1({\bf q}')\omega_2({\bf q}')m_1^2m_2^2}
d\rho(G')
\label{128}
\end{eqnarray}
where the relations between the variables $G',{\bf q}'$ and
$g_1',g_2'$ have the same form as the relations between the variables
$G,{\bf q}$ and $g_1,g_2$ given by Eqs. (\ref{12}) and (\ref{13}). It is
assumed in the last integral that $g_i$ $(i=1,2)$ are functions of
$G,{\bf q}$ and $g_i'$ are functions of $G',{\bf q}'$. To integrate
over ${\bf q}'$ we use Eq. (\ref{13}) for calculating the Jacobian
\begin{equation}
|\frac{\partial^3{\bf g}_2'}{\partial^3{\bf q}'}|=\frac{g_2^{'0}}
{m_2\omega_2({\bf q}')}
\label{129}
\end{equation}
and therefore, as follows from
Eqs. (\ref{127}-\ref{129}), the action of $J_1^{\mu}(0)$ in the
variables $G,{\bf q}$ can be written in the form
\begin{equation}
J_1^{\mu}(0)\varphi(G,{\bf q})=2m_1\int\nolimits
J_1^{\mu}(g_1,g_1')\varphi(G',{\bf q}')\frac{M({\bf q}')^3}
{\omega_1({\bf q}')}d\rho(G')
\label{130}
\end{equation}
where it is assumed that $g_1$ depends on $G,{\bf q}$ and $g_1'$
depends on $G',{\bf q}'$ as above, but ${\bf q}'$ depends on $G,G'$
and ${\bf q}$. This dependence can be found from the condition
$g_2=g_2'$ (see Eq. (\ref{128})) and therefore, as follows from
Eq. (\ref{11})
\begin{equation}
q_2'=L[\alpha(G')^{-1}\alpha(G)]q_2
\label{131}
\end{equation}
where the 4-vector $q_2$ has the components $(\omega_2({\bf q}),-{\bf
q})$ and the 4-vector $q_2'$ has the components $(\omega_2({\bf q}'),
-{\bf q}')$.

 In the case considered in Sec.~\ref{S2.1} the operator $A$ entering
into Eq. (\ref{81}) is equal to unity and the operator ${\cal U}$ is
equal to $U_{12}$ given by Eq. (\ref{17}). Therefore, as follows from
Eqs. (\ref{12}), (\ref{81}) and (\ref{130}), the contribution of
$J_1^{\mu}(0)$ to the operator $J^{\mu}$ acting in the two-particle
space $H_{int}$ is given by
\begin{eqnarray}
J_1^{\mu}{\tilde \varphi}(G,{\bf q})&=& 2m_1\int\nolimits
U_{12}(G,{\bf q})^{-1}J_1^{\mu}(L[\alpha(G)]\frac{q_1}{m_1},
L[\alpha(G')]\frac{q_1'}{m_1})\cdot\nonumber\\
&\cdot&U_{12}(G,{\bf q}')\frac{M({\bf q}')^3}{\omega_1({\bf q}')}
{\tilde \varphi}(G',{\bf q}')d\rho(G')
\label{132}
\end{eqnarray}
Then the contribution of $J_1^{\mu}(0)$ to the two-particle operator
$j^{\nu}({\bf h})$ can be easily determined from Eqs. (\ref{82}) and
(\ref{87}) and analogously we can determine the contribution of
$J_2^{\mu}(0)$.

\begin{sloppypar}
 The final results for the operator $j^{\nu}({\bf h})$ corresponding
to the operator given by Eq. (\ref{126}) can be summarized as follows.
Let ${\bf d}_1$ and ${\bf d}_2$ be the vectors defined by the
condition that the Lorentz transformation
$L[\alpha(f')^{-1}\alpha(f)]$ transfers the 4-vector
$(\omega_2({\bf q}),-{\bf q})$ to $(\omega_2({\bf d}_1),-{\bf d}_1)$
and $(\omega_1({\bf q}),{\bf q})$ to $(\omega_1({\bf d}_2),{\bf d}_2)$
(compare with Eq. (\ref{131})). Using Eq. (\ref{7}) it is easy to
show that the explicit expressions for ${\bf d}_1$ and ${\bf d}_2$
in terms of ${\bf q}$ and ${\bf h}$ are the following:
\begin{equation}
{\bf d}_1={\bf q}-\frac{2{\bf h}}{1-{\bf h}^2}
[\omega_2({\bf q})-{\bf hq}], \qquad
{\bf d}_2={\bf q}+\frac{2{\bf h}}{1-{\bf h}^2}
[\omega_1({\bf q})+{\bf hq}]
\label{133}
\end{equation}
Let $d_1=(\omega_1({\bf d}_1),{\bf d}_1)$, $d_2=(\omega_2({\bf
d}_2),-{\bf d}_2)$ and $I_i({\bf h})$ $(i=1,2)$ be the operators
defined by the conditions $I_i({\bf h})\chi({\bf q})=\chi({\bf
d}_i)$. Then
\begin{eqnarray}
j^{\nu}({\bf h})&=&\sum_{i=1}^{2}U_{12}(f,{\bf q})^{-1}J_i^{\nu}
(L[\alpha(f)]\frac{q_i}{m_i},L[\alpha(f')]\frac{d_i}{m_i})\cdot\nonumber\\
&\cdot&U_{12}(f',{\bf d}_i)\frac{m_i}{\omega_i({\bf d}_i)}
\bigl[\frac{M({\bf d}_i)}{M({\bf q})}\bigr]^{3/2}I_i({\bf h})
\label{134}
\end{eqnarray}
\end{sloppypar}

 If the particles do not interact with each other, the operator
defined by Eq. (\ref{134}) automatically satisfies all the conditions
mentioned in Sec.~\ref{S3.3} since each operator $J_i^{\mu}(0)$
satisfies these conditions and $J_i^{\mu}(0)$ acts only through the
variables of particle $i$.

 Using Eqs. (\ref{17}) and (\ref{87}) it is possible to express
$j^{\nu}({\bf h})$ directly in terms of the single-particle operators
$j_i^{\nu}({\bf h})$. Let $f_i$ and $f'_i$ be the
4-vectors constructed by means of Eq. (\ref{85}) with $G$ replaced by
$L[\alpha(f)]q_i/m_i$ and $G'$ replaced by $L[\alpha(f')]d_i/m_i$,
and ${\bf h}_i={\bf f}_i/f_i^0$ . Then
\begin{eqnarray}
&&j^{\nu}({\bf h})=\sum_{i=1}^{2}L(L[\alpha(f)]\frac{q_i}{m_i},
L[\alpha(f')]\frac{d_i}{m_i})^{\mu}_{\nu}\cdot\nonumber\\
&& D[{\bf s}_k;\alpha(\frac{q_k}{m_k})^{-1}\alpha(f)^{-1}\alpha(f')
\alpha(\frac{d_k}{m_k})]\cdot\nonumber\\
&&D[{\bf s}_i;\alpha(\frac{q_i}{m_i})^{-1}\alpha(f)^{-1}
\alpha(L[\alpha(f)]\frac{q_i}{m_i},L[\alpha(f')]\frac{d_i}{m_i})
\alpha(f_i)]j_i^{\nu}({\bf h})\cdot\nonumber\\
&&D[{\bf s}_i;\alpha(f'_i)^{-1}\alpha(L[\alpha(f)]\frac{q_i}{m_i},
L[\alpha(f')]\frac{d_i}{m_i})^{-1}
\alpha(f')\alpha(\frac{d_i}{m_i})]\nonumber\\
&&\frac{m_i}{\omega_i({\bf d}_i)}
\bigl[\frac{M({\bf d}_i)}{M({\bf q})}\bigr]^{3/2}I_i({\bf h})
\label{134a}
\end{eqnarray}
where $k=2$ if $i=1$ and $k=1$ if $i=2$.

 If ${\bf h}=0$ then (see Eq. (\ref{133})) ${\bf d}_1={\bf d}_2={\bf
q}$ and therefore $I_i(0)=1$, ${\bf h}_i=0$. Since $j_i^0(0)=e_i$ and
${\bf j}_i(0)=0$ where $e_i$ is the electric charge of particle $i$ (see
Sec.~\ref{S4.1}), it follows from Eq. (\ref{134a}) that
\begin{equation}
j^0(0)=e_1+e_2,\qquad {\bf j}(0)={\bf q}(\frac{e_1}{\omega_1({\bf
q})}-\frac{e_2}{\omega_2({\bf q})})
\label{135}
\end{equation}
The first expression obviously follows from Eq. (\ref{98}) and the
second expression shows that ${\bf j}(0)$ commutes with the free mass
operator and therefore satisfies Eq. (\ref{93f}).

 When the particles interact with each other the results of
Sec.~\ref{S3.3} imply that to solve the problem of constructing the
operator ${\hat J}^{\mu}(x)$ we should find the operators
$w^0({\bf h})$ and ${\bf w}_{\bot}({\bf h})$ such that
\begin{equation}
{\hat j}^0({\bf h})=j^0({\bf h})+w^0({\bf h}),\quad {\hat {\bf
j}}_{\bot}({\bf h})={\bf j}_{\bot}({\bf h})+{\bf w}_{\bot}({\bf h}),
\label{136a}
\end{equation}
\begin{eqnarray}
&&w^0({\bf h})=D[{\bf S};u]w^0(L(u)^{-1}{\bf h})
D[{\bf S};u]^{-1},\nonumber\\
&& w^0({\bf h})^*=w^0(-{\bf h}),\quad w^0(0)=0,\nonumber\\
&&{\bf w}_{\bot}({\bf h})=L(u)D[{\bf S};u]{\bf w}_{\bot}
(L(u)^{-1}{\bf h})D[{\bf S};u]^{-1},\nonumber\\
&& {\bf w}_{\bot}({\bf h})^*={\bf w}_{\bot}(-{\bf h}),
\label{136b}
\end{eqnarray}
and the operators $w^0({\bf h})$ and ${\bf w}_{\bot}({\bf h})$
become zero when the interaction is turned off. In addition, it is easy
to show that if the theory should be P and T invariant, then, as follows
from Eq. (\ref{112a})
\begin{eqnarray}
{\tilde U}_P(w^0({\bf h}),{\bf w}_{\bot}({\bf h}))
{\tilde U}_P^{-1}&=&(w^0(-{\bf h}),-{\bf w}_{\bot}(-{\bf
h})),\nonumber\\
 {\tilde U}_R(w^0({\bf h}),{\bf w}_{\bot}
({\bf h})){\tilde U}_R^{-1}&=&(w^0({\bf h}),{\bf w}_{\bot}({\bf h}))
\label{136c}
\end{eqnarray}
 Then the operator ${\hat J}^{\mu}(x)$ is defined by
Eqs. (\ref{78}), (\ref{81}), (\ref{82}), (\ref{87}), (\ref{93e}),
(\ref{108}) and (\ref{109}).

 We see that a possible choice for $w^0({\bf h})$ and
${\bf w}_{\bot}({\bf h})$ is $w^0({\bf h})={\bf w}_{\bot}({\bf h})=0$
and therefore ${\hat j}^0({\bf h})=j^0({\bf h})$,
${\hat {\bf j}}_{\bot}({\bf h})={\bf j}_{\bot}({\bf h})$. Let us note
that even in this case ${\hat{\bf j}}(0)\neq {\bf j}(0)$ and
${\hat j}_{||}({\bf h})\neq j_{||}({\bf h})$
since, as follows from Eqs. (\ref{93e}) and (\ref{109}), ${\hat{\bf j}}(0)$
and ${\hat j}_{||}({\bf h})$ depend not only on ${\hat j}^0({\bf
h})$ but also on ${\hat M}_{int}$.  Another reasons for the operator
${\hat J}^{\mu}(x)$ to differ from $J^{\mu}(x)$ are the presence of
${\hat M}_{int}$ in Eq. (\ref{82}) and the presence of ${\hat P}$ in
Eq. (\ref{78}).

 We conclude that there exists an infinite number of solutions for the
two-particle operator ${\hat J}^{\mu}(x)$ since there exists an infinite
number of sets $w^0({\bf h})$ and ${\bf w}_{\bot}({\bf h})$ satisfying
Eq. (\ref{136b}). In particular, a possible solution
corresponds to the choice $w^0({\bf h})={\bf w}_{\bot}({\bf h})=0$,
but there are no physical grounds to prefer this solution in
comparison with the others.

\section{ECO for systems of three particles}
\label{S4.3}

 If particles $\alpha$ and $\beta$ interact with each other and
particle $\gamma$ is free ($\alpha,\beta,\gamma=1,2,3$, $\alpha
\neq \beta\neq \gamma$) then, as follows from Eqs. (\ref{35}),
(\ref{103}) and (\ref{104}), the action of the three-particle ECO in
the three-particle space $\tilde H$ is given by
\begin{eqnarray}
J_{\alpha\beta,\gamma}^{\mu}&=&U_{123}^{-1}U_{\alpha\beta}
B_{\alpha\beta,\gamma}(M_{\alpha\beta})^{-1}B_{\alpha\beta,\gamma}
({\hat M}_{\alpha\beta}^{int})[{\hat J}_{\alpha\beta}^{\mu}+\nonumber\\
&+&J_{\gamma}^{\mu}]B_{\alpha\beta,\gamma}({\hat M}_{\alpha
\beta}^{int})^{-1}B_{\alpha\beta,\gamma}(M_{\alpha\beta})
U_{\alpha\beta}^{-1}U_{123}
\label{137}
\end{eqnarray}
where ${\hat J}_{\alpha\beta}^{\mu}$ is the ECO of the interacting
system $\alpha\beta$ in the two-particle space ${\tilde
H}_{\alpha\beta}$. In turn, as follows from Eqs. (\ref{105}) and
(\ref{106}), this action is fully determined by the operators
$j_{\alpha\beta,\gamma}^{\nu}({\bf h})$ such that
\begin{equation}
J_{\alpha\beta,\gamma}^{\mu}{\tilde \varphi}(G)=2\int\nolimits
(M_{\alpha\beta,\gamma}^{int})^{3/2}
J_{\alpha\beta,\gamma}^{\mu}(G,G')(M_{\alpha\beta,\gamma}^{int})^{3/2}
{\tilde \varphi}(G')d\rho(G')
\label{138}
\end{equation}
\begin{eqnarray}
&&J_{\alpha\beta,\gamma}^{\mu}(G,G')=L(G,G')^{\mu}_{\nu}
D[{\bf S};\alpha(G)^{-1}\alpha(G,G')\alpha(f)]\cdot\nonumber\\
&&j_{\alpha\beta,\gamma}^{\nu}({\bf h})
D[{\bf S};\alpha(G')^{-1}\alpha(G,G')\alpha(f')]^{-1}
\label{139}
\end{eqnarray}
where ${\bf S}$ is the three-particle spin operator (see Sec.
{}~\ref{S2.3}).

 Analogously, if all the three particles do not interact with each
other, the action of the three-particle ECO in the three-particle
space $\tilde H$ is fully determined by the operator $j^{\nu}({\bf
h})$ such that
\begin{eqnarray}
&&J^{\mu}=U_{123}^{-1}(\sum_{i=1}^{3}J_i^{\mu})U_{123},\nonumber\\
&&J^{\mu}{\tilde \varphi}(G)=2\int\nolimits
M^{3/2}J^{\mu}(G,G')M^{3/2}{\tilde \varphi}(G')d\rho(G'),\nonumber\\
&&J^{\mu}(G,G')=L(G,G')^{\mu}_{\nu}D[{\bf S};\alpha(G)^{-1}
\alpha(G,G')\alpha(f)]j^{\nu}({\bf h})\cdot\nonumber\\
&&D[{\bf S};\alpha(G')^{-1}\alpha(G,G')\alpha(f')]^{-1}
\label{140}
\end{eqnarray}

 It is obvious that $j_{\alpha\beta,\gamma}^{\nu}({\bf h})$ becomes
$j^{\nu}({\bf h})$ when the interaction between particles $\alpha$
and $\beta$ is turned off and therefore cluster separability for
the three-particle operator ${\hat j}^{\nu}({\bf h})$ will be
satisfied if ${\hat j}^{\nu}({\bf h})$ becomes
$j_{\alpha\beta,\gamma}^{\nu}({\bf h})$ when all interactions
involving any particle $\gamma$ are turned off (see Sec.~\ref{S3.3}).
Therefore to explicitly construct the operator ${\hat j}^{\nu}({\bf h})$
we have to determine first the explicit expressions for the operators
$j_{\alpha\beta,\gamma}^{\nu}({\bf h})$ and $j^{\nu}({\bf h})$.

 As follows from Eqs. (\ref{137}-\ref{140}),
\begin{equation}
j_{\alpha\beta,\gamma}^{\nu}({\bf h})=j_{(\alpha\beta)}^{\nu}({\bf h})
+j_{(\gamma)}^{\nu}({\bf h}),\quad j^{\nu}({\bf h})=\sum_{i=1}^{3}
{\iota}_i^{\nu}({\bf h})
\label{141}
\end{equation}
where $j_{(\alpha\beta)}^{\nu}({\bf h})$ is the contribution of
${\hat J}_{\alpha\beta}^{\mu}$ to $j_{\alpha\beta,\gamma}^{\nu}({\bf
h})$, $j_{(\gamma)}^{\nu}({\bf h})$ is the contribution of
$J_{\gamma}^{\mu}$ to $j_{\alpha\beta,\gamma}^{\nu}({\bf h})$ and
${\iota}_i^{\nu}({\bf h})$ is the contribution of $J_i^{\mu}$ to
$j^{\nu}({\bf h})$. It is important to note that all the operators in
Eq. (\ref{141}) act in the {\it three-particle} Hilbert space
$H_{int}$. We suppose that the single-particle and two-particle ECO's
have been already constructed as described in Secs.~\ref{S4.1} and
{}~\ref{S4.2}, and our goal is to express the action of the operators
in Eq. (\ref{141}) in terms of the single-particle operators
$j_{\gamma}^{\nu}({\bf h})$ and two-particle operators
${\hat j}_{\alpha\beta}^{\nu}({\bf h})$ acting in the internal
single-particle and two-particle spaces respectively.

 The calculation of the operators entering into Eq. (\ref{141}) is
tedious but all preparatory expressions have been already derived
above. Let us briefly describe the calculation of
$j_{(\alpha\beta)}^{\nu}({\bf h})$ which is most cumbersome.

 We have to calculate the action of $J_{\alpha\beta,\gamma}^{\nu}$
(see Eq. (\ref{137})) on the function ${\tilde \varphi}(G)\in {\tilde H}$
which depends additionally on ${\bf K}_{\alpha\beta}$,
${\bf k}_{\alpha\beta}$ and the spin variables of the three particles
(see Sec~\ref{S2.3}). The action of
$$B_{\alpha\beta,\gamma}({\hat M}_{\alpha
\beta}^{int})^{-1}B_{\alpha\beta,\gamma}(M_{\alpha\beta})
U_{\alpha\beta}^{-1}U_{123}$$
on this function can be calculated using Eqs. (\ref{17}), (\ref{36}),
(\ref{47}), (\ref{49}) and (\ref{56}). Then to calculate the action of
${\hat J}_{\alpha\beta}^{\mu}$ we use Eq. (\ref{82}) and write the
kernel of this operator in the form (see Eq. (\ref{91a}))
\begin{equation}
{\hat J}_{\alpha\beta}^{\mu}(G,G')=\int\nolimits\int\nolimits
d{\hat e}_{\alpha\beta}^{int}(m){\hat J}_{\alpha\beta}^{\mu}
(G,m;G',m')d{\hat e}_{\alpha\beta}^{int}(m')
\label{142}
\end{equation}
We change the integration variable from $G'_{\alpha\beta}$ to
$G'=G'(m')=(m'G'_{\alpha\beta}+m_{\gamma}g_{\gamma})/
|m'G'_{\alpha\beta}+m_{\gamma}g_{\gamma}|$ using Eq. (\ref{45}).
Then we again use Eqs. (\ref{17}), (\ref{36}), (\ref{47}),
(\ref{49}) and (\ref{56}) in order to calculate the action of
$$U_{123}^{-1}U_{\alpha\beta}B_{\alpha\beta,\gamma}(M_{\alpha\beta})^{-1}
B_{\alpha\beta,\gamma}({\hat M}_{\alpha\beta}^{int})$$
Expressing all the variables in terms of $G,{\bf K}_{\alpha\beta},
{\bf k}_{\alpha\beta}$ and using Eqs. (\ref{66}) and (\ref{82}) we can
determine the contribution of ${\hat J}_{\alpha\beta}^{\mu}$ to
$$J_{(\alpha\beta)}^{\mu}(G,G').$$ It
is very convenient to use the relation between the operators
$O_{\alpha\beta}$ in $H_{\alpha\beta}^{int}$ and
${\tilde O}_{\alpha\beta}$ in $\tilde H$ (see Eqs. (\ref{68}) and
(\ref{69})). Finally, as follows from Eq. (\ref{87}),
$j_{(\alpha\beta)}^{\mu}({\bf h})=J_{(\alpha\beta)}^{\mu}(f,f')$.

 The result of such a calculation is the following. We use
$k_{\gamma}$ and $k'_{\gamma}$ to denote the 4-vectors
$$(\omega_{\gamma}({\bf K}_{\alpha\beta}),-{\bf K}_{\alpha\beta})
\quad \mbox{and}\quad
(\omega_{\gamma}({\bf K}'_{\alpha\beta}),-{\bf K}'_{\alpha\beta})$$
respectively where ${\bf K}'_{\alpha\beta}$ is defined by the
condition (compare with Eq. (\ref{131})) $k'_{\gamma}=L[\alpha(f')^{-1}
\alpha(f)]k_{\gamma}$. Then (compare with Eq. (\ref{133}))
\begin{equation}
{\bf K}'_{\alpha\beta}={\bf K}_{\alpha\beta}-
\frac{2{\bf h}}{1-{\bf h}^2}[\omega_{\gamma}({\bf K}_{\alpha\beta})
-{\bf h}{\bf K}_{\alpha\beta}]
\label{143}
\end{equation}
We use $K_{\alpha\beta}(m)$ and $K'_{\alpha\beta}(m')$ to denote the
4-vectors
$$((m^2+{\bf K}_{\alpha\beta}^2)^{1/2},{\bf K}_{\alpha\beta})\quad
\mbox{and} \quad ((m^{'2}+{\bf K}_{\alpha\beta}^{'2})^{1/2},{\bf
K}'_{\alpha\beta})$$
respectively and use $G_{\alpha\beta}(m)$ and
$G'_{\alpha\beta}(m')$ to denote the 4-vectors
$$L[\alpha(f)]K_{\alpha\beta}(m)/m \quad
\mbox{and} \quad L[\alpha(f')]K'_{\alpha\beta}(m')/m'$$
respectively (compare with Eq. (\ref{43})). Then
\begin{eqnarray}
&&j_{(\alpha\beta)}^{\mu}({\bf h})\chi({\bf K}_{\alpha\beta},
{\bf k}_{\alpha\beta})=D[{\bf s}_{\gamma};\alpha(\frac{k_{\gamma}}
{m_{\gamma}})^{-1}\alpha(f)^{-1}\alpha(f')\alpha(\frac{k'_{\gamma}}
{m_{\gamma}})]\cdot\nonumber\\
&&\int\nolimits\int\nolimits
\bigl[\frac{mm'}{K_{\alpha\beta}^0(m)K_{\alpha\beta}^{'0}(m')}
\bigr]^{1/2} D[{\tilde {\bf S}}_{\alpha\beta};
\alpha(\frac{K_{\alpha\beta}(m)}{m})^{-1}
\alpha(f)^{-1}\alpha(G_{\alpha\beta}(m))]\cdot\nonumber\\
&&d{\hat {\tilde e}}_{\alpha \beta}^{int}(m){\hat
{\tilde J}}_{\alpha\beta}^{\mu}(G_{\alpha\beta}(m),m;
G'_{\alpha\beta}(m'),m')d{\hat {\tilde e}}_{\alpha
\beta}^{int}(m')\cdot\nonumber\\
&&D[{\tilde {\bf S}}_{\alpha\beta};\alpha(G'_{\alpha\beta}(m'))^{-1}
\alpha(f')\alpha(\frac{K'_{\alpha\beta}(m')}{m'})]
\chi({\bf K}'_{\alpha\beta},{\bf k}_{\alpha\beta})
\label{144}
\end{eqnarray}
In practical computations one may take into account that if $u\in
$SU(2) then
\begin{equation}
D[{\tilde {\bf S}}_{\alpha\beta};u]\chi({\bf K}_{\alpha\beta},
{\bf k}_{\alpha\beta})=(\prod_{i=\alpha,\beta}D[{\bf s}_i;u])
\chi({\bf K}_{\alpha\beta},L[\alpha(u)]^{-1}
{\bf k}_{\alpha\beta})
\label{144a}
\end{equation}

 Using Eq. (\ref{87}) it is possible to express
$j_{(\alpha\beta)}^{\mu}({\bf h})$ directly in terms of
$j_{\alpha\beta}^{\mu}({\bf h},m,m')$. Let $f_{\alpha\beta}(m,m')$
and $f'_{\alpha\beta}(m,m')$ be the 4-vectors constructed by means of
Eq. (\ref{85}) with $G$ replaced by $G(m)$ and $G'$ replaced by
$G'(m')$, and
$${\bf h}_{\alpha\beta}(m,m')=\frac{{\bf f}_{\alpha\beta}(m,m')}
{f_{\alpha\beta}^0(m,m')}$$
Then
\begin{eqnarray}
&&j_{(\alpha\beta)}^{\mu}({\bf h})\chi({\bf K}_{\alpha\beta},
{\bf k}_{\alpha\beta})=D[{\bf s}_{\gamma};\alpha(\frac{k_{\gamma}}
{m_{\gamma}})^{-1}\alpha(f)^{-1}\alpha(f')\alpha(\frac{k'_{\gamma}}
{m_{\gamma}})]\cdot\nonumber\\
&&\int\nolimits\int\nolimits D[{\tilde {\bf S}}_{\alpha\beta};
\alpha(\frac{K_{\alpha\beta}(m)}{m})^{-1}\alpha(f)^{-1}
\alpha(G_{\alpha\beta}(m),G'_{\alpha\beta}(m'))\cdot\nonumber\\
&& \alpha(f_{\alpha\beta}(m,m'))]
d{\hat {\tilde e}}_{\alpha \beta}^{int}(m)
L[G_{\alpha\beta}(m),G'_{\alpha\beta}(m')]^{\mu}_{\nu}\cdot\nonumber\\
&& {\hat {\tilde j}}^{\nu}_{\alpha\beta}
({\bf h}_{\alpha\beta}(m,m');m,m')
d{\hat {\tilde e}}_{\alpha \beta}^{int}(m')
D[{\tilde {\bf S}}_{\alpha\beta};
\alpha(f'_{\alpha\beta}(m,m'))^{-1} \cdot\nonumber\\
&& \alpha(G_{\alpha\beta}(m),G'_{\alpha\beta}(m'))^{-1}
\alpha(f')\alpha(\frac{K'_{\alpha\beta}(m')}{m'})]\cdot\nonumber\\
&&\bigl[\frac{mm'}{K_{\alpha\beta}^0(m)K_{\alpha\beta}^{'0}(m')}
\bigr]^{1/2}\chi({\bf K}'_{\alpha\beta},{\bf k}_{\alpha\beta})
\label{145}
\end{eqnarray}

 The result for $j_{(\gamma)}^{\mu}({\bf h})$ is the following. We again
use $k_{\gamma}$, $k'_{\gamma}$, $K_{\alpha\beta}(m)$ and
$K'_{\alpha\beta}(m)$ to denote the quantities expressed in terms of
${\bf K}_{\alpha\beta}$ and ${\bf K}'_{\alpha\beta}$ as above, but
now ${\bf K}'_{\alpha\beta}={\bf K}'_{\alpha\beta}(m)$ is defined by
the condition $L[\alpha(f)]K_{\alpha\beta}(m)/m=L[\alpha(f')]
K'_{\alpha\beta}(m')/m'$ and therefore (compare with the second
expression in Eq. (\ref{133}))
\begin{equation}
{\bf K}'_{\alpha\beta}(m)={\bf K}_{\alpha\beta}+
\frac{2{\bf h}}{1-{\bf h}^2}[(m^2+{\bf K}_{\alpha\beta}^2)^{1/2}+
{\bf h}{\bf K}_{\alpha\beta}]
\label{146}
\end{equation}
Then
\begin{eqnarray}
&&j_{(\gamma)}^{\mu}({\bf h})\chi({\bf K}_{\alpha\beta},
{\bf k}_{\alpha\beta})=D[{\bf s}_{\gamma};\alpha(\frac{k_{\gamma}}
{m_{\gamma}})^{-1}\alpha(f)^{-1}\alpha(L[\alpha(f)]\frac{k_{\gamma}}
{m_{\gamma}})]\cdot\nonumber\\
&&\int\nolimits d{\hat {\tilde e}}_{\alpha \beta}^{int}(m)
D[{\tilde {\bf S}}_{\alpha\beta};\alpha(\frac{K_{\alpha\beta}(m)}
{m})^{-1}\alpha(f)^{-1}\alpha(f')
\alpha(\frac{K'_{\alpha\beta}(m)}{m})]\cdot\nonumber\\
&&\frac{m_{\gamma}}{\omega_{\gamma}({\bf K}'_{\alpha\beta})}
[\frac{K^{'0}_{\alpha\beta}(m)}{K^0_{\alpha\beta}(m)}]^{1/2}
j_{\gamma}^{\mu}(L[\alpha(f)]\frac{k_{\gamma}}{m_{\gamma}},
L[\alpha(f')]\frac{k'_{\gamma}(m)}{m_{\gamma}})\cdot\nonumber\\
&&D[{\bf s}_{\gamma};\alpha(L[\alpha(f')]\frac{k'_{\gamma}(m)}
{m_{\gamma}})^{-1}\alpha(f')\alpha(\frac{k'_{\gamma}(m)}{m_{\gamma}})]
\chi({\bf K}'_{\alpha\beta},{\bf k}_{\alpha\beta})
\label{147}
\end{eqnarray}

 Using Eq. (\ref{87}) it is possible to express
$j_{(\gamma)}^{\mu}({\bf h})$ directly in terms of
$j_{\gamma}^{\mu}({\bf h})$. Let $f_{\gamma}(m)$ and $f'_{\gamma}(m)$
be the 4-vectors constructed by means of Eq. (\ref{85}) with $G$
replaced by $L[\alpha(f)]k_{\gamma}/m_{\gamma}$ and $G'$
replaced by $L[\alpha(f')]k'_{\gamma}(m)/m_{\gamma}$, and
${\bf h}_{\gamma}(m)={\bf f}_{\gamma}(m)/f_{\gamma}^0(m)$. Then
\begin{eqnarray}
&&j_{(\gamma)}^{\mu}({\bf h})\chi({\bf K}_{\alpha\beta},
{\bf k}_{\alpha\beta})=\int\nolimits d{\hat
{\tilde e}}_{\alpha\beta}^{int}(m)L(L[\alpha(f)]\frac{k_{\gamma}}
{m_{\gamma}},L[\alpha(f')]\frac{k'_{\gamma}(m)}
{m_{\gamma}})_{\nu}^{\mu}\cdot\nonumber\\
&&D[{\bf s}_{\gamma};\alpha(\frac{k_{\gamma}}{m_{\gamma}})^{-1}
\alpha(f)^{-1}\alpha(L[\alpha(f)]\frac{k_{\gamma}}{m_{\gamma}},
L[\alpha(f')]\frac{k'_{\gamma}(m)}{m_{\gamma}})
\alpha(f_{\gamma}(m))]\cdot\nonumber\\
&&D[{\tilde {\bf S}}_{\alpha\beta};\alpha(\frac{K_{\alpha\beta}(m)}
{m})^{-1}\alpha(f)^{-1}\alpha(f')\alpha(\frac{K'_{\alpha\beta}(m)}{m})]
j_{\gamma}^{\nu}({\bf h}_{\gamma}(m))\cdot\nonumber\\
&&D[{\bf s}_{\gamma};\alpha(f'_{\gamma}(m))^{-1}
\alpha(L[\alpha(f)]\frac{k_{\gamma}}{m_{\gamma}},
L[\alpha(f')]
\frac{k'_{\gamma}(m)}{m_{\gamma}})^{-1}\alpha(f')\cdot\nonumber\\
&&\alpha(\frac{k'_{\gamma}(m)}{m_{\gamma}})]
\frac{m_{\gamma}}{\omega_{\gamma}
({\bf K}'_{\alpha\beta})}[\frac{K^{'0}_{\alpha\beta}(m)}
{K^0_{\alpha\beta}(m)}]^{1/2}
\chi({\bf K}'_{\alpha\beta}(m),{\bf k}_{\alpha\beta})
\label{148}
\end{eqnarray}

 The result for $\iota_{\gamma}^{\mu}({\bf h})$ follows from
Eq. (\ref{148}):
\begin{eqnarray}
&&\iota_{\gamma}^{\mu}({\bf h})\chi({\bf K}_{\alpha\beta},
{\bf k}_{\alpha\beta})=L(L[\alpha(f)]\frac{k_{\gamma}}{m_{\gamma}},
L[\alpha(f')]\frac{k'_{\gamma}}{m_{\gamma}})_{\nu}^{\mu}
\cdot\nonumber\\
&&D[{\bf s}_{\gamma};\alpha(\frac{k_{\gamma}}{m_{\gamma}})^{-1}
\alpha(f)^{-1}\alpha(L[\alpha(f)]\frac{k_{\gamma}}{m_{\gamma}},
L[\alpha(f')]\frac{k'_{\gamma}}{m_{\gamma}})
\alpha(f_{\gamma})]\cdot\nonumber\\
&&D[{\tilde {\bf S}}_{\alpha\beta};\alpha(\frac{K_{\alpha\beta}}
{M_{\alpha\beta}})^{-1}\alpha(f)^{-1}\alpha(f')
\alpha(\frac{K'_{\alpha\beta}}{M_{\alpha\beta}})]
j_{\gamma}^{\nu}({\bf h}_{\gamma})\cdot\nonumber\\
&&D[{\bf s}_{\gamma};\alpha(f'_{\gamma})^{-1}
\alpha(L[\alpha(f)]\frac{k_{\gamma}}{m_{\gamma}},
L[\alpha(f')]\frac{k'_{\gamma}}{m_{\gamma}})^{-1}\alpha(f')
\alpha(\frac{k'_{\gamma}}{m_{\gamma}})]\cdot\nonumber\\
&&\frac{m_{\gamma}}{\omega_{\gamma}
({\bf K}'_{\alpha\beta})}[\frac{K^{'0}_{\alpha\beta}}
{K^0_{\alpha\beta}}]^{1/2}
\chi({\bf K}'_{\alpha\beta},{\bf k}_{\alpha\beta})
\label{149}
\end{eqnarray}
where the quantities ${\bf K}'_{\alpha\beta}$, $k'_{\gamma}$,
$K_{\alpha\beta}$, $K'_{\alpha\beta}$, $f_{\gamma}$ and $f'_{\gamma}$
depend on $M_{\alpha\beta}=M_{\alpha\beta}({\bf k}_{\alpha\beta})$ in
the same manner as they depend on $m$ in Eq. (\ref{148}).

 Equations (\ref{141}), (\ref{144a}), (\ref{145}), (\ref{146}),
(\ref{148}) and (\ref{149}) explicitly describe the operators
$j_{\alpha\beta,\gamma}^{\nu}({\bf h})$ and $j^{\nu}({\bf h})$. As
follows from the results of Sec.~\ref{S3.2},
$j_{\alpha\beta,\gamma}^0(0)=j^0(0)=e_{\alpha}+e_{\beta}+e_{\gamma}$.
This can be verified directly since, as follows from Eqs. (\ref{90}),
(\ref{148}) and (\ref{149}), $j_{(\gamma)}^0(0)=\iota_{(\gamma)}^0(0)=
e_{\gamma}$ since $j_{\gamma}^0(0)=e_{\gamma}$, and if ${\bf h}=0$
then ${\bf K}'_{\alpha\beta}={\bf K}_{\alpha\beta}$ in Eq. (\ref{144})
and
\begin{eqnarray}
j_{(\alpha\beta)}^0(0)&=&\int\nolimits\int\nolimits
\bigl[\frac{mm'}{K_{\alpha\beta}^0(m)K_{\alpha\beta}^{'0}(m')}
\bigr]^{1/2}{\hat {\tilde e}}_{\alpha \beta}^{int}(m){\hat
{\tilde J}}_{\alpha\beta}^0(\frac{K_{\alpha\beta}(m)}{m},m;\nonumber\\
&&\frac{K'_{\alpha\beta}(m')}{m'},m')d{\hat {\tilde e}}_{\alpha
\beta}^{int}(m')
\label{150}
\end{eqnarray}
As follows from Eq. (\ref{91b}),
$${\hat {\tilde J}}_{\alpha\beta}^0(\frac{K_{\alpha\beta}(m)}{m},m;
\frac{K'_{\alpha\beta}(m')}{m'},m')=0$$
if $m\neq m'$. Therefore only the values of $m'$ in an infinitely
small vicinity of the point $m$ may contribute to Eq. (\ref{150}). Then
Eq. (\ref{150}) is a special case of Eq. (\ref{94}) for $\tau=0$ and
$\lambda=(1,0,0,0)$. Therefore
$j_{(\alpha\beta)}^0(0)=e_{\alpha}+e_{\beta}$ as it should be.

 As follows from the results of Sec.~\ref{S3.3}, we can seek the
three-particle operators ${\hat j}^0({\bf h})$ and
${\hat {\bf j}}_{\bot}({\bf h})$ in the form (compare with
Eq. (\ref{71}))
\begin{eqnarray}
&&{\hat j}^0({\bf h})=j_{12,3}^0({\bf h})+j_{13,2}^0({\bf h}) +
j_{23,1}^0({\bf h})-2j^0({\bf h})+w^0({\bf h})=\nonumber\\
&&=j_{(12)}^0({\bf h})+j_{(13)}^0({\bf h})+j_{(23)}^0({\bf h})+
j_{(1)}^0({\bf h})+j_{(2)}^0({\bf h})+j_{(3)}^0({\bf h})-\nonumber\\
&&-2\iota_1^0({\bf h})-2\iota_2^0({\bf h})-2\iota_3^0({\bf h})+
w^0({\bf h})
\label{152}
\end{eqnarray}
\begin{eqnarray}
&&{\hat {\bf j}}_{\bot}({\bf h})={\bf j}_{\bot 12,3}({\bf h})+
{\bf j}_{\bot 13,2}({\bf h})+{\bf j}_{\bot 23,1}({\bf h})-
2{\bf j}_{\bot}({\bf h})+{\bf w}_{\bot}({\bf h})=\nonumber\\
&&={\bf j}_{\bot(12)}({\bf h})+{\bf j}_{\bot(13)}({\bf h})+
{\bf j}_{\bot(23)}({\bf h})+{\bf j}_{\bot(1)}({\bf h})+
{\bf j}_{\bot(2)}({\bf h})+\nonumber\\
&&+{\bf j}_{\bot(3)}({\bf h})-
2{\mbox {\boldmath $\iota$} }_{\bot 1}({\bf h})-
2{\mbox {\boldmath $\iota$} }_{\bot 2}({\bf h})-
2{\mbox {\boldmath $\iota$} }_{\bot 3}({\bf h})+{\bf w}_{\bot}({\bf h})
\label{153}
\end{eqnarray}
where the operators $w^0({\bf h})$ and ${\bf w}_{\bot}({\bf h})$
describe the contribution of {\it three-body interactions} to
${\hat j}^0({\bf h})$ and ${\hat {\bf j}}_{\bot}({\bf h})$. In other
words, $w^0({\bf h})$ and ${\bf w}_{\bot}({\bf h})$ must be zero if
any particle does not interact with the others. These operators must
satisfy the relations which can be written in the form of
Eq. (\ref{136b}).

 We conclude  that a possible choice of $w^0({\bf h})$ and
${\bf w}_{\bot}({\bf h})$ in Eqs. (\ref{152}) and (\ref{153}) is
$w^0({\bf h})={\bf w}_{\bot}({\bf h})=0$, but in the general case
there exists an infinite number of solutions satisfying the
conditions described in Sec.~\ref{S1.1}.

\section{ECO for systems with any number of particles}
\label{S4.4}

 As follows from the results of Sec.~\ref{S3.3}, we can seek the
operators ${\hat j}^0({\bf h})$ and ${\hat {\bf j}}_{\bot}({\bf h})$ in
the form (compare with Eq. (\ref{77}))
\begin{eqnarray}
&&{\hat j}^0({\bf h})=\sum_{k=2}^{N} (-1)^k(k-1)!j_{(k)}^0({\bf h})
+ w^0({\bf h}),\nonumber\\
&& j_{(k)}^0({\bf h})=\sum_{\alpha_1...\alpha_k}
j_{\alpha_1...\alpha_k}^0({\bf h})
\label{154}
\end{eqnarray}
\begin{eqnarray}
&&{\hat {\bf j}}_{\bot}({\bf h})=\sum_{k=2}^{N}
(-1)^k(k-1)!{\bf j}_{\bot (k)}({\bf h})
+  {\bf w}_{\bot}({\bf h}),\nonumber\\
&& {\bf j}_{\bot (k)}({\bf h})=\sum_{\alpha_1...\alpha_k}
{\bf j}_{\bot\alpha_1...\alpha_k}({\bf h})
\label{155}
\end{eqnarray}
where $w^0({\bf h})$ and ${\bf w}_{\bot}({\bf h})$ describe the
contribution of {\it N-body interactions} to ${\hat j}^0({\bf h})$
and ${\hat {\bf j}}_{\bot}({\bf h})$ respectively. The operators
$w^0({\bf h})$ and ${\bf w}_{\bot}({\bf h})$ must satisfy the relations
which can be written in the form of Eq. (\ref{136b}).

 We conclude  that a possible choice for $w^0({\bf h})$ and
${\bf w}_{\bot}({\bf h})$ in Eqs. (\ref{154}) and (\ref{155}) is
$w^0({\bf h})={\bf w}_{\bot}({\bf h})=0$, but in the general case
there exists an infinite number of solutions satisfying the
conditions described in Sec.~\ref{S1.1}.

 To explicitly describe the ECO for a system of $N$
particles we have to calculate the explicit expressions for the
operators $j_{\alpha_1...\alpha_k}^{\mu}({\bf h})$. For this purpose
we have to know the explicit expressions for the operators
$A_{\alpha_1...\alpha_k}$ (see Sec.~\ref{S3.2}). As noted in
Sec.~\ref{S2.4}, the problem of constructing such operators is solved
in principle, but explicit expressions for the operators
$A_{\alpha_1...\alpha_k}$ in the point form have not been written so
far for the case $N>3$.

\section{Example: relativistic correction to the magnetic moment of
the deuteron}
\label{S4.5}

 It is well known that in the nonrelativistic approximation
the magnetic moment of the deuteron is given by
\begin{equation}
\mu_d^{nr}=\mu_p+\mu_n-\frac{3}{2}P_D(\mu_p+\mu_n-\frac{1}{2})
\label{e}
\end{equation}
where $\mu_d$, $\mu_p$, and $\mu_n$ are the magnetic moments of the
deuteron, proton and neutron in nuclear magnetons, "nr" means
nonrelativistic and $P_D$ is the probability of the D-state in the
deuteron. For most realistic nucleon-nucleon potentials the value of
$\mu_d^{nr}$ is less than the experimental value $\mu_d^{exp}=0.857$.
For example, $\mu_d^{nr}$ is equal to 0.843, 0.845, 0.847 and 0.849
for the Reid soft core, Argonne, Paris and Nijmegen potentials
respectively\cite{CKC}. For different versions of the Bonn potential
the value of $\mu_d^{nr}$ is in the range 0.852-0.856 \cite{CKC}, but
the attitude of physicists to the Bonn potential is controversial, and
we shall not discuss this question.

 In our approach the magnetic moment of the deuteron can be
calculated using the fact that $F_M(0)$ in Eq. (\ref{123}) is the
deuteron magnetic moment in units $e/2m_d$ where $e$ and $m_d$ are
the deuteron electric charge and mass respectively (see
Sec.~\ref{S4.1}). Therefore we have to calculate the matrix element
$\langle \chi"({\bf q})|{\hat {\bf j}}_{\bot}({\bf h})
|\chi'({\bf q})\rangle$ between two deuteron states in first order in
${\bf h}$ and compare the result with the quantity $$-\imath eF_M(0)
\langle \chi"({\bf q})|{\bf h}\times {\bf S}|\chi'({\bf q})\rangle .$$

\begin{sloppypar}
 Let us choose for ${\hat {\bf j}}_{\bot}({\bf h})$ the solution
corresponding to ${\bf w}_{\bot}({\bf h})=0$ (see Sec.~\ref{S4.2}).
Then ${\hat {\bf j}}_{\bot}({\bf h})={\bf j}_{\bot}({\bf h})$
and ${\bf j}_{\bot}({\bf h})$ is given by
Eq. (\ref{134}). We use $m$ to denote the nucleon mass and neglect the
difference between the masses of the proton and neutron. Since the
deuteron wave function is symmetrical under the interchange of spatial
and spin variables of the proton and neutron, we can calculate the
contribution of only the first term in Eq. (\ref{134}) where
$J_1^{\nu}(g_1,g_1')$ is taken from Eq. (\ref{121}) but for
the nucleon form factors $F_E$ and
$F_M$ in these expressions we must take the sum of the corresponding
quantities for the proton and neutron at zero momentum transfer ,
i.e.  $F_E^0=1$, $F_M^0=\mu_p+\mu_n$.
\end{sloppypar}

 As follows from the "minimal relativity principle"
{}~\cite{BrKuo,CoPi,Ko}, we can use for $\chi'({\bf q})$ and
$\chi"({\bf q})$ the deuteron wave functions calculated for the usual
phenomenological potentials (see also refs.~\cite{KoSt,CKC}).
Therefore we shall use the functions normalized as
\begin{equation}
\int\nolimits ||\chi({\bf q})||^2 d^3{\bf q} =
\int\nolimits[\varphi_0(q)^2+\varphi_2(q)^2]d^3{\bf q}= 1
\label{eee}
\end{equation}
where $\varphi_0(q)$ and $\varphi_2(q)$ are the radial wave function of
the S and D states in momentum representation. Then a direct
calculation using Eqs. (\ref{7}), (\ref{14}), (\ref{17}), (\ref{121}),
(\ref{133}) and (\ref{134}) shows that in first order in ${\bf h}$ the
action of ${\bf j}({\bf h})$ is given by
\begin{eqnarray}
&&{\bf j}({\bf h})\chi({\bf q})=\frac{1}{\omega}\{-4\imath F_M^0[({\bf
h}\omega -\frac{{\bf q}({\bf q}{\bf h})}{\omega+m})\times
{\bf s}_1-\nonumber\\
&&-\frac{\omega{\bf q}}{m(\omega+m)}
(({\bf h}\times{\bf q}){\bf s}_1)]+
2F_E^0({\bf q}-{\bf h}\omega)[1-\nonumber\\
&&-\imath\frac{({\bf h}\times{\bf q})}
{\omega+m}({\bf s}_1-{\bf s}_2)-
2\imath\omega \frac{({\bf h}\times{\bf q})}{m(\omega+m)}
{\bf s}_1]\}\chi({\bf q}-2{\bf h}\omega)
\label{4.43}
\end{eqnarray}
where $\omega=\omega({\bf q})=(m^2+{\bf q}^2)^{1/2}$.
We use $t_1$ and $t_2$ to denote the spin projections of the proton and
neutron. Then
the internal deuteron wave function describing the deuteron with the
polarization vector ${\bf e}$ has the form
\begin{equation}
\chi({\bf q},t_1,t_2)=\frac{1}{\sqrt{2}}[\varphi_0(q)\delta_{ik}-
\frac{1}{\sqrt{2}}(\delta_{ik}-3\frac{q_iq_k}{q^2})\varphi_2(q)]
e_k(\sigma_i\sigma_2)_{t_1t_2}
\end{equation}
where $q=|{\bf q}|$ and a sum over repeated indices $i,k=1,2,3$ is
assumed.

 The result of our calculation is the following. If $\epsilon$ is the
deuteron binding energy and $\mu_d=\mu_d^{nr}+\delta\mu_d$ where
$\delta\mu_d$ is the relativistic correction to the deuteron magnetic
moment in nuclear magnetons then
\begin{eqnarray}
\delta\mu_d&=&\frac{\epsilon}{m_d}\mu_d^{nr}-\frac{2m}{3m_d}\int\nolimits
\frac{q^2d^3{\bf q}}{\omega(q)+m}[\varphi_0(q)^2+\nonumber\\
&+&\frac{1}{\sqrt{2}}\varphi_0(q)\varphi_2(q)-\varphi_2(q)^2][\frac{F_M^0}
{\omega(q)}+\frac{F_E^0-F_M^0}{m}]
\label{ee}
\end{eqnarray}
This quantity is negligible for all the realistic potentials mentioned
above. For example, $\delta\mu_d=6\cdot 10^{-4}$ for the Reid soft core
potential.

 The relativistic correction to the deuteron magnetic moment was
considered by many authors. Usually $\delta\mu_d$ turned out to be
negative, but in the framework of approach considered in
refs.\cite{KoSt,CKC} $\delta\mu_d$ is of about a half of the quantity
$\mu_d^{exp}-\mu_d^{nr}$. However, the ECO's used in these approaches
do not satisfy the properties given by Eqs. (\ref{1}) and (\ref{2}). The
result given by Eq. (\ref{ee}) is obtained by using the ECO which does
satisfy these properties but among the solutions discussed in
Sec.~\ref{S4.2} the special solution with
${\bf w}_{\bot}({\bf h})=0$ has been chosen. We conclude that this choice
also does not solve the problem of the deuteron magnetic moment.

\chapter{Electromagnetic current operator in the instant form of
dynamics}
\label{C5}

\section{Systems of interacting particles in the instant form of
dynamics}
\label{S5.1}

 In the instant form it is convenient to use the realization of UIR's
of the Poincare group in the space of functions $\phi(g)$ with the
range in ${\cal D}(\bf s)$ and such that
\begin{equation}
(\phi,\phi)=\int\nolimits ||\phi({\bf
p})||^2\frac{d^3{\bf p}}{(2\pi)^3}\quad < \infty
\label{5.1}
\end{equation}
Then the generators of the UIR have the well known form (see, for
example, refs.~\cite{Wig,Nov,CP,mutze})
\begin{eqnarray}
{\bf P}={\bf p},\quad E=(m^2+{\bf p}^2)^{1/2},\quad
{\bf M}={\bf l}({\bf p})+{\bf s}, \nonumber\\
{\bf N}=-\imath(m^2+{\bf p}^2)^{1/4}\frac{\partial}{\partial
{\bf p}}(m^2+{\bf p}^2)^{1/4}+\frac{{\bf s}\times{\bf p}}{m+(m^2+
{\bf p}^2)^{1/2}}
\label{5.2}
\end{eqnarray}

 The Hilbert space $H_I$ for the representations describing systems
of $N$ free or interacting particles can be realized as the space of
functions $\phi({\bf p}_1,...{\bf p}_N)$ with the range in ${\cal
D}({\bf s}_1)\bigotimes ...\bigotimes {\cal D}({\bf s}_N)$ and such
that
\begin{equation}
\int\nolimits ||\phi({\bf p}_1,...{\bf p}_N)||^2
\prod_{i=1}^{N} d\rho_i^I({\bf p}_i) \quad <\quad \infty
\label{5.3}
\end{equation}
where (compare with Eq. (\ref{5}))
\begin{equation}
d\rho_i^I({\bf p}_i)=\frac{d^3{\bf p}_i}{2(2\pi)^3\omega_i({\bf
p}_i)}=m^2d\rho_i(g_i)
\label{5.4}
\end{equation}
(the subscript "I" means "instant").

 Instead of the variables ${\bf p}_1,...{\bf p}_N$ we introduce the
variables ${\bf P},{\bf k}_1,...{\bf k}_N$ where ${\bf P}={\bf
p}_1+\cdots +{\bf p}_N$ and ${\bf k}_i$ is the spatial part of the
4-vector $k_i=L[\alpha(P/M)]^{-1}p_i$ where $p_i$ is the momentum of the
$i$-th particle, $P=p_1+...+p_N$ and $M=|P|$. It is obvious that
$k_i$ defined in such a way is the same as in Eq. (\ref{51}).

 By analogy with Eq. (\ref{2.42}) we can show that
\begin{eqnarray}
&&\prod_{i=1}^{N}d\rho_i^I({\bf p}_i)=
\frac{d^3{\bf P}}{2(2\pi)^3(1+{\bf P}^2/M^2)^{1/2}}d\rho^I(int),\nonumber\\
&&d\rho^I(int)=(2\pi)^3\delta^{(3)}({\bf k}_1+\cdots +{\bf k}_N)
\prod_{i=1}^{N}d\rho_i^I({\bf k}_i)
\label{5.5}
\end{eqnarray}

 Let us define the "internal' space $H_{int}^I$ as the space of
functions $\chi({\bf k}_1,...{\bf k}_N)$ with the range in ${\cal D}
({\bf s}_1)\bigotimes\cdots \bigotimes {\cal D}({\bf s}_N)$ and such that
\begin{equation}
||\chi||^2=\int\nolimits ||\chi({\bf k}_1,...{\bf k}_N)||^2
d\rho^I(int)\quad < \infty
\label{5.6}
\end{equation}
and the space ${\tilde H}_I$ as the space of functions
${\tilde \phi}({\bf P})$ with the range in $H_{int}^I$ and such that
\begin{equation}
\int\nolimits ||\phi({\bf P})||_{int}^2d^3{\bf P}\quad < \infty
\label{5.7}
\end{equation}
Let $\cal U$ be the same operator as in Eq. (\ref{2.44}) but with $g_i$
replaced by $p_i/m_i$. Then ${\cal U}^I=(1+{\bf P}^2/M^2)^{1/4}{\cal
U}$ is the unitary operator from ${\tilde H}_I$ to $H_I$.

 The method of Sokolov packing operators in the instant form implies
that the generators $\hat {\Gamma_I^i}$ $(i=1,...10)$ of the
representation describing a system of $N$ interacting particles
should be written in the form $\hat {\Gamma_I^i}=A_I{\cal U}_I{\hat
{\tilde \Gamma_I^i}}{\cal U}_I^{-1}A_I^{-1}$ where the packing operator
$A_I$ commutes with ${\bf P}$ and ${\bf M}$ and the generators
${\hat {\tilde \Gamma_I^i}}$ in ${\tilde H}_I$ have the following
"canonical" form (compare with Eq. (\ref{5.2}))
\begin{eqnarray}
&&{\bf P}_I={\bf P},\quad {\hat E}_I=(({\hat M}_{int}^I)^2+{\bf
P}^2)^{1/2}, \quad {\bf M}_I={\bf l}({\bf P})+{\bf S}, \nonumber\\
&&{\hat {\bf N}}_I=-\imath(({\hat M}_{int}^I)^2+{\bf P}^2)^{1/4}
\frac{\partial}{\partial {\bf P}} (({\hat M}_{int}^I)^2+
{\bf P}^2)^{1/4}+\nonumber\\
&&+\frac{{\bf S}\times{\bf P}}{{\hat M}_{int}^I+
(({\hat M}_{int}^I)^2+ {\bf P}^2)^{1/2}}
\label{5.8}
\end{eqnarray}
Here the first expression implies that the momentum operator is
equal to the operator of multiplication by the variable ${\bf P}$
defined above, the spin operator ${\bf S}$ is the same as in
Eq. (\ref{73}) and the mass operator ${\hat M}_{int}^I$ in the instant
form acts only in $H_{int}^I$ and commute with ${\bf S}$. As already
noted, the problem of constructing the operators $A_I$ and
${\hat M}_{int}^I$ has been solved by Coester and Polyzou \cite{CP}
and Mutze \cite{mutze} (see also refs. \cite{KP,lev4,mutze1,lev2}).

\section{Unitary equivalence of the point and instant forms of
dynamics}
\label{S5.2}

 As already noted, the unitary equivalence of the three basic forms
of dynamics has been proved by Sokolov and Shatny \cite{SoSh}. The
key element of their approach is the construction of unitary
operators relating the "canonical" forms of the generators in these
forms of dynamics. In particular, to prove the unitary equivalence of
the point and instant forms we have to construct a unitary operator
$\Theta$ from $\tilde H$ to ${\tilde H}_I$ which transform the
generators defined by Eq. (\ref{73}) to the generators defined by
Eq. (\ref{5.8}). Following ref. \cite{SoSh} we shall seek $\Theta$ in
the form
\begin{equation}
\Theta =U_{IP}\xi (M)^{-1}\xi ({\hat M}_{int})
\label{5.9}
\end{equation}
where the unitary operators $\xi (M)$ and
$\xi ({\hat M}_{int})$ in $\tilde H$ are defined by the spectral
integrals (compare with Eq. (\ref{36}))
\begin{equation}
\xi ({\hat M}_{int})=\int\nolimits \xi (m)d{\hat e}_{int}(m),
\quad \xi (M)=\int\nolimits \xi (m)de(m)
\label{5.10}
\end{equation}
over the spectral measures of the operators ${\hat M}_{int}$ and $M$
respectively and $U_{IP}$ is the unitary operator from $\tilde H$ to
${\tilde H}_I$ defined as
\begin{equation}
{\tilde \phi}({\bf P})=U_{IP} {\tilde \varphi}({\bf G}) =
\frac{{\tilde \varphi}({\bf P}/M)}{(1+{\bf P}^2/M^2)^{1/4}m_1\cdots
m_N}
\label{5.11}
\end{equation}
The fact that $U_{IP}$ is unitary easily follows from Eqs. (\ref{2.43}),
(\ref{5.5}-\ref{5.7}). The operator $U_{IP}^{-1}$ is given by
\begin{equation}
{\tilde \varphi}({\bf G})=U_{IP}^{-1} {\tilde \phi}({\bf P}) =
m_1\cdots m_N(1+{\bf G}^2)^{1/4} {\tilde \phi}(M{\bf G})
\label{5.12}
\end{equation}

 As follows from Eqs. (\ref{73}), (\ref{5.8}), (\ref{5.9}), (\ref{5.11})
and (\ref{5.12}), the condition $\Theta {\hat M}_{int}{\bf
G}\Theta^{-1}={\bf P}$ implies that
\begin{equation}
\xi ({\hat M}_{int}){\hat M}_{int}{\bf G}\xi ({\hat M}_{int})=
\xi (M) M{\bf G} \xi (M)^{-1}
\label{5.13}
\end{equation}
This expression will be satisfied if for all $m$ belonging to the
spectra of the operators ${\hat M}_{int}$ and $M$
\begin{equation}
\xi(m)m{\bf G}\xi(m)^{-1}=m_0{\bf G}
\label{5.14}
\end{equation}
where $m_0$ is some constant \cite{SoSh}. This condition can be
satisfied if the actions of the operators $\xi(m)$ and $\xi(m)^{-1}$
are given by \cite{SoSh}
\begin{eqnarray}
&&\xi(m){\tilde \varphi}({\bf G})=(\frac{m}{m_0})^{3/2}\bigl(\frac{1+
{\bf G}^2}{1+m_0^2{\bf G}^2/m^2}\bigr)^{1/4}
{\tilde \varphi}(\frac{m_0}{m}{\bf G})\nonumber\\
&&\xi(m)^{-1}{\tilde \varphi}({\bf G})=(\frac{m}{m_0})^{3/2}
\bigl(\frac{1+ {\bf G}^2}{1+m^2{\bf G}^2/m_0^2}\bigr)^{1/4}
{\tilde \varphi}(\frac{m}{m_0}{\bf G})
\label{5.15}
\end{eqnarray}
As follows from Eqs. (\ref{2.43}), (\ref{5.5}) and (\ref{5.15}), the
operators $\xi(m)$ and $\xi(m)^{-1}$ are unitary and commute with the
operators ${\hat e}_{int}(m')$ and $e(m')$. Therefore the operators
defined by Eq. (\ref{5.10}) are unitary.

 Let $O$ be an operator in $H_{int}$. We symbolically represent the
action of $O$ in the form
\begin{equation}
O{\tilde \varphi}({\bf G},int)=\int\nolimits O(int,int')
{\tilde \varphi}({\bf G},int')d\rho(int')
\label{5.16}
\end{equation}
where $O(int,int')$ is the kernel of the operator $O$. Let us
introduce the operator
\begin{equation}
F_I\{O\}\equiv U_{IP}\xi(M)^{-1}O\xi(M)U_{IP}^{-1}
\label{5.17}
\end{equation}
Then, as follows from Eqs. (\ref{2.43}), (\ref{5.5}-\ref{5.7}),
(\ref{5.15}) and (\ref{5.16}), $F_I\{O\}$ is the operator in
$H_{int}^I$. If we write symbolically
\begin{equation}
F_I\{O\}{\tilde \phi}({\bf P},int)=\int\nolimits F_I\{O\}(int,int')
{\tilde \phi}({\bf P},int')d\rho^I(int')
\label{5.18}
\end{equation}
then the kernels of the operators $O$ and $F_I\{O\}$ are related as
\begin{equation}
F_I\{O\}(int,int')=\frac{2(MM')^{3/2}}{m_1^2\cdots m_N^2}O(int,int')
\label{5.19}
\end{equation}

 Since $\xi({\hat M}_{int})$ obviously commutes with ${\hat M}_{int}$,
then as follows from Eqs. (\ref{5.9}) and (\ref{5.17}), the mass
operators in the point and instant forms are related as
\begin{equation}
{\hat M}_{int}^I=\Theta {\hat M}_{int}\Theta^{-1}=
F_I\{{\hat M}_{int}\}
\label{5.20}
\end{equation}
In particular, using Eq. (\ref{5.19}) we can verify the fact mentioned
in Sec.~\ref{S2.3} that the mass operators found in that section and
in ref.\cite{lev5} are unitarily equivalent if $A_{ij}=1$.

 Using Eqs. (\ref{5.9}-\ref{5.12}), (\ref{5.15}) and (\ref{5.20}) one
can explicitly verify that not only the momentum generators, but also
the remaining 7 generators from the sets defined by Eqs. (\ref{73}) and
(\ref{5.8}) are related as
\begin{equation}
{\hat E}_I=\Theta{\hat E}\Theta^{-1},\quad {\bf M}_I=\Theta{\bf M}
\Theta^{-1},\quad {\hat {\bf N}}_I=\Theta {\hat {\bf N}} \Theta^{-1}
\label{5.21}
\end{equation}

 As shown in ref.\cite{SoSh}, if the operators $A$, ${\hat M}_{int}$,
and $\Theta$ are known then it is possible to determine $A_I$ and
conversely, if $A_I$, ${\hat M}_{int}^I$, and $\Theta$ are known then
it is possible to determine $A$. Therefore the operator ${\hat {\cal
U}}_{IP}=A_I{\cal U}\Theta{\cal U}^{-1}A^{-1}$ realizes the unitary
equivalence of the point and instant forms.

\section{Explicit construction of the ECO in the instant form of
dynamics}
\label{S5.3}

 If ${\hat J}_I^{\mu}(x)$ is the ECO in the instant form, then, by
analogy with Eq. (\ref{78}) we can write
\begin{equation}
{\hat J}_I^{\mu}(x)=exp(\imath {\hat P}_Ix){\hat J}_I^{\mu}(0)
exp(-\imath {\hat P}_Ix)
\label{5.22}
\end{equation}
and, by analogy with Eq (\ref{79}), one can easily show that
${\hat J}_I^{\mu}(0)$ must satisfy the properties
\begin{equation}
{\hat U}_I(l)^{-1}{\hat J}_I^{\mu}(0){\hat U}_I(l)=L(l)^{\mu}_{\nu}
{\hat J}_I^{\nu}(0), \qquad [{\hat P}_{I\mu},{\hat J}_I^{\mu}(0)]=0
\label{5.23}
\end{equation}
In turn, by analogy with Eq. (\ref{81}), we shall seek
${\hat J}_I^{\mu}(0)$ in the form
\begin{equation}
{\hat J}_I^{\mu}(0)=A_I{\cal U}_I{\hat J}_I^{\mu}{\cal U}_I^{-1}A_I^{-1}
\label{5.24}
\end{equation}
where ${\hat J}_I^{\mu}$ acts in ${\tilde H}_I$. Then it is obvious
from the results of Sec.~\ref{S5.2} that the operator
${\hat J}_I^{\mu}(x)$ will satisfy all the properties described in
Sec.~\ref{S1.1} if
\begin{equation}
{\hat J}_I^{\mu}=\Theta {\hat J}^{\mu} \Theta^{-1}
\label{5.29}
\end{equation}

 The action of ${\hat J}_I^{\mu}$ can be written in the form
\begin{equation}
{\hat J}_I^{\mu}{\tilde \phi}({\bf P})=\int\nolimits
{\hat J}_I^{\mu}({\bf P},{\bf P}'){\tilde \phi}({\bf P}')
\frac{d^3{\bf P}'}{(2\pi)^3}
\label{5.30}
\end{equation}
where the kernel ${\hat J}_I^{\mu}({\bf P},{\bf P}')$ is an operator
in $H_{int}^I$ for each fixed values of ${\bf P}$ and ${\bf P}'$. In
turn, by analogy with Eq. (\ref{91a}), the operator
${\hat J}_I^{\mu}({\bf P},{\bf P}')$ can be defined by the set of
operators ${\hat J}_I^{\mu}({\bf P},m;{\bf P}',m')$ such that
\begin{equation}
{\hat J}_I^{\mu}({\bf P},{\bf P}')=\int\nolimits \int\nolimits
d{\hat e}_{int}^I(m){\hat J}_I^{\mu}({\bf P},m;{\bf P}',m')
d{\hat e}_{int}^I(m')
\label{5.31}
\end{equation}
where ${\hat e}_{int}^I(m)$ is the spectral function of the operator
${\hat M}_{int}^I$.

 A direct calculation using Eqs. (\ref{82}), (\ref{91a}),
(\ref{5.9}-\ref{5.12}), (\ref{5.15}), (\ref{5.17}), (\ref{5.20}) and
(\ref{5.29}-\ref{5.31}) shows that if the operator
$${\hat J}^{\mu}({\bf G},m;{\bf G}',m')$$ in the point form is known
then
\begin{eqnarray}
&&{\hat J}_I^{\mu}({\bf P},m;{\bf P}',m')=\frac{(mm')^{1/2}}{[(m^2+{\bf
P}^2)(m^{'2}+{\bf P}^{'2})]^{1/4}}\cdot\nonumber\\
&&\cdot F_I\{{\hat J}^{\mu}(\frac{{\bf P}}{m},m;\frac{{\bf P}'}{m'},m')\}
\label{5.32}
\end{eqnarray}
Therefore Eqs. (\ref{5.19}), (\ref{5.31}) and (\ref{5.32}) make
it possible to explicitly determine the ECO in the instant form if the
problem of constructing the ECO is solved in the point form.

\section{Matrix elements of the ECO in the instant form of dynamics}
\label{S5.4}

 Since we require that one-particle states with a definite momentum
should be normalized as in Eq. (\ref{113}), then, as follows from
Eq. (\ref{5.1}), such states should be chosen in the form
\begin{equation}
|p',\sigma'\rangle_I=(2\pi)^3[2\omega({\bf p}')]^{1/2}\delta^{(3)}
({\bf p}-{\bf p}')\delta_{\sigma\sigma'}
\label{5.33}
\end{equation}

 To define the scattering states in the N-particle case we have to
solve the eigenvalue problem for the operator ${\hat M}_{int}^I$ in
the space $H_{int}^I$. Let $\chi'_I\in H_{int}^I$ be the internal
wave function of a bound state with the mass $M'$ and ${\bf P}'$ be
the momentum of this bound state. Then, as follows from
Eq. (\ref{5.8}), the wave function of such a state in the space $H_I$
can be written as
\begin{equation}
|P',\chi'\rangle_I =A_I{\cal U}_I(2\pi)^3(2E')^{1/2}
\delta^{(3)} ({\bf P}-{\bf P}')\chi'_I
\label{5.34}
\end{equation}
where $P^{'0}=E'=(M^{'2}+{\bf P}^{'2})^{1/2}$ is the total energy of
the bound state. It is clear from Eq. (\ref{5.1}) that such states will
be normalized as in Eq. (\ref{3.44}). As in the point form,
Eq. (\ref{5.34}) can be written not only if $\chi'_I$ is a bound state,
but also if $\chi'_I$ is a generalized eigenfunction of the
operator ${\hat M}_{int}^I$ with the eigenvalue $M'$. Therefore we
conclude that all matrix elements of the operator ${\hat
J}_I^{\mu}(x)$ can be expressed in terms of the quantities $_I\langle
P",\chi_I"|{\hat J}_I^{\mu}(x)|P',\chi_I'\rangle_I$.

As follows from
Eqs. (\ref{5.22}), (\ref{5.24}), (\ref{5.30}), and (\ref{5.34})
\begin{eqnarray}
&&_I\langle P",\chi_I"|{\hat J}_I^{\mu}(x)|P',\chi'_I\rangle_I=
2[(M^{"2}+{\bf P}^{"2})(M^{'2}+{\bf P}^{'2})]^{1/2}\cdot\nonumber\\
&&\cdot exp(\imath \Delta x)
_I\langle \chi_I"|{\hat J}_I^{\mu}({\bf P}",{\bf P}')|\chi'_I\rangle_I
\label{5.35}
\end{eqnarray}
where the matrix element on the right-hand-side must be calculated only
in the space $H_{int}^I$. If the problem of constructing the ECO is
solved in the point form then, as follows from Eq. (\ref{5.32}), we can
calculate the matrix elements of the ECO in the instant form. On the
other hand, as follows from Eqs. (\ref{5.9}) and (\ref{5.20}), if
$\chi'$ is an eigenfunction of the operator ${\hat M}_{int}$ with the
eigenvalue $M'$ then $\chi_I"=U_{IP}\xi(M)^{-1}\chi'$ is an
eigenfunction of the operator ${\hat M}_{int}^I$ with the same
eigenvalue. Therefore, as follows from Eqs. (\ref{5.17}) and
(\ref{5.32})
\begin{eqnarray}
_I\langle \chi_I"|{\hat J}_I^{\mu}({\bf P}",{\bf P}')|
\chi'_I\rangle_I&=&\frac{(M"M')^{1/2}}{[(M^{"2}+{\bf P}^{"2})(M^{'2}+
{\bf P}^{'2})]^{1/2}}\cdot\nonumber\\
&\cdot& \langle\chi"|{\hat J}^{\mu}(\frac{{\bf
P}"}{M"},\frac{{\bf P}'}{M'})|\chi'\rangle
\label{5.36}
\end{eqnarray}
Expressing ${\hat J}^{\mu}({\bf G}",{\bf G}')$ in terms of
$j^{\nu}({\bf h})$ according to Eq. (\ref{87}) and comparing
Eq. (\ref{3.45}) with Eqs. (\ref{5.35}) and (\ref{5.36}) we conclude that
\begin{equation}
\langle P",\chi"|{\hat J}^{\mu}(x)|P'\chi'\rangle =
_I\langle P",\chi_I"|{\hat J}_I^{\mu}(x)|P',\chi'_I\rangle_I
\label{5.37}
\end{equation}
This result shows that the matrix elements of the ECO do not depend on
the choice of the form of dynamics as it should be.

\chapter{Electromagnetic current operator in the front form of
dynamics}
\label{C6}
\section{Systems of interacting particles in the front form of
dynamics}
\label{S6.1}

 In the front form it is convenient to use the realization of UIR's
of the Poincare group in the space of functions $\phi({\bf p}_{\bot},p^+)$
with the range in ${\cal D}(\bf s)$ and such that
\begin{eqnarray}
&&(\phi,\phi)=\int\nolimits ||\phi({\bf
p}_{\bot},p^+)||^2d\rho^F({\bf p}_{\bot},p^+)<\infty,\nonumber\\
&&d\rho^F({\bf p}_{\bot},p^+)=\frac{d^2{\bf p}_{\bot}dp^+}{2(2\pi)^3p^+}
\label{6.1}
\end{eqnarray}
where the superscript "F" means "front". The spin variables in the front
form are defined assuming that the boosts are described not by the
matrices $\alpha(p/m)$ (see Eq. (\ref{7})) but by the matrices
$\beta(p/m)=\alpha(p/m)v(p/m)$ where
\begin{equation}
v(p/m)=exp\bigl(\frac{2\imath\epsilon_{jl}p^js^l}
{p_{\bot}}artg\frac{p_{\bot}}{m+\omega({\bf p})+p^z}\bigr)
\label{6.2}
\end{equation}
is the Melosh matrix \cite{Mel}. Here $p_{\bot}=|{\bf p}_{\bot}|$, a
sum over $j,l=x,y$ is assumed and
$\epsilon_{jl}$ has the components $\epsilon_{xy}=-\epsilon_{yx}=1, \,
\epsilon_{xx}=\epsilon_{yy}=0$. Then the generators of the UIR have the
well known form (see, for example, refs.~\cite{BarHal,Ter,LeySt,Fuda})
\begin{eqnarray}
&&P_F^{+} = p^{+},\quad{\bf P}_{F\bot}={\bf p}_{\bot},\quad P_F^-=p^-=
\frac{m^2+ {\bf p}_{\bot}^2}{2p^+}, \nonumber\\
&&M_F^{+-} = \imath p^+\frac{\partial}{\partial p^+},\quad
M_F^{+j}=-\imath p^+\frac{\partial}{\partial
p^j},\quad M_F^{xy}=l^z({\bf p}_{\bot})+s^z, \nonumber\\
&&M_F^{-j}=-\imath(p^j\frac{\partial}{\partial p^+}+
p^-\frac{\partial}
{\partial p^j})-\frac{\epsilon_{jl}}{p^+}(ms^l+p^ls^z)
\label{6.3}
\end{eqnarray}

 The Hilbert space $H_F$ for the representation of the Poincare group
describing a system of $N$ free or interacting
particles is realized in the space of functions
$\phi({\bf p}_{1\bot},p_1^+,...{\bf p}_{N\bot},p_N^+)$ with the range in
${\cal D}({\bf s}_1)\bigotimes\cdots \bigotimes {\cal D}({\bf s}_N)$
and such that
\begin{equation}
\int\nolimits ||\phi({\bf p}_{1\bot},p_1^+,...{\bf p}_{N\bot},p_N^+)||^2
\prod_{i=1}^{N} d\rho^F({\bf p}_{i\bot},p_i^+) \, <\, \infty
\label{6.4}
\end{equation}

  Instead of the variables ${\bf p}_{1\bot}$, $p_1^+$,...
${\bf p}_{N\bot}$, $p_N^+$ we introduce the variables ${\bf P}_{\bot}$,
$P^+$, ${\bf q}_1$,...${\bf q}_N$ where ${\bf P}_{\bot}={\bf p}_{1\bot}+
\cdots +{\bf p}_{N\bot}$,
$P^+=p_1^+\cdots p_N^+$ and ${\bf q}_i$ is the spatial part of the
4-vector $q_i=L[\beta(P/M)]^{-1}p_i$ where $P=p_1+...+p_N$ and $M=|P|$.
It is easy to see that ${\bf k}_i=L[v(P/M)]{\bf q}_i$.

 By analogy with Eq. (\ref{2.42}) one can show that
\begin{eqnarray}
&&\prod_{i=1}^{N}d\rho_i^F({\bf p}_{i\bot},p_i^+)=
d\rho^F({\bf P}_{\bot},P^+)d\rho^F(int),\nonumber\\
&&d\rho^F(int)=2(2\pi)^3
M\delta^{(3)}({\bf q}_1+\cdots +{\bf q}_N)
 \prod_{i=1}^{N}d\rho_i^F({\bf q}_{i\bot},q_i^+)
\label{6.5}
\end{eqnarray}

  Let us define the "internal' space $H_{int}^F$ as the space of
functions $\chi({\bf q}_1,...{\bf q}_N)$ with the range in
${\cal D}({\bf s}_1)\bigotimes\cdots \bigotimes {\cal D}({\bf s}_N)$
and such that
\begin{equation}
||\chi||^2=\int\nolimits ||\chi({\bf q}_1,...{\bf q}_N)||^2
d\rho^F(int)\quad < \infty
\label{6.6}
\end{equation}
and the space ${\tilde H}_F$ as the space of functions
${\tilde \phi}({\bf P}_{\bot},P^+)$ with the range in $H_{int}^F$ and
such that
\begin{equation}
\int\nolimits ||\phi({\bf P}_{\bot},P^+)||_{int}^2
d\rho^F({\bf P}_{\bot},P^+)\quad < \infty
\label{6.7}
\end{equation}

 The main reason for choosing $\beta(g)$ instead of $\alpha(g)$ (see,
for example, ref. \cite{KoTer}) is that $\beta(g)$ satisfies the
following important property
\begin{equation}
\beta(g)l\beta(L(l)^{-1}g)=1
\label{6.8}
\end{equation}
Owing to this property the front analog of the operator ${\cal U}$
given by Eq. (\ref{2.44}) is equal to unity and the space $H_F$ coincides
with ${\tilde H}_F$.

 The method of Sokolov packing operators in the front form implies
that the generators $\hat {\Gamma_F^i}$ $(i=1,...10)$ of the
representation describing a system of $N$ interacting particles
should be written in the form $\hat {\Gamma_F^i}=A_F{\hat
{\tilde \Gamma_F^i}}A_F^{-1}$ where the packing operator
$A_F$ commutes with $P^+,P^j,M^{+-},M^{+j},M^{xy}$ $(j=x,y)$, and the
generators ${\hat {\tilde \Gamma_F^i}}$ in $H_F$ have the following
"canonical" form (compare with Eq. (\ref{6.3}))
\begin{eqnarray}
&&P_F^+=P^+,\quad {\bf P}_{F\bot}={\bf P}_{\bot},\quad {\hat P}_F^-=
\frac{({\hat M}_{int}^F)^2+
{\bf P}_{\bot}^2}{2P^+}, \nonumber\\
&&M_F^{+-}=\imath P^+\frac{\partial}{\partial P^+},\quad
M_F^{+j}=-\imath P^+\frac{\partial}{\partial P^j},\quad
 M_F^{xy}=l^z({\bf P}_{\bot})+S_F^z, \nonumber\\
&&{\hat M}_F^{-j}=-\imath(P^j\frac{\partial}{\partial P^+}+
{\hat P}_F^-\frac{\partial}{\partial P^j})-
\frac{\epsilon_{jl}}{P^+}({\hat M}_{int}^FS_F^l+P^lS^z)
\label{6.9}
\end{eqnarray}
Here the first expression implies that the generator $P_F^+$ is equal to
the operator of multiplication by the variable $P^+$ defined above and
the second expression should be understood analogously.

The expressions
for the system spin and mass operators ${\bf S}_F$ and
${\hat M}_{int}^F$ differ from the corresponding expressions in the
point and instant forms. The explicit expressions for ${\bf S}_F$ and
${\hat M}_{int}^F$ have been first derived by Terentiev \cite{Ter} for
systems of two particles and by Berestetskii and Terentiev \cite{BerTer}
for systems of three particles. As pointed out in ref. \cite{BKT}, the
result of ref. \cite{BerTer} is inaccurate, and the authors of ref.
\cite{BKT} have derived a correct result (see also refs.
\cite{Coest3,lev2}). The explicit expressions defining the operator
$A_F$ have been derived in refs. \cite{lev6,Coest3,lev2}.

 The investigation of the electromagnetic properties of mesons and
baryons in the framework of constituent quark model in the front form
was carried out by several authors (see, for example, refs.
\cite{Azn,Jaus,Schl} and references cited therein). However it was
usually assumed that the ECO is the sum of the quark ECO's.

\begin{sloppypar}
\section{Unitary equivalence of the point and front forms of dynamics}
\label{S6.2}
\end{sloppypar}

 In Chap.~\ref{C2} we have constructed the RQM in the point form
assuming that the spin variables are defined using the matrix
$\alpha(g)$ (see Eq. (\ref{7})) and the internal momenta are defined by
Eq. (\ref{51}). However it is possible to construct the description in
the point form in terms of the light-front variables. This can be done
by choosing the matrix $\beta(g)$ for the definition
of the spin variables, the quantities ${\bf q}_i$ for the definition
of the internal momentum variables and the 4-velocity $G$ as the
"external" variable. In this case it is convenient to choose
$(G^+,{\bf G}_{\bot})$ as three independent quantities defining $G$.

 In particular, the UIR can be realized
in the space of functions $\varphi({\bf g}_{\bot},g^+)$
with the range in ${\cal D}(\bf s)$ and such that
\begin{equation}
(\varphi,\varphi)=\int\nolimits ||\phi({\bf
g}_{\bot},g^+)||^2d\rho^F({\bf g}_{\bot},g^+)\,<\,\infty
\label{6.10}
\end{equation}
while the generators of the UIR have the form (compare with
Eq. (\ref{6.3}))
\begin{eqnarray}
&&P^+=mg^+,\quad{\bf P}_{\bot}=m{\bf g}_{\bot},\quad P^-=mg^-=
m\frac{1+{\bf g}_{\bot}^2}{2g^+}, \nonumber\\
&&M^{+-} = \imath g^+\frac{\partial}{\partial g^+},\quad
M^{+j}=-\imath g^+\frac{\partial}{\partial g^j},\quad
M^{xy}=l^z({\bf g}_{\bot})+s^z, \nonumber\\
&&M^{-j}=-\imath(g^j\frac{\partial}{\partial g^+}+g^-
\frac{\partial}{\partial g^j})-\frac{\epsilon_{jl}}{g^+}(s^l+g^ls^z)
\label{6.11}
\end{eqnarray}

 As follows from Eq. (\ref{6.5})
\begin{eqnarray}
&&\prod_{i=1}^{N}d\rho_i^F({\bf g}_{i\bot},g_i^+)=
d\rho^F({\bf G}_{\bot},G^+)d\rho^P(int),\nonumber\\
&&d\rho^P(int)=\frac{M^2}{m_1^2\cdots m_n^2}d\rho^F(int)
\label{6.12}
\end{eqnarray}

   Let $H_{int}^P$ be the space of
functions $\chi({\bf q}_1,...{\bf q}_N)$ with the range in
${\cal D}({\bf s}_1)\bigotimes\cdots \bigotimes {\cal D}({\bf s}_N)$
and such that
\begin{equation}
||\chi||^2=\int\nolimits ||\chi({\bf q}_1,...{\bf q}_N)||^2
d\rho^P(int)\, <\, \infty
\label{6.13}
\end{equation}
and ${\tilde H}_P$ be the space of functions
${\tilde \varphi}({\bf G}_{\bot},G^+)$ with the range in $H_{int}^P$
and such that
\begin{equation}
\int\nolimits ||\varphi({\bf G}_{\bot},G^+)||_{int}^2 d\rho^F({\bf
G}_{\bot},G^+)\, <\, \infty
\label{6.14}
\end{equation}
Then by analogy with the results of the preceding section it can be
shown that the space of the tensor product of the UIR's describing
particles 1,...N is just the space $H_P={\tilde H}_P$.

 In the given case the method of Sokolov packing operators implies
that the generators $\hat {\Gamma_i}$ $(i=1,...10)$ of the
representation describing a system of $N$ interacting particles
should be written in the form $\hat {\Gamma_i}=A_P{\hat
{\tilde \Gamma}_i}A_P^{-1}$ where the packing operator
$A_P$ commutes with $U(l)$ (as it should be in the point form) and the
generators ${\hat {\tilde \Gamma}^i}$ in $H_P$ have the following
"canonical" form (compare with Eq. (\ref{6.9}))
\begin{eqnarray}
&&{\hat P}^+={\hat M}_{int}^PG^+,\quad {\hat{\bf P}}_{\bot}=
{\hat M}_{int}^P{\bf G}_{\bot},\quad
{\hat P}^-={\hat M}_{int}^P G^-=\nonumber\\
&&={\hat M}_{int}^P\frac{1+{\bf G}_{\bot}^2}{2G^+},\quad
M^{+-}=\imath G^+\frac{\partial}{\partial G^+},
\quad M^{+j}=-\imath G^+\frac{\partial}{\partial G^j},\nonumber\\
&& M^{xy}=l^z({\bf G}_{\bot})+S_F^z, \quad
M^{-j}=-\imath(G^j\frac{\partial}{\partial G^+}+
G^-\frac{\partial}{\partial G^j})-\nonumber\\
&&-\frac{\epsilon_{jl}}{G^+}(S_F^l+G^lS^z)
\label{6.15}
\end{eqnarray}
Here the spin operator ${\bf S}_F$ has the same form as in Eq.
(\ref{6.9}) and ${\hat M}_{int}^P$ is the mass operator acting in
$H_{int}^P$.

 The operators ${\bf S}_F$, ${\hat M}_{int}^P$ and $A_P$ must be unitarily
equivalent to the operators ${\bf S}$, ${\hat M}_{int}$ and $A$
respectively (see Sec.~\ref{S2.4}) since both these sets describe
representations in the point form for different choices of the spin
and momentum variables. The explicit expressions for the operators
$({\bf S}_F$, ${\hat M}_{int}^P$, $A_P)$ and the unitary operator relating
the sets (${\bf S}_F$, ${\hat M}_{int}^P$) and
(${\bf S}$, ${\hat M}_{int}$)
in the cases $N=2$ and $N=3$ have been derived in refs.
\cite{lev1,lev2}.

 The remainder of this section is an analog of Sec.~\ref{S5.2}.
According to Sokolov and Shatny \cite{SoSh}, the key element in
proving the unitary equivalence of the point and front forms is the
construction of the unitary operator $\Theta$ such that
\begin{equation}
\Theta {\hat {\tilde \Gamma}^i} \Theta^{-1}=
{\hat {\tilde \Gamma}_F^i}
\label{6.16}
\end{equation}

 Let $U_{FP}$ be  the unitary operator from $H_P$ to $H_F$ such that
\begin{eqnarray}
&&U_{FP} \varphi({\bf G}_{\bot},G^+) =
\frac{\varphi({\bf P}_{\bot}/M,P^+/M)}{m_1\cdots m_N},\nonumber\\
&&U_{FP}^{-1} \phi({\bf P}_{\bot},P^+)=
m_1\cdots m_N{\tilde \phi}(M{\bf G}_{\bot},MG^+)
\label{6.17}
\end{eqnarray}
The fact that $U_{FP}$ is unitary easily follows from Eqs. (\ref{6.6}),
(\ref{6.7}), (\ref{6.12}-\ref{6.14}). By analogy with ref.~\cite{SoSh}
we shall seek $\Theta$ in the form
\begin{equation}
\Theta =U_{FP}\xi (M)^{-1}\xi ({\hat M}_{int}^P)
\label{6.18}
\end{equation}
where the unitary operators $\xi (M)$ and $\xi ({\hat M}_{int}^P)$ in
$H_P$ are defined by the spectral integrals analogous to those in
Eq. (\ref{5.10}). To avoid misunderstanding we note that the operators
$\Theta$ and $\xi(m)$ in this section obviously differ from the
corresponding operators in Sec.~\ref{S5.2}. However we use the same
notations as in Sec.~\ref{S5.2} in order to demonstrate that both
construction are analogous.

 Following ref. \cite{SoSh} we define the operators $\xi (m)$ and
$\xi (m)^{-1}$ as (compare with Eq. (\ref{5.15}))
\begin{eqnarray}
&&\xi (m)\varphi({\bf G}_{\bot},G^+)=\frac{m_0}{m}
\varphi(\frac{m_0}{m}{\bf G}_{\bot},\frac{m_0}{m}G^+)\nonumber\\
&&\xi (m)^{-1}\varphi({\bf G}_{\bot},G^+)=\frac{m}{m_0}
\varphi(\frac{m}{m_0}{\bf G}_{\bot},\frac{m}{m_0}G^+)
\label{6.19}
\end{eqnarray}
where $m_0>0$ is some constant. The fact that these operators are
unitary easily follows from Eq. (\ref{6.12}).

  Let $O$ be an operator in $H_{int}$. We symbolically represent the
action of $O$ in the form
\begin{equation}
O \varphi({\bf G}_{\bot},G^+,int)=\int\nolimits O(int,int')
\varphi({\bf G}_{\bot},G^+,int')d\rho^P(int')
\label{6.20}
\end{equation}
where $O(int,int')$ is the kernel of the operator $O$. Let us
introduce the operator
\begin{equation}
F_F\{O\}\equiv U_{FP}\xi(M)^{-1}O\xi(M)U_{FP}^{-1}
\label{6.21}
\end{equation}
Then, as follows from Eqs. (\ref{6.6}), (\ref{6.7}),
(\ref{6.12}-\ref{6.14}), (\ref{6.17}) and (\ref{6.19}-\ref{6.21}),
$F_F\{O\}$ is the operator in $H_{int}^F$. If we write symbolically
\begin{eqnarray}
&&F_F\{O\}\phi({\bf P}_{\bot},P^+,int)=\nonumber\\
&&=\int\nolimits F_F\{O\}(int,int')
\phi({\bf P}_{\bot},P^+,int')d\rho^F(int')
\label{6.22}
\end{eqnarray}
then the kernels of the operators $O$ and $F_f\{O\}$ are related as
\begin{equation}
F_F\{O\}(int,int')=\frac{MM'}{m_1^2\cdots m_N^2}O(int,int')
\label{6.23}
\end{equation}

 Now using Eqs. (\ref{6.19}) and (\ref{6.21}) we can explicitly verify
that Eq. (\ref{6.16}) is satisfied if
\begin{equation}
{\hat M}_{int}^F=F_F\{{\hat M}_{int}^P\}
\label{6.24}
\end{equation}
In other words, the operator defined by Eq. (\ref{6.18}) is the
operator realizing the unitary equivalence of the sets defined by
Eqs. (\ref{6.15}) and (\ref{6.9}).

 As shown in ref.\cite{SoSh}, if the operators $A_P$, ${\hat M}_{int}^P$,
and $\Theta$ are known then it is possible to determine $A_F$ and
conversely, if $A_F$, ${\hat M}_{int}^F$, and $\Theta$ are known then
it is possible to determine $A_P$. Therefore the operator ${\hat {\cal
U}}_{FP}=A_F\Theta A_P^{-1}$ realizes the unitary
equivalence of the point and front forms.

 The advantages of the front form in the Feynman diagram approach
were first pointed out by Weinberg \cite{Wein} (see also refs.
\cite{LepBr,Namysl}). At present different physicists believe that
important problems in the strong interaction theory can be solved by
using quantum field theory in the front form (see, for example, refs.
\cite{GlPer,GlWil,Br} and references cited therein). The major
difference between this approach and conventional ones is in treating
the vacuum problem.

 In RQM the vacuum problem does not arise and there are no principal
differences between the front form and other ones. One might think
that the front form has technical advantages since the generators for
the two-body problem can be easily transformed to the form
(\ref{6.9}) without using Wigner rotations. However in cases of three
and more particles the front form has serious practical disadvantages
since the spin operators for the system as a whole necessarily depend
on interactions (see, for example, ref. \cite{lev2}). In any case,
the results of ref. \cite{SoSh} show that in RQM the three basic
forms are unitarily equivalent, and therefore we can stress once more
that in the framework of RQM the choice of the form is only the
matter of convenience but not the matter of principle.

\section{Explicit construction of the ECO in the front form of
dynamics} \label{S6.3}

 If ${\hat J}_F^{\mu}(x)$ is the ECO in the front form, then, by
analogy with Eq. (\ref{78}) we can write
\begin{equation}
{\hat J}_F^{\mu}(x)=exp(\imath {\hat P}_Fx){\hat J}_F^{\mu}(0)
exp(-\imath {\hat P}_Fx)
\label{6.25}
\end{equation}
and, by analogy with Eq. (\ref{79}), one can easily show that
${\hat J}_F^{\mu}(0)$ must satisfy the properties
\begin{equation}
{\hat U}_F(l)^{-1}{\hat J}_F^{\mu}(0){\hat U}_F(l)=L(l)^{\mu}_{\nu}
{\hat J}_F^{\nu}(0), \qquad [{\hat P}_{F\mu},{\hat J}_F^{\mu}(0)]=0
\label{6.26}
\end{equation}
In turn, by analogy with Eq. (\ref{81}), we shall seek
${\hat J}_F^{\mu}(0)$ in the form
\begin{equation}
{\hat J}_F^{\mu}(0)=A_F{\hat J}_F^{\mu}A_F^{-1}
\label{6.27}
\end{equation}
Then it is obvious from the results of Sec.~\ref{S6.2} that the
operator
${\hat J}_F^{\mu}(x)$ will satisfy all the properties described in
Sec.~\ref{S1.1} if
\begin{equation}
{\hat J}_F^{\mu}=\Theta {\hat J}^{\mu} \Theta^{-1}
\label{6.28}
\end{equation}

 Let us note however that in order to explicitly calculate the ECO in
the front form using such a prescription we have to determine first
the action of the operator ${\hat J}^{\mu}$ in the point form in
terms of light front variables. It is easy to see that the
scheme of constructing the operator ${\hat J}^{\mu}$ described in
Chaps.~\ref{C3} and ~\ref{C4} can be also used for constructing the
operator ${\hat J}^{\mu}$ in these variables. Let us briefly discuss
some aspects of such a construction.

 The one-particle ECO in the light front variables can be constructed
by analogy with the construction in the usual variables (see
Sec.~\ref{S4.1}). For example, the expression defining the ECO for
a spin 1/2 particle is the same as Eq. (\ref{121}) but the usual Dirac
byspinor $u(p,\sigma)$ should be replaced by the light front byspinor
$w(p,\sigma)$. In the spinorial representation of the Dirac $\gamma$
matrices $u(p,\sigma)$ is the byspinor with the components
$\sqrt{m}(\alpha(p/m)\chi(\sigma),\alpha(p/m)^{-1*}\chi(\sigma))$
where $\chi(\sigma)$ is the usual spinor with the $z$ projection of
the spin equal to $\sigma$ while
$w(p,\sigma)$ is the byspinor with the components
$$\sqrt{m}(\beta(p/m)\chi(\sigma),\beta(p/m)^{-1*}\chi(\sigma)).$$

 In the general case the action of ${\hat J}^{\mu}$ can be again
written in the form of Eq. (\ref{82}) where ${\hat M}_{int}$ is the
mass operator in the light front variables and ${\hat J}^{\mu}(G,G')=
{\hat J}^{\mu}({\bf G}_{\bot},G^+;{\bf G}_{\bot}',G^{'+})$ (note that
$d\rho(G)=d\rho^F({\bf G}_{\bot},G^+)$). Then Eq. (\ref{84}) will be
valid if $\alpha$ is replaced by $\beta$. However since $\beta(G)$
satisfies Eq. (\ref{6.8}), Eq. (\ref{84}) in the light front variables
has a much more simple form
\begin{equation}
{\hat J}^{\mu}(G,G')=L(l)^{\mu}_{\nu}
{\hat J}^{\nu}(L(l)^{-1}G,L(l)^{-1}G')
\label{6.29}
\end{equation}

 Now we use $\beta(G,G')$ to denote $\beta((G+G')/|G+G'|)\in$SL(2,C)
and $L_F(G,G')$ to denote the Lorentz transformation
$L[\beta(G,G')]$. Instead of $f$ and $f'$ defined by Eq. (\ref{85})
we introduce
\begin{equation}
f_F=L_F(G,G')^{-1}G, \qquad f_F'=L_F(G,G')^{-1}G'
\label{6.30}
\end{equation}
Then it is easy to see that
\begin{equation}
f_F^0=f_F^{'0}=f^0, \quad {\bf h}_F=-{\bf h}_F'=
L[v(\frac{G+G'}{|G+G'|})]^{-1}{\bf h}
\label{6.31}
\end{equation}

 By analogy with Eq. (\ref{87}) we can express ${\hat J}^{\mu}(G,G')$
in terms of the operator ${\hat j}^{\nu}({\bf h}_F)$ depending on one
three-dimensional vector ${\bf h}_F$. However as follows from
Eq. (\ref{6.29}), the corresponding expression has a much more simple
form than Eq. (\ref{87}):
\begin{equation}
{\hat J}^{\mu}(G,G')=L_F(G,G')^{\mu}_{\nu}{\hat j}^{\nu}({\bf h}_F)
\label{6.32}
\end{equation}

 On the other hand, the choice of the light front variables has some
serious practical disadvantages. In particular, the expressions
describing the transformation of ${\hat j}^{\nu}({\bf h}_F)$ relative
to the spatial rotations and the space reflection have a much
more complicated form than Eqs. (\ref{89}) and (\ref{112a}).

 Let us note that the quantity ${\hat j}^0(0)$ in the light front
variables also must be equal to the electric charge of the system
under consideration, and a possible solution of the problem of
constructing the ECO can be obtained assuming that the operators
$w^0({\bf h}_F)$ and ${\bf w}_{\bot}({\bf h}_F)$ in the light front
variables are equal to zero. Here ${\bf w}_{\bot}({\bf h}_F)$ is the
part of ${\bf w}({\bf h}_F)$ orthogonal to ${\bf h}_F$.

 Now we return to the front form. Here the action of ${\hat
J}_F^{\mu}$ can be written as
\begin{equation}
{\hat J}_F^{\mu}\phi({\bf P}_{\bot},P^+)=\int\nolimits
{\hat J}_F^{\mu}({\bf P}_{\bot},P^+;{\bf P}_{\bot}',P^{'+})
\phi({\bf P}_{\bot}',P^{'+})d\rho^F({\bf P}_{\bot}',P^{'+})
\label{6.33}
\end{equation}
In turn, by analogy with Eq. (\ref{91a}), the operator
$${\hat J}_F^{\mu}({\bf P}_{\bot},P^+;{\bf P}_{\bot}',P^{'+})$$ can be
defined by the set of operators
$${\hat J}_F^{\mu}({\bf P}_{\bot},P^+,m;{\bf P}_{\bot}',P^{'+},m')$$
such that
\begin{eqnarray}
&&{\hat J}_F^{\mu}({\bf P}_{\bot},P^+;{\bf P}_{\bot}',P^{'+})=
\int\nolimits \int\nolimits d{\hat e}_{int}^F(m)\cdot\nonumber\\
&&\cdot {\hat J}_F^{\mu}({\bf P}_{\bot},P^+,m;{\bf P}_{\bot}',P^{'+},m')
d{\hat e}_{int}^F(m')
\label{6.34}
\end{eqnarray}
where ${\hat e}_{int}^F(m)$ is the spectral function of the operator
${\hat M}_{int}^F$.

 On the other hand, by analogy with Eq. (\ref{91a}), the kernel
$${\hat J}^{\mu}({\bf G}_{\bot},G^+;{\bf G}_{\bot}',G^{'+})$$ can be
defined by the set of operators
$${\hat J}^{\mu}({\bf G}_{\bot},G^+,m;{\bf G}_{\bot}',G^{'+},m')$$
such that
\begin{eqnarray}
&&{\hat J}^{\mu}({\bf G}_{\bot},G^+;{\bf G}_{\bot}',G^{'+})=
\int\nolimits \int\nolimits d{\hat e}_{int}^P(m)\cdot\nonumber\\
&&\cdot {\hat J}^{\mu}({\bf G}_{\bot},G^+,m;{\bf G}_{\bot}',G^{'+},m')
d{\hat e}_{int}^P(m')
\label{6.35}
\end{eqnarray}
where ${\hat e}_{int}^P(m)$ is now the spectral function of the operator
${\hat M}_{int}^P$ in the light front variables. Then a direct
calculation using Eqs. (\ref{6.12}), (\ref{6.17}-\ref{6.24}),
(\ref{6.28}) and (\ref{6.33}-\ref{6.35}) shows that
\begin{eqnarray}
&&{\hat J}_F^{\mu}({\bf P}_{\bot},P^+,m;
{\bf P}_{\bot}',P^{'+},m')=\nonumber\\
&&=2(mm')^{1/2}F_F\{{\hat J}^{\mu}(\frac{{\bf P}_{\bot}}{m},\frac{P^+}{m},m;
\frac{{\bf P}_{\bot}'}{m'},\frac{P^{'+}}{m'},m')\}
\label{6.36}
\end{eqnarray}

 We conclude that if the ECO in the point form is constructed in
terms of the light front variables then Eqs. (\ref{6.25}), (\ref{6.27}),
(\ref{6.33}), (\ref{6.34}), and (\ref{6.36}) (where $F_F$ is defined by
Eq. (\ref{6.23})) make it possible to explicitly construct the ECO in
the front form.

\section{Matrix elements of the ECO in the front form of dynamics}
\label{S6.4}

 The one-particle states in the light front variables are usually
chosen in the form
\begin{equation}
|p',\sigma'\rangle_F=2(2\pi)^3p^{'+}\delta^{(2)}({\bf p}_{\bot}-
{\bf p}_{\bot}')\delta(p^+-p^{'+})\delta_{\sigma\sigma'}
\label{6.37}
\end{equation}
Then as follows from Eq. (\ref{6.1})
\begin{equation}
_F\langle p",\sigma"|p',\sigma'\rangle_F=2(2\pi)^3p^{'+}
\delta^{(2)}({\bf p}_{\bot}"-
{\bf p}_{\bot}')\delta(p^{"+}-p^{'+})\delta_{\sigma"\sigma'}
\label{6.38}
\end{equation}
It is easy to see that this normalization is equivalent to the
normalization in the usual variables given by Eq. (\ref{113}).

 To define the scattering states in the N-particle case we have to
solve the eigenvalue problem for the operator ${\hat M}_{int}^F$ in
the space $H_{int}^F$. Let $\chi'_F\in H_{int}^F$ be the internal
wave function describing a scattering state. If $\chi_F'$ is the
eigenfunction of the operator ${\hat M}_{int}^F$ with the eigenvalue
$M'$ and the scattering state is the eigenstate of the operators
${\bf P}_{\bot}$ and $P^+$ with the eigenvalues
${\bf P}_{\bot}'$ and $P^{'+}$ respectively then, as follows from
Eq. (\ref{6.9}), the wave function of such a state in the space $H_F$
can be written as
\begin{equation}
|P',\chi_F'\rangle_F=A_F2(2\pi)^3P^{'+}\delta^{(2)}({\bf P}_{\bot}-
{\bf P}_{\bot}')\delta(P^+-P^{'+})\chi_F'
\label{6.39}
\end{equation}
It is obvious from Eq. (\ref{6.9}) that this state is the eigenstate of
the operator ${\hat P}^-$ with the eigenvalue $P^{'-}=(M^{'2}+
{\bf P}_{\bot}^{'2})/2P^{'+}$. As follows from Eq. (\ref{6.5}), the
normalization condition for the wave function given by Eq. (\ref{6.39})
has the form
\begin{eqnarray}
&&_F\langle P",\chi_F"|P',\chi_F'\rangle_F=
2(2\pi)^3P^{'+}\delta^{(2)}({\bf P}_{\bot}"-{\bf P}_{\bot}')\cdot\nonumber\\
&&\cdot \delta(P^{"+}-P^{'+})\,_F\langle\chi_F"|\chi_F'\rangle_F
\label{6.40}
\end{eqnarray}
where the last scalar product should be calculated only in the space
$H_{int}^F$.

 By analogy with the considerations in Secs.~\ref{S3.4} and
{}~\ref{S5.4} we conclude that all matrix elements of the operator ${\hat
J}_F^{\mu}(x)$ can be expressed in terms of the quantities $_F\langle
P",\chi_F"|{\hat J}_F^{\mu}(x)|P',\chi_F'\rangle_F$.
As follows from Eqs. (\ref{6.25}-\ref{6.28}), (\ref{6.33}) and
(\ref{6.34})
\begin{eqnarray}
&&_F\langle P",\chi_F"|{\hat  J}_F^{\mu}(x)|P',\chi'_F\rangle_F
=exp(\imath \Delta x)\, _F\langle \chi_F"|
\int\nolimits \int\nolimits d{\hat e}_{int}^F(m)\cdot\nonumber\\
&&\cdot {\hat J}_F^{\mu}({\bf P}_{\bot}",P^{"+},m;{\bf P}_{\bot}',
P^{'+},m')d{\hat e}_{int}^F(m')|\chi'_F\rangle_F
\label{6.41}
\end{eqnarray}
where $\Delta=P"-P'$ and the matrix element on the
right-hand-side must be calculated only in the space $H_{int}^F$.

 On the other hand, as follows from Eqs. (\ref{82}), (\ref{3.45}) and
(\ref{6.35}), in the point form the matrix elements of the ECO in
terms of the light front variables can be written as
\begin{eqnarray}
&&_P\langle P",\chi_P"|{\hat  J}^{\mu}(x)|P',\chi'_P\rangle_P=
2(M"M')^{1/2}exp(\imath \Delta x)\cdot\nonumber\\
&&\cdot _P\langle \chi_P"|
\int\nolimits \int\nolimits d{\hat e}_{int}^P(m)
{\hat J}^{\mu}({\bf G}_{\bot}",G^{"+},m;
{\bf G}_{\bot}',G^{'+},m')\cdot\nonumber\\
&&\cdot d{\hat e}_{int}^P(m')|\chi'_P\rangle_P
\label{6.42}
\end{eqnarray}
If $\chi_P'\in H_{int}^P$ is the eigenfunction of the operator ${\hat
M}_{int}^P$ with the eigenvalue $M'$ then, as follows from
Eqs. (\ref{6.21}) and (\ref{6.24}),
$$\chi_F'=U_{FP}\xi(M)^{-1}\chi_P'$$
is the eigenfunction of the operator ${\hat M}_{int}^P$ with the same
eigenvalue. Taking into account the definition of the operation $F_F$
(see Eq. (\ref{6.21})) and Eq. (\ref{6.36}) we conclude that
\begin{equation}
_F\langle P",\chi_F"|{\hat  J}_F^{\mu}(x)|P',\chi'_F\rangle_F
=_P\langle P",\chi_P"|{\hat  J}^{\mu}(x)|P',\chi'_P\rangle_P
\label{6.43}
\end{equation}
This result shows that the matrix elements of the ECO do not depend on
the choice of the form of dynamics as it should be.

\chapter{Discussion and conclusions}
\label{C7}

 Let us discuss the main results of the present paper.

 In Chap.~\ref{C2} we explicitly construct the description of systems
of two and three particles in the point form of relativistic
dynamics. For this purpose we derive explicit expressions for the
Sokolov packing operators in the case of particles with arbitrary
spin. We also note that it is possible to explicitly describe systems
with any number of particles.

 In Chap.~\ref{C3} the problem of constructing the operator
${\hat J}^{\mu}(x)$ is reduced to the problem of constructing the
operator ${\hat j}^{\nu}({\bf h})$. The results of this chapter are
based on the assumption that the representation of the Poincare group
for the system under consideration is realized in the point form.
However we do not assume that the system is described in the
framework of RQM.  Therefore the results can be in principle applied
also to the problem of constructing the ECO for systems of quantized
fields.

 It has been known for a long time that the matrix elements of the
one-particle ECO become especially simple in the Breit frame, i.e. in
the reference frame where the momenta of the initial and final states
${\bf p}'$ and ${\bf p}"$ satisfy the relation ${\bf p}"+{\bf p}'=0$.
This relation can also be written as ${\bf g}"+{\bf g}'=0$ where $g'$
and $g"$ are the 4-velocities in the initial and final states. There
exists an obvious analogy between the Breit frame and the c.m.frame
of two interacting particles. The latter is defined by the condition
${\bf p}_1+{\bf p}_2=0$, where ${\bf p}_i$ is the momentum of particle
$i$, and the interaction between the particles has the simplest form
just in the c.m.frame. The difference between the Breit frame and the
c.m.frame is that the former is defined for one and the same particle
in the initial and final states while the latter is defined for two
different particles.

 If the system under consideration has the mass spectrum consisting
of more than one point then the conditions ${\bf P}"+{\bf P}'=0$ and
${\bf G}"+{\bf G}'=0$ are not equivalent. The Lorentz transformation
which transfers the 4-vectors $P'$ and $P"$ to the reference frame
where ${\bf P}"+{\bf P}'=0$ depends not only on ${\bf P}'$ and ${\bf
P}"$ but also on the masses $M'$ and $M"$ in the initial and final
states. Therefore for different pairs $(M',M")$ there exist different
Lorentz transformations transferring $P'$ and $P"$ to the reference
frame where ${\bf P}"+{\bf P}'=0$. On the other hand, the
Lorentz transformations transferring $G'$ and $G"$ to the reference
frame where ${\bf G}"+{\bf G}'=0$ depends only on ${\bf G}'$ and
${\bf G}"$. Taking also into account that the representation
operators of the Lorentz group do not depend on interactions in the
point form we conclude that the transition from the operator
${\hat J}^{\mu}(G",G')$ to the operator ${\hat j}^{\nu}({\bf h})$
discussed in Sec.~\ref{S3.1} is much more simple than the analogous
transitions in the instant and front forms.

 The vector ${\bf f}$ defined by Eq. (\ref{85}) is the velocity analog
of the momentum in the c.m.frame. We have seen in Chap.~\ref{C2} that
the role of the external variables in the point form is played by
velocities, not momenta, but the role of the internal variables may
play momenta, as usual. It is possible to consider such a
version of the point form where the role of the internal variables is
also played by velocities. Such a version was first considered by
Ruijgrok and coauthors \cite{Ru,RuGr}. It is easy to show that the
ordinary approach to RQM can be formulated in terms of the Ruijgrok
approach and {\it vica versa}. However we see that in the problem of
constructing the ECO the analog of the Ruijgrok approach has
considerable advantages.

 Combining the results of Chaps.~\ref{C2} and \ref{C3} we derive in
Chap.~\ref{C4} the exact solution for the ECO in the framework of
RQM. This solution is not unique since it is easy to write down a
variety of operators $w^0({\bf h})$ and ${\bf w}_{\bot}({\bf h})$
satisfying Eq. (\ref{136b}) and cluster separability.
The fact that relativity and current conservation do not impose
considerable restrictions on the operator ${\hat J}^{\mu}(x)$ was
noted by several authors \cite{Bentz,GrRi,KorSh,pol1}. For example, in
ref.\cite{pol1} the problem of constructing the operator
${\hat J}^{\mu}(x)$ was studied on the language of matrix elements of
this operator and it was shown that after taking into account the
constraints imposed by relativity and current conservation, there would
exist a minimal (usually infinite) set of unconstrained matrix elements.

 Though the solution of the problem under consideration is not unique,
the very fact that the ECO satisfying the properties specified in
Sec.~\ref{S1.1}  can be explicitly constructed seems very important.
In particular we expect that for various electromagnetic processes
the theoretical predictions obtained by using a correct ECO will
considerably differ from the predictions obtained by using ECO's not
satisfying the above properties. Our solution for the ECO is most
general in the sense that if we assume that interactions between the
constituents can be
described in the framework of RQM then any model for the ECO should
correspond to a certain choice of the operators $w^0({\bf h})$ and
${\bf w}_{\bot}({\bf h})$. At the same time it is necessary to
investigate what additional constraints should be imposed on the ECO in
order to make the solution unique.

 Our solution for the ECO is described in the point form of dynamics
while, as noted in Sec.~\ref{S1.2}, the most popular forms are the
instant and front ones. One might try to construct the ECO in this
forms using only the representation operators of the Poincare group
in the corresponding form. However, as argued above, such a
construction is expected to be more difficult than in the point form.
In Chaps.  ~\ref{C5} and ~\ref{C6} we show that these difficulties
can be bypassed with the help of the results by Sokolov and Shatny
{}~\cite{SoSh} who have proved that all three basic forms of dynamics
are unitarily equivalent. Namely, using the unitary operators
relating these forms we can explicitly construct the ECO in the
instant and the front forms if the solution in the point form has
been already constructed. Analogously, using the unitary operators of
ref.~\cite{Fuda1} relating the $\xi$-picture and the front form we
can construct the ECO in the $\xi$-picture considered in
ref.~\cite{Fuda1}.

 We conclude that the results of this paper can be the basis for
systematic calculations of different electromagnetic observables in
nuclear physics and constituent quark models. Such calculations are
necessary for explaining the existing experimental data and planning
future experiments on powerful electron accelerators.

\vskip 1em
\begin{center} {\bf Acknowledgments} \end{center}
\vskip 1em

\begin{sloppypar}
 The author is very grateful to L.A.Kondratyuk and S.N.Sokolov for
numerous valuable discussions and to V.A.Karmanov, A.Yu.Korchin,
A.I.Machavariani and H.J.Weber for useful remarks. This work was
supported by grant No. 93-02-3754 from the Russian Foundation for
Fundamental Research.
\end{sloppypar}


\begin{thebibliography}{200}
\bibitem{Sieg} A.F.Siegert, Phys.Rev. 52 (1937) 787.
\bibitem{sok1} S.N.Sokolov, Teor.Mat.Fiz. 23 (1975) 355.
\bibitem{sok2} S.N.Sokolov, Teor.Mat.Fiz. 36 (1978) 193.
\bibitem{CP} F.Coester and W.N.Polyzou, Phys.Rev. D26 (1982) 1348.
\bibitem{mutze} U.Mutze, Habilitationssrift Univ. Munchen (1982);
Phys.Rev. D29 (1984) 2255.
\bibitem{lev1} F.M.Lev, Fortshr.Phys. 31 (1983) 75.
\bibitem{Wig1} E.Wigner, Got.Nachr. (1932) 546.
\bibitem{Sch} J.Schwinger, Phys.Rev. 82 (1951) 914.
\bibitem{Bentz} W.Bentz, Nucl.Phys. A446 (1985) 678.
\bibitem{GrRi} F.Gross and D.O.Riska, Phys.Rev. C36 (1987) 1928.
\bibitem{NaKo} H.W.L.Naus and J.H.Koch, Phys.Rev. C39 (1989) 1907.
\bibitem{Ohta} K.Ohta, Nucl.Phys. A495 (1989) 564; Phys.Rev. C40
(1989) 1335.
\bibitem{KorSh} A.Yu.Korchin and A.V.Shebeko, Yad.Fiz. 54 (1991) 357.
\bibitem{Anis} V.V.Anisovich, M.N.Kobrinsky, D.I.Melikhov and
A.V.Sarantsev, Nucl.Phys. A544 (1992) 747.
\bibitem{GarHu} H.Hyuga and M.Gari, Nucl.Phys. A274 (1976) 333.
\bibitem{Ris} D.O.Riska, Phys.Rep. 181 (1989) 207.
\bibitem{GrHen} F.Gross and H.Henning, Nucl.Phys. A537 (1992) 344.
\bibitem{HenSa} H.Henning, P.U.Sauer and W.Theis, Nucl.Phys. A537
(1992) 367.
\bibitem{UnkHof} M.Unkelbach and H.M.Hoffman, Nucl.Phys. A549 (1992)
550.
\bibitem{Gian} M.M.Giannini, Rept.Progr.Phys. 54 (1991) 453.
\bibitem{Friar} J.L.Friar, Ann.Phys. (NY) 104 (1977) 380; Phys.Rev.
C22 (1980) 796.
\bibitem{TsuRi} K.Tsushima,D.O.Riska and P.G.Blunden, Nucl.Phys. A559
(1993) 543.
\bibitem{HuTj1} E.Hummel and J.A.Tjon, Phys.Rev. C42 (1990) 423.
\bibitem{HuTj2} M. van der Schaar et.al. Phys.Rev.Lett. 68 (1992) 776.
\bibitem{Adam} Jr.Adam, E.Truhlik and D.Adamova, Nucl.Phys. A492
(1991) 556.
\bibitem{ArSa} H.Arenhovel and M.Sanzone, Few-Body Syst.Suppl. 3 (1991) 1.
\bibitem{GolAr} H.Goller and H.Arenhovel, Few-Body Syst. 13 (1992) 117.
\bibitem{Tamura} K.Tamura, T.Niwa, T.Sato and H.Ohtsubo, Nucl.Phys.
A536 (1992) 597.
\bibitem{MoRi} B.Mosconi and P.Ricci, Few-Body Syst.Suppl. 5 (1992) 36.
\bibitem{MoPaRi} B.Mosconi, J.Pauschenwein and P.Ricci, Few-Body
Syst.Suppl. 6 (1992) 223; Phys.Rev. C48 (1993) 332.
\bibitem{KleBr} D.J.Klepacki, Y.E.Kim and R.A.Brandenburg, Nucl.Phys. A550
(1992) 53.
\bibitem{SchSa} R.W.Schulze and P.U.Sauer, Phys.Rev. C48 (1993) 38.
\bibitem{Giusti} C.Giusti and F.D.Pacati, Nucl.Phys. A336 (1980) 427.
\bibitem{CoGlLee} W.Glockle, T.S.-H.Lee and F.Coester, Phys.Rev. C33
(1986) 709.
\bibitem{GodIs} Godfrey and N.Isgur, Phys.Rev. D32 (1985) 189.
\bibitem{CaIs} S.Capstick and N.Isgur, Phys.Rev. D34 (1986) 2809.
\bibitem{IsSm} N.Isgur and C.H.Llevellyn Smith, Phys.Rev.Lett. 52
(1984) 1080; Nucl.Phys. B317 (1989) 526.
\bibitem{JaKi} O.C.Jacob and L.S.Kisslinger Phys.Lett. B243 (1990) 323.
\bibitem{Ito} H.Ito, W.W.Back and F.Gross, Phys.Lett. B248 (1990) 28.
\bibitem{KrTr} A.F.Krutov and V.E.Troitsky, J.Phys. 19 (1993) L127.
\bibitem{Dir} P.A.M.Dirac, Rev.Mod.Phys. 21 (1949) 392.
\bibitem{KP} B.D.Keister and W.Polyzou, Adv.Nucl.Phys. 21 (1991) 225.
\bibitem{lev2} F.M.Lev, Rivista Nuovo Cimento 16 (1993) No.2, 1.
\bibitem{pol1} W.N.Polyzou and W.H.Klink, Ann.Phys. 185 (1988) 369.
\bibitem{O1} H.Osborn, Nucl.Phys. B38 (1972) 429.
\bibitem{CO} F.Coester and A.Ostebee, Phys. Rev. C11 (1975) 1836.
\bibitem{C} F.Coester, Lect. Notes Phys. 162 (1982); PHY-6747-TH-90,
Argonne (1990).
\bibitem{O2} H.Osborn, Phys. Rev. 176 (1968) 1523.
\bibitem{Ger} S.B.Gerasimov, Yad.Fiz. 2 (1965) 598.
\bibitem{DH} D.Drell and A.S.Hearn, Phys.Rev.Lett. 16 (1966) 908.
\bibitem{BP} S.J.Brodsky and J.Primack, Ann.Phys. (N.Y) 52 (1969) 315.
\bibitem{ClCo} F.E.Close and L.A.Copley, Nucl.Phys. B19 (1970) 477.
\bibitem{ClOs} F.E.Close and H.Osborn, Phys.Rev. D2 (1970) 2127.
\bibitem{FK} L.L.Foldy and R.A.Krajcik, Phys.Rev. D10 (1974) 1777.
\bibitem{Fr} J.L.Friar, Ann.Phys. (N.Y) 81 (1973) 332.
\bibitem{OF} R.K.Osborn and L.L.Foldy, Phys. Rev. 79 (1950) 795.
\bibitem{G} F.Gross, Phys.Rev. 142 (1966) 1025.
\bibitem{FS} L.L.Frankfurt  and  M.I.Strikman, Phys.Rep. 76 (1981) 215.
\bibitem{GrKo} I.L.Grach and L.A.Kondratyuk, Yad.Fiz. 39 (1984) 316.
\bibitem{CCKP} P.L.Chung, F.Coester, B.D.Keister and W.N.Polyzou,
Phys.Rev. C37 (1988) 2000.
\bibitem{GrKoFS} I.L.Grach, L.A.Kondratyuk, M.I.Strikman and
L.L.Frankfurt, Phys.Rev.Lett. 62 (1989) 387.
\bibitem{Keister} B.D.Keister, Phys.Rev. C43 (1991) 229.
\bibitem{CCP1} P.L.Chung, F.Coester and W.N.Polyzou, Phys.Lett. B205
(1988) 545.
\bibitem{CC} P.L.Chung and F.Coester, Phys.Rev. D44 (1991) 229.
\bibitem{W} H.J.Weber, Phys. Lett. (1992) 14; Ann. Phys. (N.Y.) (1991)
417.
\bibitem{KS} V.A.Karmanov and A.V.Smirnov, Nucl.Phys. A546 (1992)
691; Preprint Lebedev Physical Inst., Moscow (October,1992).
\bibitem{Gl} S.Glazek, Acta Phys. Pol. B14 (1983) 893.
\bibitem{WS1} M.Warns,
H.Schroeder, W.Pfeil and H.Rollnik, Z. Phys. C45 (1990) 613.
\bibitem{CL} F.Close and Zhenping Li, Phys. Rev. D42 (1990) 2194.
\bibitem{WS2} M.Warns, H.Schroeder, W.Pfeil and H.Rollnik,
Z.Phys. 45 (1990) 627; M.Warns, W.Pfeil and H.Rollnik, Phys.
Rev. D42 (1990) 2215.
\bibitem{LC} Zhenping Li and F.Close, Phys. Rev. D42 (1990) 2207.
\bibitem{CAP1} S.Capstick, Phys.Rev. D46 (1992) 1965; Phys. Rev. D46
(1992) 2864.
\bibitem{CAP2} S.Capstick and B.D.Keister, Phys. Rev. D46 (1992) 84.
\bibitem{Li} Zhenping Li, Phys.Rev. D47 (1993) 4114.
\bibitem{lev3} F.M.Lev, Nucl.Phys.A, to be published.
\bibitem{sok3} S.N.Sokolov, "Coordinates in relativistic mechanics".
In: "{\it Proceedings of the 7th seminar onproblems in high energy
physics and quantum field theory}" (1984) 85. IHEP, Protvino, 1984.
\bibitem{SoSh} S.N.Sokolov and A.M.Shatny, Teor.Mat.Fiz. 37 (1978) 291.
\bibitem{Karm} V.A.Karmanov, Nucl.Phys. B166 (1980) 378; Nucl.Phys. A362
(1981) 331; Sov.J.Part.Nucl. 19 (1988) 525.
\bibitem{Fuda1} M.G.Fuda, Phys.Rev. D44 (1991) 1880.
\bibitem{lev4} F.M.Lev, J.Phys. 17 (1984) 2047; Nucl.Phys. A433
(1985) 605.
\bibitem{Wig} E.P.Wigner, Ann. of Math. 40 (1939) 1; Proc. Amer.
Phil.Soc. 93 (1949) 521; Rev.Mod.Phys. 29 (1957) 255.
\bibitem{Nov} Yu.V.Novozhilov, {\it Introduction to the Theory of
Elementary Particles} (Nauka, Moscow, 1972).
\bibitem{BT} Bakamdjian and L.H.Thomas, Phys.Rev. 92 (1953) 1300.
\bibitem{lev5} F.M.Lev, Phys.Rev. D49 (1994) 603.
\bibitem{Coest2} F.Coester, Helv.Phys.Acta 38 (1965) 7.
\bibitem{BKT} B.L.G.Bakker, L.A.Kondratyuk and M.V.Terent'ev,
Nucl.Phys. B158 (1979) 497.
\bibitem{GKM} Gudavadze, T.Kopaleishvili and A.Machavariani, Proc.
Tbilisi University 242 (1983) 153.
\bibitem{Coest3} F.Coester, In: "{\it Constraint's theory and
relativistic dynamics}". Proc. Workshop Firenze (1986) 157. Eds.
G.Longhi and L.Lusanna --- World Scientific, Singapore, 1987.
\bibitem{sok4} S.N.Sokolov, Doklady Akademii Nauk SSSR 233 (1977) 575.
\bibitem{sok5} S.N.Sokolov, Doctoral dissertation (1975), IHEP, Protvino.
\bibitem{lev6} F.M.Lev, Yad. Fiz. 37 (1983) 1056.
\bibitem{Dix} J.Dixmier. {\it Les algebres operateurs dans l'espace
hilbertien}.  Gauthier-Villars (1969) Paris.
\bibitem{Naim} M.A.Naimark. {\it Normalized rings}. Nauka (1968) Moscow.
\bibitem{CKC} P.L.Chung, B.D.Keister and F.Coester,
Phys.Rev. C39 (1989) 1544.
\bibitem{BrKuo} G.E.Brown, A.D.Jackson and T.T.S.Kuo, Nucl.Phys. A133
(1969) 481.
\bibitem{CoPi} F.Coester, S.C.Pieper and F.J.D.Serduke, Phys.Rev.
C11 (1974) 1.
\bibitem{Ko} L.A.Kondratyuk, Doctoral dissertation (1985) ITEP, Moscow.
\bibitem{KoSt} L.A.Kondratyuk and M.I.Strikman, Nucl.Phys. A426 (1984) 575.
\bibitem{mutze1} U.Mutze, Private communication of January 25, 1986.
{\it The solution of the N-body problem presented in this communication is
described in ref.\cite{lev2}}.
\bibitem{Mel} H.Melosh, Phys.Rev. D9 (1974) 1095.
\bibitem{BarHal} K.Bardakci and M.B.Halpern, Phys.Rev. 176 (1968) 1686.
\bibitem{Ter} M.V.Terentiev, Yad.Fiz. 24 (1976) 207.
\bibitem{LeySt} H.Leytwiller and J.Stern, Ann.Phys. (N.Y) 112 (1978) 94;
Nucl.Phys. B133 (1978) 115.
\bibitem{Fuda} M.G.Fuda, Phys.Rev. C36 (1987) 1489; Ann.Phys. (N.Y)
197 (1990) 265.
\bibitem{KoTer} L.A.Kondratyuk and M.V.Terentiev, Yad.Fiz. 31 (1980)
1087.
\bibitem{BerTer} V.B.Berestetskii and M.V.Terentiev, Yad.Fiz. 24
(1976) 1044.
\bibitem{Azn} I.G.Aznauryan, A.S.Bagdasaryan and N.L.Ter-Isaakyan,
Sov.J.Nucl.Phys. 39 (1984) 66; I.G.Aznauryan and K.A.Oganessyan,
Phys.Lett. B249 (1991) 309.
\bibitem{Jaus} W.Jaus, Phys.Rev. D41 (1990) 3394; Phys.Rev. D44 (1991)
2851.
\bibitem{Schl} F.Schlumpf, Phys.Rev. D47 (1993) 4114.
\bibitem{Wein} S.Weinberg, Phys.Rev. 150 (1966) 1313.
\bibitem{LepBr} G.P.Lepage and S.J.Brodsky, Phys.Rev. D22 (1980) 2157.
\bibitem{Namysl} J.M.Namyslowski, Prog.Part.Nucl.Phys. 14 (1984) 49.
\bibitem{GlPer} S.D.Glasek and R.J.Perry, Phys.Rev. D45 (1992) 3734;
3740.
\bibitem{GlWil} S.D.Glasek and K.G.Wilson, Phys.Rev. D47 (1993) 4657.
\bibitem{Br} S.J.Brodsky, G.McCartor, H.C.Pauli and S.S.Pinsky,
Particle World 3 (1993) 109.
\bibitem{Ru} Th.W.Ruijgrok, In: {\it Quantum theory of particles and
fields} (1983) 117. Eds. B.Jancewicz and J.Lukierski. World
Scientific, Singapore, 1983.
\bibitem{RuGr} E.H. de Groot and Th.W.Ruijgrok. The work presented to
the 25th International School on Theoretical Physics in Zakopane
(Poland), 1985.
\end{thebibliography}
\end{document}